\documentclass[article]{JHEP3}
\usepackage{epsfig,subfigure}
\usepackage{graphicx}
\usepackage{amssymb}
\usepackage{amsmath}
\usepackage{booktabs,multirow,tabularx}
\usepackage{slashed}
\usepackage{cite}
\usepackage{caption}
\usepackage{float}
\usepackage{placeins}
\usepackage{rotating}
\usepackage{lscape}
\usepackage[Q=yes,pverb-linebreak=no]{examplep}

\DeclareGraphicsExtensions{.eps}
\graphicspath{{./}}

\newcommand\sss{\scriptscriptstyle}

\newcommand{\gev}{\,\textrm{GeV}}

\newcommand{\TO}{\rightarrow}

\newcommand{\tth}{t\bar{t}H}
\newcommand{\tthc}{\lambda_{\tth}}
\newcommand{\pptth}{pp\TO\tth}

\newcommand{\asa}[2]{\alpha_s^{#1}\alpha^{#2}}
\newcommand{\ord}{\mathcal{O}}

\def\beq{\begin{equation}}
\def\beqn{\begin{eqnarray}}
\def\eeq{\end{equation}}
\def\eeqn{\end{eqnarray}}
\def\beal{\begin{align}}
\def\endal{\end{align}}

\newcommand\mydot{\!\cdot\!}

\newcommand\bb{\bar{b}}
\newcommand\tb{\bar{t}}

\newcommand\as{\alpha_{\sss S}}

\newcommand\aem{\alpha}

\newcommand\muF{\mu_{\sss F}}
\newcommand\muR{\mu_{\sss R}}

\newcommand\bt{\bar{t}}
\newcommand\bq{\bar{q}}
\newcommand\bqp{\bar{q}^\prime}
\newcommand\aNLO{{\sc\small MadGraph5\_aMC@NLO}}
\newcommand\UFO{{\sc\small UFO}}
\newcommand\MLf{{\sc\small MadLoop5}}
\newcommand\ML{{\sc\small MadLoop}}
\newcommand\CutTools{{\sc\small CutTools}}
\newcommand\OL{{\sc\small OpenLoops}}
\newcommand\MadFKS{{\sc\small MadFKS}}
\newcommand{\pt}{p_{\sss T}}
\newcommand{\Ht}{H_{\sss T}}

\author{S.~Frixione$^a$, V.~Hirschi$^b$, D.~Pagani$^c$, H.-S.~Shao$^d$, 
M.~Zaro$^{ef}$\\
$^a$ PH Department, TH Unit, CERN, CH-1211 Geneva 23, Switzerland\\
$^b$ SLAC, National Accelerator Laboratory,\\
$\phantom{^b}$ 2575 Sand Hill Road, Menlo Park, CA 94025-7090, USA\\
$^c$ Centre for Cosmology,  Particle Physics and Phenomenology (CP3),\\
$\phantom{^c}$ Universit\'e Catholique de Louvain, B-1348 Louvain-la-Neuve, 
Belgium\\
$^d$ Department of Physics and State Key Laboratory of Nuclear Physics 
and Technology,\\ 
$\phantom{^d}$ Peking University, Beijing 100871, China\\
$^e$ Sorbonne Universit\'es, UPMC Univ. Paris 06, UMR 7589, LPTHE,\\
$\phantom{^e}$ F-75005, Paris, France\\
$^f$ CNRS, UMR 7589, LPTHE, F-75005, Paris, France
}
\abstract{We present the calculation of the next-to-leading contribution
of order $\alpha_S^2\alpha^2$ to the production of a Standard Model Higgs
boson in association with a top-quark pair at hadron colliders.
All effects of weak and QCD origin are included, whereas those of QED 
origin are ignored. We work in the {\sc\small MadGraph5\_aMC@NLO} 
framework, and discuss sample phenomenological applications at a 8, 13, 
and 100~TeV $pp$ collider, including the effects of the dominant 
next-to-leading QCD corrections of order $\alpha_S^3\alpha$.
}
\title{Weak corrections to Higgs hadroproduction in association with a 
top-quark pair}
\preprint{
 CERN-PH-TH-2014-123\\
 CP3-14-49\\
 }
\begin{document}

\section{Introduction\label{sec:intro}} 
After the discovery of a new particle with a mass of 125 GeV at the 
LHC~\cite{Aad:2012tfa,Chatrchyan:2012ufa}, the determination of its physical
properties has become one of the main priorities in high-energy particle
physics. The recent results of the ATLAS and CMS collaborations strongly
suggest that this particle is the Higgs scalar boson emerging from the
Brout--Englert--Higgs mechanism in the Standard Model 
(SM)~\cite{Englert:1964et,Higgs:1964ia,Higgs:1964pj}. In particular, the
analyses of the distributions of its decay products point to a dominantly 
CP-even scalar~\cite{Chatrchyan:2012jja, Aad:2013xqa}, and the fitted values 
for its couplings are compatible with those predicted by the 
SM~\cite{ATLAS:2013sla,Chatrchyan:2013lba}.
However, the precision of the current measurements still leaves room for Beyond
the Standard Model (BSM) scenarios. Thus, more accurate measurements, and
their theoretical-result counterparts with matching precision, are necessary 
to fully understand the nature of this new particle.

In this context, an accurate determination of the $\tth$ coupling $\tthc$ 
is of great interest; among other things, it might also help shed light on 
the possible interplay of the Higgs boson and the top quark in the Electroweak
Symmetry Breaking (EWSB) mechanism.  To this purpose, the associated production
of a Higgs boson and a top-quark pair at the LHC $(\pptth)$ offers an unique
opportunity, since its cross section is directly proportional to $\tthc^2$
at the leading order (LO). While a direct measurement of this production 
mode has not been achieved so far, mostly because of its small cross section 
and large background contamination, several searches have already been 
published by ATLAS and CMS~\cite{TheATLAScollaboration:2013mia,ATLAS:2012cpa,
Chatrchyan:2013yea,CMS:2013tfa,CMS:2013sea,CMS:2013fda,CMS:2012qaa}, 
which use a variety of decay channels.

As is the case for all processes involving the Higgs, the effects of the
radiative corrections to $\tth$ production must be taken into account in
order to achieve a realistic phenomenological description. Next-to-leading
order (NLO) QCD effects, which were calculated more than ten years
ago~\cite{Beenakker:2001rj,Beenakker:2002nc,Dawson:2002tg,Dawson:2003zu},
increase the total cross section by a factor of about $1.3$ (at a 13 
TeV LHC). Moreover, they significantly diminish the dependence of the cross
section on the factorization and renormalization scales. More recently,
in refs.~\cite{Frederix:2011zi,Garzelli:2011vp} NLO QCD corrections have 
also been matched to parton showers, and the differences with 
respect to fixed-order results are generally found to be small for
inclusive and infrared-insensitive observables. Computations with the 
same level of perturbative accuracy have also been performed
for the dominant background process to $\tth$ production at the LHC, 
namely the production of a top-quark pair in association with a 
bottom-quark pair, without~\cite{Bredenstein:2009aj,Bevilacqua:2009zn,
Bredenstein:2010rs} or with~\cite{Kardos:2013vxa,Cascioli:2013era}
parton-shower matching.

Besides QCD, electroweak (EW) effects might also lead to significant
modifications of the LO predictions, particularly 
for differential distributions. 
So far, EW NLO corrections have been calculated for all of the other main
Higgs production channels: gluon fusion~\cite{Djouadi:1994ge, Aglietti:2004nj,
Degrassi:2004mx, Actis:2008ug}, vector-boson
fusion~\cite{Ciccolini:2007ec,Ciccolini:2007jr} and
$VH$ associated production~\cite{Ciccolini:2003jy}. For the case of $\tth$,
they are currently not known. The purpose of this work is to amend this
situation, and to present the first calculation of such corrections;
similarly to what has been done as a first step in the case of $t\bt$ 
hadroproduction~\cite{Beenakker:1993yr,Bernreuther:2005is,Kuhn:2005it,
Bernreuther:2006vg,Kuhn:2006vh}, we do not include in our results effects 
of QED origin (dealt with in later papers~\cite{Hollik:2007sw,
Bernreuther:2010ny,Hollik:2011ps,Kuhn:2011ri} for $t\bt$). Our computations
are performed in the automated \aNLO\ framework~\cite{Alwall:2014hca}.

The motivation for separating weak and QED corrections 
to the $\pptth$ cross section is twofold. Firstly, it is only
weak corrections which can induce effects whose size may be of the 
same order as the QCD ones in those regions of the phase space associated 
with large invariants, owing to the possible presence of Sudakov 
logarithms (see e.g.~refs.~\cite{Ciafaloni:1998xg,Ciafaloni:2000df,
Denner:2000jv,Denner:2001gw}),
which compensate the stronger suppression of $\aem$ w.r.t.~that
of $\as$. Secondly, weak corrections spoil the trivial dependence of 
LO and NLO QCD cross sections on $\tthc$. This is because they also depend 
on the couplings of the Higgs to the $W$ and $Z$ bosons, and on the
Higgs self-coupling, while QED corrections do not involve any of these
additional couplings. Thus, if one wants to assess possible contaminations due
to higher-order effects in the extraction of $\tthc$, one may start by 
focusing on weak-only corrections. 

From a technical viewpoint, by excluding QED corrections one also simplifies
the structure of the calculation, and in particular that relevant to the
subtraction of the infrared singularities. We note, however, that such
a simplification is not particularly significant in the context of an
automated approach that is already able to deal with the more complicated
situation of QCD-induced subtractions, as is the case for \aNLO. It is indeed
weak corrections that introduce several elements of novelty in our automated
approach (see e.g.~sect.~4.3 of ref.~\cite{Alwall:2014hca}); 
the possibility of testing them in $t\bt H$ production is yet another 
motivation to pursue the computation we are presenting in this paper.

We point out that, in all cases where both QCD and EW effects are
relevant, the structure of the cross section at any given perturbative
order (LO, NLO, and so forth) is a linear combination of terms, each
of which factorises a coupling-constant factor of the type $\asa{n}{m}$,
with $n+m$ a constant.
Owing to the numerical hierarchy $\aem\ll\as$, it is natural to organise
this combination in decreasing powers of $\as$. The leading term 
has the largest power of $\as$ and the smallest of $\aem$, and at the NLO 
it is identified with QCD corrections. The next term has one power less
in $\as$, and an extra one in $\aem$: it is what is often
identified with EW corrections. This is something of a misnomer, because QCD
effects contribute to this term as well, and because it renders difficult
the classification of the remaining terms (i.e., beyond the second) in the 
linear combination mentioned before. Although slightly annoying, this 
is not a major problem, being a question 
of (naming) conventions and, especially, because the computations
of terms beyond the second require a massive effort which one {\em assumes}
not to be justified in view of the coupling hierarchy. However, if
such computations can be performed automatically, no effort will be
required, and the validity of that assumption can be explicitly checked.
At present, we are facing precisely the situation in which the automated
calculation of all the $\asa{n}{m}$-proportional terms, both at the LO and
the NLO, is becoming feasible. It is therefore useful to reconsider
the general structure of a cross section that involve both strong and EW
interactions, and to define more precisely what is dealt with in
the context of higher-order computations.

The rest of this paper is organised as follows. In sect.~\ref{sec:org}
we discuss the implications of having to treat both QCD and EW effects
as small perturbations; although the ideas we introduce are general,
we concentrate on $t\bt H$ production to be definite, with further
details on its calculation with \aNLO\ given in sect.~\ref{sec:calc}. 
In sect.~\ref{sec:results} we present our phenomenological results; in
particular, we compare EW and QCD effects at the NLO. We conclude 
and give our outlook in sect.~\ref{sec:concl}.

\section{Organisation of the calculation\label{sec:org}}
\begin{figure}[h]
  \begin{center}
    \epsfig{figure=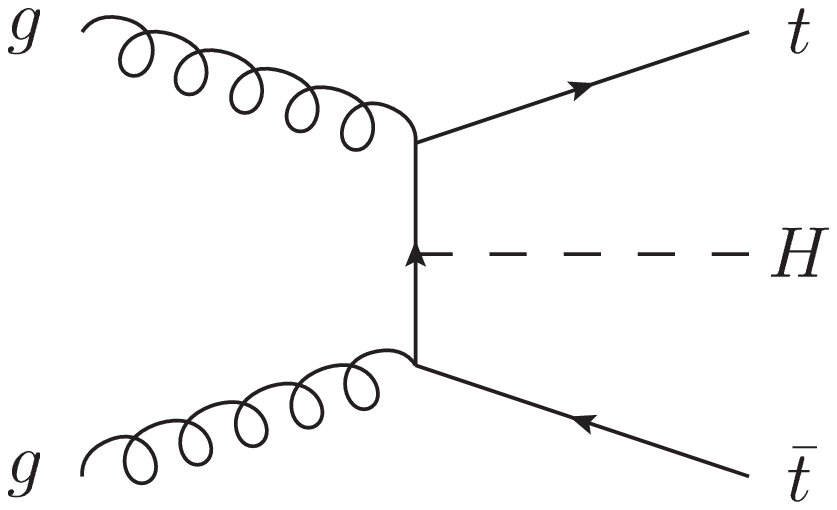,width=0.35\textwidth}
$\phantom{aaaaaaa}$
    \epsfig{figure=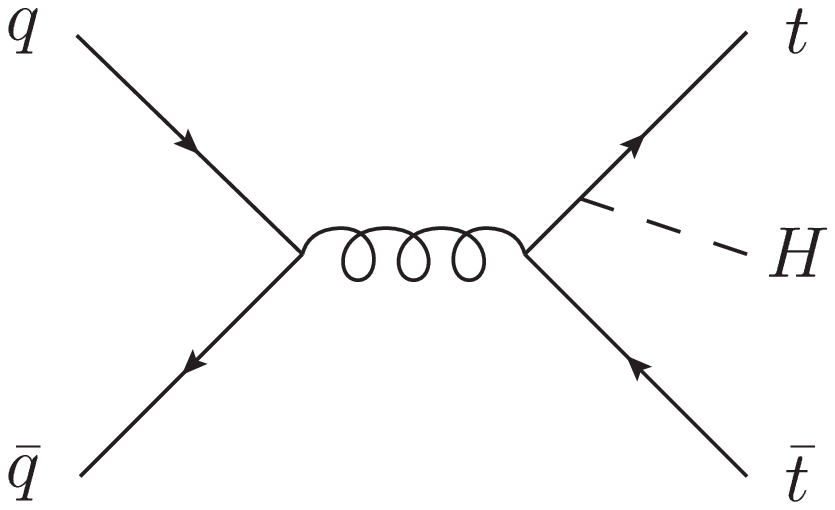,width=0.35\textwidth}
\caption{\label{fig:diagtree1}
Representative $\ord(\asa{1}{1/2})$ Born-level diagrams.
}
  \end{center}
\end{figure}

\begin{figure}[h]
  \begin{center}
    \epsfig{figure=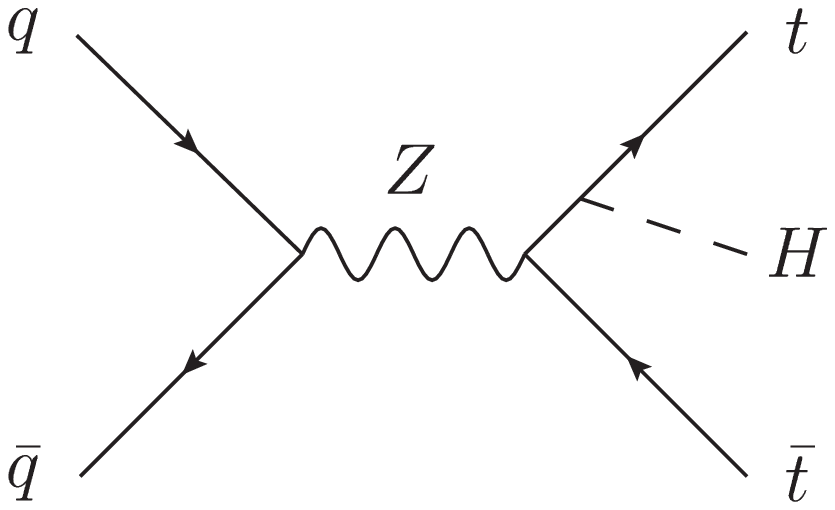,width=0.35\textwidth}
$\phantom{aaaaaaa}$
    \epsfig{figure=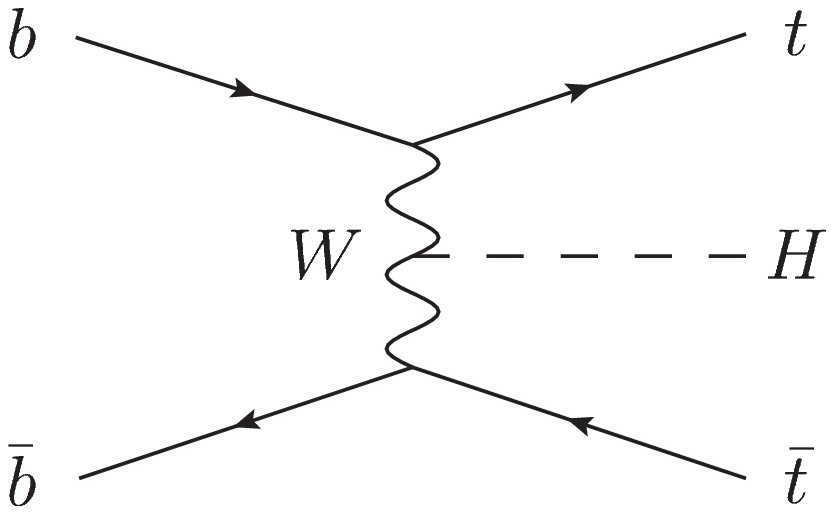,width=0.35\textwidth}
\caption{\label{fig:diagtree2}
Representative $\ord(\aem^{3/2})$ Born-level diagrams.
}
  \end{center}
\end{figure}

The calculation we are carrying out is one where we expand simultaneously
in the strong ($\as$) and weak ($\aem$) coupling constants; this scenario
has been called mixed-coupling expansion in ref.~\cite{Alwall:2014hca},
a paper whose notation, and in particular that of sect.~2.4, we shall 
adopt in what follows. Denoting by $\Sigma(\as,\aem)$ a generic observable 
(e.g., a cross section within cuts, or a histogram bin), in $t\bt H$ 
production we have, at the Born level:
\beq
\Sigma_{t\bt H}^{\rm (Born)}(\as,\aem)=
\as^2\aem\,\Sigma_{3,0}
+\as\aem^2\,\Sigma_{3,1}
+\aem^3\,\Sigma_{3,2}\,,
\label{SigB}
\eeq
which is a direct consequence of the coupling-constant factors 
associated with the amplitudes relevant 
to the three classes of contributing partonic processes, which we list 
in table~\ref{ord_tree}; samples of the corresponding Feynman diagrams 
are depicted in figs.~\ref{fig:diagtree1} and~\ref{fig:diagtree2}. 
From table~\ref{ord_tree} one also sees that $\Sigma_{3,1}=\Sigma_{3,2}=0$
in the case of the $gg$-initiated process, while $\Sigma_{3,1}=0$ for
the $q\bq$-initiated process with $q\ne b$, owing to the colour structure 
(proportional to the trace of a single Gell-Mann matrix) of this
interference term. When $q=b$, $\Sigma_{3,1}\ne 0$ because of the
contribution of diagrams such as the second one of fig.~\ref{fig:diagtree2},
which induce a different colour structure. It has to be pointed out that
diagrams of that kind would be present when $q\ne b$ as well, if the CKM
matrix featured off-diagonal terms in the third generation; in this
work, we have assumed this matrix to be diagonal. At the NLO, 
we have:
\beq
\Sigma_{t\bt H}^{\rm (NLO)}(\as,\aem)=
\as^3\aem\,\Sigma_{4,0}
+\as^2\aem^2\,\Sigma_{4,1}
+\as\aem^3\,\Sigma_{4,2}
+\aem^4\,\Sigma_{4,3}\,,
\label{SigNLO}
\eeq
which follows from eq.~(\ref{SigB}), since in a QCD-EW mixed-coupling 
expansion the coupling-constant factors at the NLO are obtained from 
those relevant to the LO by multiplying them by one power of either 
$\as$ or $\aem$ (see eq.~(2.23) of ref.~\cite{Alwall:2014hca}).
\begin{table}[h]
\begin{center}
\begin{tabular}{l|ll}
Process	& $\ord(\mathcal{A})$ & $\ord(\Sigma)$\\
\hline
$gg\TO\tth$ & $\asa{1}{1/2}$ & $\asa{2}{1}$\\
\hline
$q\bar{q}\TO\tth,\quad q\neq b$	& 
 $\asa{1}{1/2},\;\alpha^{3/2}$$\phantom{aaaa}$ & 
 $\asa{2}{1},\;\alpha^{3}$ \\
\hline
$q\bar{q}\TO\tth,\quad q=b$ & 
 $\asa{1}{1/2},\;\alpha^{3/2}$	& 
 $\asa{2}{1},\;\asa{1}{2},\;\alpha^{3}$ \\
\hline
\end{tabular}
\end{center}
\caption{\label{ord_tree}
Born-level partonic processes relevant to $t\bt H$ production.
For each of them, we report the coupling-constant factors in
front of the non-null contributions, both at the amplitude (middle column)
and at the amplitude squared (rightmost column) level.
}
\end{table}

The notation for the generic short-distance coefficient $\Sigma_{k,q}$ 
has the following motivation. The integer $k$ is the sum of the powers 
of $\as$ and $\aem$ at any given perturbative order; in $t\bt H$ production, 
$k=3$ at the LO (eq.~(\ref{SigB})) and $k=4$ at the NLO (eq.~(\ref{SigNLO})).
This immediately shows that it is also convenient to write 
$\Sigma_{k,q}\equiv\Sigma_{k_0+p,q}$, with $p\ge 0$, for the N$^p$LO 
coefficients; $k_0$ is then a fixed, process-specific integer associated 
with the Born cross section, equal to $3$ in $t\bt H$ production.
The integer $q$ identifies the various terms of eqs.~(\ref{SigB})
and~(\ref{SigNLO}). We have conventionally chosen to associate
increasing values of $q$ with $\Sigma_{k_0+p,q}$ coefficients (at fixed $p$) 
which are increasingly suppressed in terms of the hierarchy of the coupling
constants, $\aem\ll\as$. Thus, $q=0$ corresponds to the coefficient
with the largest (smallest) power of $\as$ ($\aem$), and conversely
for $q=q_{\max}$. This maximum value $q_{\max}$ that can be assumed
by $q$ is process- and perturbative-order-dependent, and it grows with 
the number of amplitudes that interfere and that factorise different 
coupling-constant combinations; in the case of $t\bt H$ production at 
the LO, this can be seen by comparing the two rightmost columns of 
table~\ref{ord_tree}.

We propose that the coefficient $\Sigma_{k_0+p,q}$ be called the
leading (when $q=0$), or the $(q+1)^{th}$-leading (when $q\ge 1$, 
i.e.~second-leading, third-leading, and so forth), term of the N$^p$LO 
contribution to the cross section\footnote{This classification is 
the same as that one obtains by counting the powers of $\lambda$ after 
rescaling $\as\to\lambda\as$, $\aem\to\lambda^2\aem$. Both can be 
generalised to the case of a mixed-coupling expansion in more than 
two couplings.}. The word ``term'' may be replaced by any suitable
synonymous, and in particular by ``correction'' at the NLO and beyond.
We explicitly emphasise that the above convention implies that
expressions such as ``QCD corrections'' or ``EW corrections''
should be avoided to identify the coefficients $\Sigma_{k_0+p,q}$.
The key point is that while $\Sigma_{k_0+p,q}$ is a well-defined
quantity in perturbation theory, QCD corrections or EW corrections
are ambiguous concepts (except in two cases, as we shall explain below),
which might lead to some confusion.

In order to further the points above, which are valid independently
of the process considered, let us restrict to the case of $t\bt H$
production for definiteness. The goal of this paper is that of computing 
the so-far unknown second-leading NLO correction $\Sigma_{4,1}$ (with 
some restrictions, to be discussed in sect.~\ref{sec:calc}), 
and to use it, together with the leading LO 
and NLO terms, $\Sigma_{3,0}$ and $\Sigma_{4,0}$ respectively, 
for a sample phenomenology study. The coefficient $\Sigma_{4,0}$
has been available in the literature for a while~\cite{Beenakker:2001rj,
Beenakker:2002nc,Dawson:2002tg,Dawson:2003zu,Dittmaier:2003ej,
Frederix:2011zi,Garzelli:2011vp}, and is traditionally referred to as 
NLO {\em QCD} corrections; the analogue of the coefficient $\Sigma_{4,1}$,
available in the literature for other processes such as $t\bt$
production~\cite{Beenakker:1993yr,Bernreuther:2005is,Kuhn:2005it,
Bernreuther:2006vg,Kuhn:2006vh} is often referred to as 
NLO {\em EW} corrections. These naming conventions, in their explicit
use of ``QCD'' and ``EW'', is what we suggest
to avoid in the context of a mixed-coupling expansion, and the
reason is particularly clear in the case when $\Sigma_{4,1}$ is
identified with EW corrections. When doing so, in fact, one
implicitly assumes that these are EW corrections just to the 
leading Born term; furthermore, such corrections cannot be disentangled 
unambiguously from QCD corrections to the second-leading Born term.

\begin{figure}[h]
  \begin{center}
    \epsfig{figure=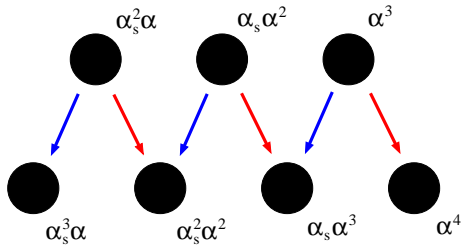,width=0.4\textwidth}
\caption{\label{fig:corr}
QCD (blue, right-to-left arrows) corrections and
EW (red, left-to-right arrows) corrections
to $t\bt H$ hadroproduction. See the text for details.
}
  \end{center}
\end{figure}
The situation is depicted schematically in fig.~\ref{fig:corr}
(which is adapted from ref.~\cite{Alwall:2014hca}): each blob in 
the upper or lower row corresponds to one of the $\Sigma_{3,q}$
or $\Sigma_{4,q}$ coefficients of eq.~(\ref{SigB}) or eq.~(\ref{SigNLO}),
respectively. We propose that they keywords ``QCD corrections''
and ``EW corrections'' be used only to identify the computations
that lead to an NLO contribution given a Born contribution, according
to the scheme:
\beqn
&&\as^n\aem^m\;\;\;\stackrel{\rm QCD}{\longrightarrow}
\;\;\;\as^{n+1}\aem^m\,,
\label{QCDcorr}
\\
&&\as^n\aem^m\;\;\;\stackrel{\rm EW}{\longrightarrow}
\;\;\;\as^n\aem^{m+1}\,.
\label{QEDcorr}
\eeqn
These definitions correspond to the arrows that appear in fig.~\ref{fig:corr}:
from right to left for QCD corrections, and from left to right for
EW corrections. We point out that this terminology is consistent
with that typically used in the literature. It {\em only} becomes 
misleading when it is {\em also} applied to the coefficients 
$\Sigma_{k_0+1,q}$, because this is equivalent to giving the same name
to two different classes of objects in fig.~\ref{fig:corr}: the blobs 
and the arrows. If the roles of these two classes are kept distinct,
no ambiguity is possible. Consider, for example, the coefficient
$\Sigma_{4,1}$ in which we are interested here: it is the second-leading
NLO term, which receives contributions both from the EW corrections to the
leading Born term $\Sigma_{3,0}$, and from the QCD corrections to the
second-leading Born term $\Sigma_{3,1}$. 

We note that the discussion given above explains why there is no ambiguity
when one works in a single-coupling perturbative expansion. In the case
of QCD, for example, the only relevant quantities of fig.~\ref{fig:corr}
are the two leftmost blobs (one for each row), and the leftmost arrow.
There is thus a one-to-one correspondence between the arrow and the
leftmost blob in the lower row: therefore, no confusion arises
even if one calls the latter (the leading NLO correction) with the
name of the former (the QCD corrections), which is what is usually done.
The case of the single-coupling EW expansion is totally analogous,
and applies to the quantities that in fig.~\ref{fig:corr} are to the
extreme right (namely, $\Sigma_{3,2}$, $\Sigma_{4,3}$, and the rightmost
left-to-right arrow. Note that $\Sigma_{4,1}$ is {\em not} involved).

\begin{figure}[h]
  \begin{center}
    \epsfig{figure=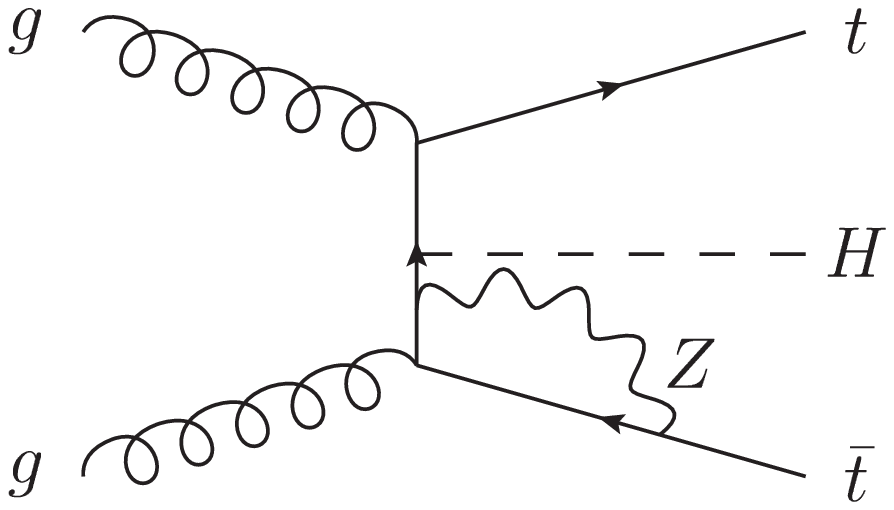,width=0.25\textwidth}
$\phantom{aaaaaaa}$
    \epsfig{figure=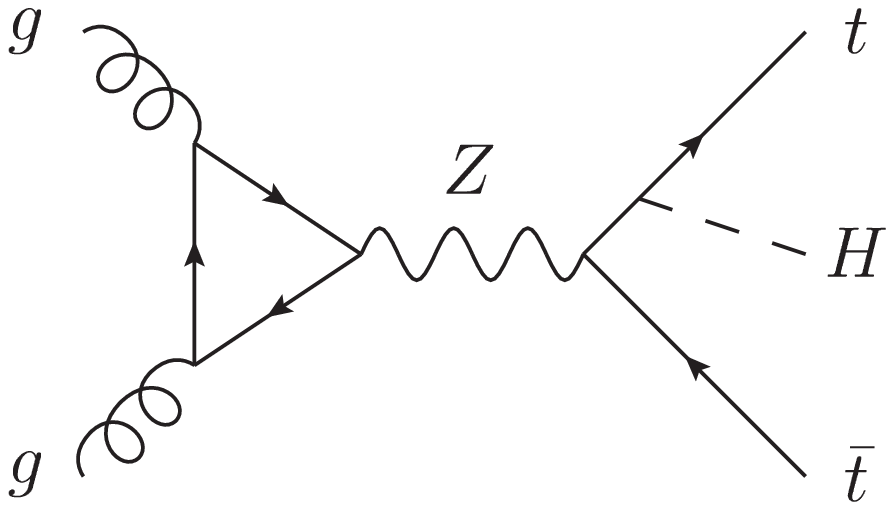,width=0.25\textwidth}
$\phantom{aaaaaaa}$
    \epsfig{figure=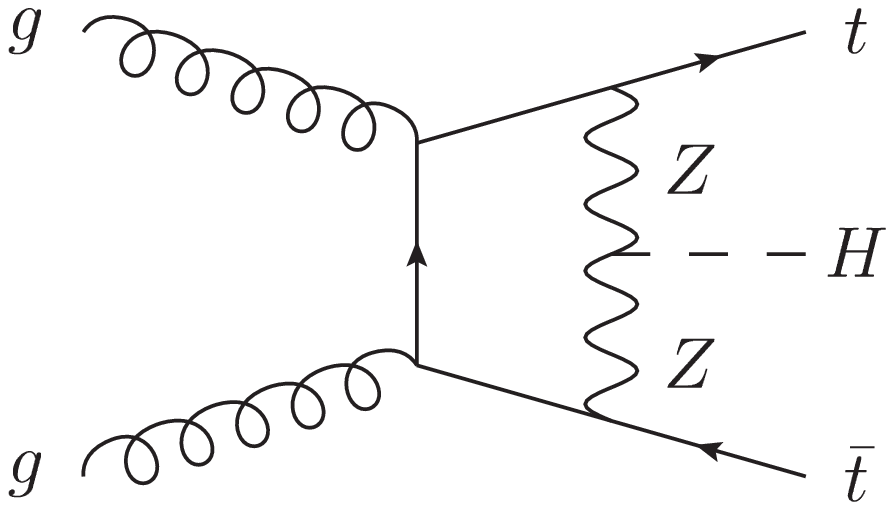,width=0.25\textwidth}
\caption{\label{fig:diagloop1}
Representative $\ord(\asa{1}{3/2})$ one-loop diagrams for
the $gg$ channel.
}
  \end{center}
\end{figure}
We would like now to elaborate further on the keywords
``QCD corrections'' and ``EW corrections'', stressing again
the fact that they do not have any deep physical meaning,
but may be useful in that they are intuitive, and can be given
an operational sense. The best way to do so is that of a constructive
bottom-up approach that starts at the level of amplitudes (we note
that eqs.~(\ref{QCDcorr}) and~(\ref{QEDcorr}) are at the level
of amplitude squared) in order to figure out which contributions 
each of the coefficients $\Sigma_{4,q}$ receives.
While doing so, one needs to bear in mind
that, at the NLO, there are two classes of such contributions:
those due to real-emission amplitudes (eventually squared), and those due
to one-loop amplitudes (eventually contracted with Born amplitudes).
Since here we are solely interested in figuring out the general
characteristics of the contributions to any given $\Sigma_{4,q}$
(as opposed to performing a complete and explicit computation,
which is rather done automatically), the easiest procedure
is that of taking representative Born-level diagrams, such as
those of figs.~\ref{fig:diagtree1} and~\ref{fig:diagtree2},
and turn them either into one-loop graphs through the insertion of
a virtual particle, or into real-emission graphs by emitting
one further final-state particle. It is clear that in general 
it is not possible to obtain all one-loop and real-emission Feynman
diagrams in this way (see e.g.~the second and third graphs
in fig.~\ref{fig:diagloop1}), but this is irrelevant for the
sake of the present exercise. What is of crucial importance
is that, in the context of a mixed QCD-EW expansion, the
virtual or final-state particle mentioned before must be chosen
in a set larger than the one relevant to a single-coupling series.
In particular, for the case of $t\bt H$ production with stable
top quarks and Higgs, such a set is:
\beq
\Big\{g,q,t,Z,W^\pm,H,\gamma\Big\}\,,
\label{corrset}
\eeq
where the light quark $q$ may also be a $b$ quark, and the top
quark enters only one-loop contributions. 
In the case of such contributions, the particles in the set of 
eq.~(\ref{corrset}) are fully analogous to the L-cut particles
(see sect.~3.2.1 of ref.~\cite{Hirschi:2011pa}), and we
understand ghosts and Goldstone bosons.
When the extra particle added to the Born diagram (be it virtual 
or real) is strongly interacting, it is then natural to classify the
resulting one-loop or real-emission diagram as a QCD-type contribution, 
and a EW-type contribution otherwise\footnote{An alternative classification
(equivalent to that used here when restricted to $t\bt H$ production
and to processes of similar characteristics, but otherwise more general) 
is one that determines the type of contribution according to
the nature of the vertex involved.}. The idea
of this amplitude-level classification is that QCD-type and EW-type
contributions will generally lead to QCD and EW corrections at 
the amplitude-squared level, respectively. However, this correspondence,
in spite of being intuitively appealing, is not exact, as we shall
show in the following; this is one of the reasons why ``QCD corrections''
and ``EW corrections'' must not be interpreted literally.
\begin{table}
\begin{center}
\begin{tabular}{cc||cc}
 &
Born & $B_0=\ord(\asa{1}{1/2})$ & $B_1=\ord(\aem^{3/2})$ \\
\hline
\multirow{2}{*}{QCD} &
Virtual & $V_{{\rm\sss QCD},0}=\ord(\asa{2}{1/2})$ & 
          $V_{{\rm\sss QCD},1}=\ord(\asa{1}{3/2})$ \\
 &
Real & $R_{{\rm\sss QCD},0}=\ord(\asa{3/2}{1/2})$ & 
       $R_{{\rm\sss QCD},1}=\ord(\asa{1/2}{3/2})$ \\
\hline
\multirow{2}{*}{EW} &
Virtual & $V_{{\rm\sss EW},0}=\ord(\asa{1}{3/2})$ & 
          $V_{{\rm\sss EW},1}=\ord(\aem^{5/2})$ \\
 &
Real & $R_{{\rm\sss EW},0}=\ord(\asa{1}{1})$ & 
       $R_{{\rm\sss EW},1}=\ord(\aem^{2})$\\
\end{tabular}
\end{center}
\caption{\label{tab:BVR}
Coupling-constant factors relevant to Born, one-loop, and real-emission
amplitudes; see the text for more details.
}
\end{table}
The classification just introduced is
used in table~\ref{tab:BVR}: for a given Born-level amplitude
$B_i$ associated with a definite coupling-constant factor, the corresponding 
one-loop and real-emission quantities are denoted by $V_{{\rm\sss QCD},i}$
and $R_{{\rm\sss QCD},i}$ in the case of QCD-type contributions, and 
by $V_{{\rm\sss EW},i}$ and $R_{{\rm\sss EW},i}$ in the case
of EW-type contributions. We can finally consider all possible combinations
$B_i\cdot V_{*,j}$, $R_{{\rm\sss QCD},i}\cdot R_{{\rm\sss QCD},j}$,
and $R_{{\rm\sss EW},i}\cdot R_{{\rm\sss EW},j}$ and associate
them with the relevant amplitude-squared quantities $\Sigma_{4,q}$. Note that 
one must not consider the $R_{{\rm\sss QCD},i}\cdot R_{{\rm\sss EW},j}$
combinations, owing to the fact that the two amplitudes here
are relevant to different final states\footnote{For generic 
processes, this is not necessarily the case, the typical situation
being that where some massless particles in the set of eq.~(\ref{corrset})
are present at the Born level.\label{ft:two}}. 

We now observe that this bottom-up
construction leads to redundant results. Here, the case in point is
that of $V_{{\rm\sss QCD},1}$ and $V_{{\rm\sss EW},0}$: the one-loop
diagram (which enters $V_{{\rm\sss QCD},1}$) obtained by exchanging 
a gluon between the $\bq$ and $\bt$ legs of the diagram to the left
of fig.~\ref{fig:diagtree2} is the same diagram as that 
(which enters $V_{{\rm\sss EW},0}$) obtained by exchanging 
a $Z$ between the $q$ and intermediate-$t$ legs of the diagram to 
the right of fig.~\ref{fig:diagtree1}. This fact does not pose any
problem in the context of the classification exercise we are carrying
out here. Conversely, it is instructive because it shows directly that
such contributions cannot be unambiguously given a ``QCD correction''
or an ``EW correction'' tag at the level of the cross section, where
they will always appear together. We note that a fully similar
situation would be that of $R_{{\rm\sss QCD},i}\cdot R_{{\rm\sss EW},j}$,
if such quantities were relevant to the present computation;
as mentioned in footnote~\ref{ft:two}, their analogues will in 
general contribute to the cross sections of other processes.

\noindent
The final results of our classification exercise are given in 
eqs.~(\ref{SvsVR0})--(\ref{SvsVR3}):
\beqn
\Sigma_{4,0} & \;\;\longleftrightarrow\;\; & 
B_0\mydot V_{{\rm\sss QCD},0}\,,\;\,
R_{{\rm\sss QCD},0}\mydot R_{{\rm\sss QCD},0}\,,
\label{SvsVR0}
\\
\Sigma_{4,1} & \;\;\longleftrightarrow\;\; & 
B_0\mydot \left(V_{{\rm\sss QCD},1}\oplus
                V_{{\rm\sss EW},0}\right)\,,\;\,
B_1\mydot V_{{\rm\sss QCD},0}\,,\;\,
R_{{\rm\sss QCD},0}\mydot R_{{\rm\sss QCD},1}\,,\;\,
R_{{\rm\sss EW},0}\mydot R_{{\rm\sss EW},0}\,,
\label{SvsVR1}
\\
\Sigma_{4,2} & \;\;\longleftrightarrow\;\; & 
B_0\mydot V_{{\rm\sss EW},1}\,,\;\,
B_1\mydot \left(V_{{\rm\sss QCD},1}\oplus
                V_{{\rm\sss EW},0}\right)\,,\;\,
R_{{\rm\sss QCD},1}\mydot R_{{\rm\sss QCD},1}\,,\;\,
R_{{\rm\sss EW},0}\mydot R_{{\rm\sss EW},1}\,,
\label{SvsVR2}
\\
\Sigma_{4,3} & \;\;\longleftrightarrow\;\; & 
B_1\mydot V_{{\rm\sss EW},1}\,,\;\,
R_{{\rm\sss EW},1}\mydot R_{{\rm\sss EW},1}\,.
\label{SvsVR3}
\eeqn
These equations help summarise the points made above in an
explicit manner. For example, we may classify as QCD corrections
the second and third terms on the r.h.s.~of eq.~(\ref{SvsVR1}),
and as EW corrections the fourth term there; as discussed above,
the first term is neither of the two. Equations~(\ref{SvsVR0})
and~(\ref{SvsVR3}) receive only QCD and EW corrections, respectively,
and indeed they correspond to the results of a single-coupling
expansion (in $\as$ and $\aem$, respectively). The case of 
eq.~(\ref{SvsVR2}) is fully analogous to that of eq.~(\ref{SvsVR1}).
Note finally that, in the equations above and in table~\ref{tab:BVR}, 
all of the contributions with index ``1' vanish identically for the 
$gg$-initiated process. 

The classification procedure carried out above can be extended to
any process. This is useful not so much in order to determine which
corrections are QCD and which are EW, given the irrelevance of this
from the physics viewpoint, but to understand in a quick manner
which contributions each of the $\Sigma_{k_0+1,q}$ coefficient receives.

\subsection{Calculation of the $\ord(\asa{2}{2})$ contribution
with \aNLO\label{sec:calc}}
As was stated in sect.~\ref{sec:intro}, we are interested in
computing $\Sigma_{4,1}$, with the sole exclusion of contributions
of QED origin, and thus including in particular weak-only effects.
For a generic process or coefficient $\Sigma_{k_0+1,q}$, 
a gauge-invariant separation of NLO EW effects into a QED and a 
weak subset is not always possible. However, in our case LO 
diagrams subject to EW corrections do not feature any $W$ boson
(note that this is not true for $\Sigma_{4,2}$ and $\Sigma_{4,3}$); 
furthermore, no triple gauge-vector vertex appears in one-loop diagrams. 
Thus, weak-only NLO corrections to $t\bt H$ hadroproduction are 
well defined.

The first implication of restricting oneself to weak-only contributions
is that of removing the photon from the set of eq.~(\ref{corrset}).
At the level of real-emission diagrams, this implies that
no graphs with external photons will contribute to
our results; this also simplifies
the structure of the subtractions, which is identical to that of
a pure-QCD computation, in view of the absence of soft and collinear
singularities of QED origin. The removal of the photon contributions
from real-emission matrix elements must have a consistent counterpart
at the level of one-loop amplitudes. In order to discuss this matter,
we remind the reader that all computations performed by \aNLO\ are
based on a \UFO~\cite{Degrande:2011ua} model, that encodes the basic
information on the Lagrangian of the relevant theory. For NLO computations,
in particular, on top of the usual Feynman rules one also needs those
for the $R_2$ counterterms~\cite{Ossola:2008xq} and for the UV 
counterterms. Two \UFO\ models are available that allow one to
perform QCD+EW corrections in the SM; our default (used for the majority
of the results to be presented in sect.~\ref{sec:results}) is that which
adopts the $\aem(m_Z)$ renormalisation scheme~\cite{Dittmaier:2001ay}
(and thus $\aem(m_Z)$, $m_Z$, and $m_W$ as input parameters), 
while an alternative one implements the $G_{\mu}$ 
scheme~\cite{Dittmaier:2001ay,Denner:1991kt}
(where the input parameters are $G_F$, measured in $\mu$ decays,
$m_Z$, and $m_W$); masses and wave functions
are renormalised on-shell. For both models, the $R_2$ rules have been 
taken from refs.\cite{Garzelli:2009is,Garzelli:2010qm,Shao:2011tg}.
In view of the complexity of the models, all counterterms have
been cross-checked with an independent 
{\sc\small Mathematica} package. Having the full QCD+EW corrections
available in the models, one can rather easily exclude the photon
contributions to loop diagrams at generation time, as well as from
masses and wave-function UV counterterms, and from $R_2$ counterterms,
thanks to the extreme flexibility of \MLf\ (see sect.~2.4.2 of 
ref.~\cite{Alwall:2014hca}, and in particular the concept
of loop-content filtering there, for more informations). The result of
this procedure has been validated by computing with \aNLO\ the
complete weak-only contributions to $pp\to t\bt$ production,
and by comparing it (at the level of differential distributions)
to what we have obtained for this process 
with {\sc\small Feynarts}~\cite{Hahn:2000kx}, 
{\sc\small Formcalc}~\cite{Agrawal:2011tm},
and {\sc\small LoopTools}~\cite{Hahn:1998yk}. Furthermore,
these tools have also been used for computing the virtual weak
contributions to $HH\to t\bt$ production, as a way to cross-check the
renormalisation of the $t\bt H$ Yukawa and its use in \aNLO;
again, an excellent agreement has been found.

\begin{table}[t]
\begin{center}
\begin{tabular}{c|c}
\multicolumn{2}{c}{Virtual corrections} \\
\hline
\multirow{2}{*}{
$B_0\mydot\left(V_{{\rm\sss QCD},1}\oplus
                V_{{\rm\sss EW},0}\right)$} &
  $gg\to t\bt H$\\
& $q\bq\to t\bt H$\\
\hline
$B_1\mydot V_{{\rm\sss QCD},0}$ &
 $q\bq\to t\bt H$\\
\hline\hline
\multicolumn{2}{c}{Real-emission corrections} \\
\hline
\multirow{2}{*}{
$R_{{\rm\sss QCD},0}\mydot R_{{\rm\sss QCD},1}$} &
  $q\bq\to t\bt Hg$\\
& $qg\to t\bt Hq$\\
\hline
\multirow{5}{*}{
$R_{{\rm\sss EW},0}\mydot R_{{\rm\sss EW},0}$} &
  $gg\to t\bt HZ$\\
& $q\bq\to t\bt HZ$\\
& $q\bqp\to t\bt HW$\\
& $gg\to t\bt HH$\\
& $q\bq\to t\bt HH$\\
\end{tabular}
\end{center}
\caption{\label{tab:as2a2}
List of partonic processes that contribute to the second-leading NLO
term $\Sigma_{4,1}$, according to the classification given in 
table~\protect\ref{tab:BVR} and eq.~(\protect\ref{SvsVR1}). 
See the text for more details.
}
\end{table}
In table~\ref{tab:as2a2} we list explicitly all the partonic
processes that contribute to the $\ord(\asa{2}{2})$ coefficient
$\Sigma_{4,1}$. Each process understands the computation of the
corresponding amplitudes in the left column of the table, according
to the classification given in table~\ref{tab:BVR}. So for example
for the first real-emission process of table~\ref{tab:as2a2},
the contribution to $\Sigma_{4,1}$ is given by the $\ord(\asa{3/2}{1/2})$
$q\bq\to t\bt Hg$ tree-level amplitude times the $\ord(\asa{1/2}{3/2})$
$q\bq\to t\bt Hg$ tree-level amplitude.

Loop diagrams of $\ord(\asa{2}{1/2})$ ($V_{{\rm\sss QCD},0}$) enter both
the first- and the second-leading NLO terms, $\Sigma_{4,0}$ and $\Sigma_{4,1}$.
However, in the latter case the interference with the $\ord(\aem^{3/2})$
Born amplitude $B_1$ (see eq.~(\ref{SvsVR1})) is such that self-energy and 
vertex corrections vanish owing to the colour structure; thus,
only boxes and pentagons contribute, and UV divergencies of QCD origin are
not present at $\ord(\asa{2}{2})$. Furthermore, the $gg$-initiated
virtual corrections are also soft- and collinear-finite; consistently, as
one can see from table~\ref{tab:as2a2}, there is no real-emission
counterpart which might cancel such divergencies. 
This is not the case for the $q\bq$ process, where cancellations of
singularities do occur between the virtual and real-emission processes. 
When $q\ne b$, such singularities are only of soft origin, owing again
to the colour structure of the diagrams involved, which implies that
emissions of the gluon from an initial-state leg both on the left
and on the right of the Cutkosky cut give a vanishing contribution.
This is consistent with the fact that for such light quark the 
$\ord(\asa{1}{2})$ Born-level cross section is zero (see 
table~\ref{ord_tree}), since this cross section would have to factorise 
(times the relevant Altarelli-Parisi kernel~\cite{Altarelli:1977zs}) 
in the case of collinear singularities. Similar considerations
(and the absence of virtual contributions) lead to the conclusion 
that the $qg$-initiated real-emission process is also soft- and 
collinear-finite. Finally, the real-emission contributions of 
weak origin ($R_{{\rm\sss EW},0}\cdot R_{{\rm\sss EW},0}$)
are finite everywhere in the phase space.

We conclude this section by outlining the ingredients that enter
the results that we shall present in sect.~\ref{sec:results},
and which are mainly based on the coefficients $\Sigma_{3,0}$
(at the LO), $\Sigma_{4,0}$, and $\Sigma_{4,1}$ (at the NLO).
The calculation of the former two coefficients is the same as that 
which has already appeared in refs.~\cite{Frederix:2011zi,Alwall:2014hca},
and is fully automated in \aNLO. 
We remind the reader that \aNLO\ contains all ingredients relevant
to the computations of LO and NLO cross sections, with or without
matching to parton showers. NLO results not matched to parton showers
are obtained by adopting the
FKS method~\cite{Frixione:1995ms,Frixione:1997np} for the subtraction 
of the singularities of the real-emission matrix elements (automated 
in the module \MadFKS~\cite{Frederix:2009yq}), and the OPP
integral-reduction procedure~\cite{Ossola:2006us} for the computation 
of the one-loop matrix elements (automated in the module 
\ML~\cite{Hirschi:2011pa}, which makes use of
\CutTools~\cite{Ossola:2007ax} and of an in-house implementation 
of the representation proposed in ref.~\cite{Cascioli:2011va} (\OL)).
The automation of the mixed-coupling expansions
has not been completely validated\footnote{This validation consists
only in addressing bookkeeping issues, given that QED subtractions 
are a simpler version of their QCD counterparts, as was already pointed
out in sect.~\ref{sec:intro}, and that QCD subtractions relevant to the
beyond-leading $\Sigma_{k_0+1,q}$ coefficients ($q\ge 1$) are fully analogous
to those, already automated, relevant to the leading term $\Sigma_{k_0+1,0}$.} 
yet in \MadFKS, but is fully operational in \MLf. Thus, the calculation
of $\Sigma_{4,1}$ has been achieved by constructing ``by hand'', for the 
specific process we are considering, the IR counterterms relevant to the
subtraction of QCD singularities; this operation will serve as
a benchmark when the automation of the subtractions in a mixed-coupling
scenario will be achieved. Apart from this, all of the other relevant
procedures, and in particular the generation of the matrix elements and of 
the ${\cal S}$ functions (which achieve the dynamic phase-space partition
needed in FKS), are automated. 

Given that the subtraction of the IR singularities that 
affect $\Sigma_{4,1}$ is not completely automated, we have simplified
the calculation by ignoring the contribution to this coefficient
due to $b\bb$-initiated partonic processes (as was discussed before,
this process is the only one where initial-state collinear singularities
appear, and thus no collinear subtractions are needed in our computation). 
This approximation is fully justified numerically, in view of
the extremely small $b\bb\to t\tb H$ cross section at the LO, which
we shall report in sect.~\ref{sec:results}. We shall also present
the contributions of the $R_{{\rm\sss EW},0}\cdot R_{{\rm\sss EW},0}$
processes (see table~\ref{tab:as2a2}) separately from the rest,
in keeping with what is usually done in the context of EW computations.
We emphasise that, as the general derivation presented before shows,
there is no real motivation for ignoring such contributions completely.
The argument that an extra final-state boson can be tagged
might be made, but only in the context of a fully realistic
analysis (since bosons cannot be seen directly in a detector),
which is beyond the scope of the present paper. We note that 
the corresponding cross section is not negligible,
as we shall document in sect.~\ref{sec:results}; our
results, being inclusive in the extra boson, represent an upper
bound for those obtained by applying proper acceptance cuts.

\begin{table}
\begin{center}
\begin{tabular}{c|c|c}
Label & Meaning & Restrictions \\
\hline
LO or Born & $\as^2\aem\,\Sigma_{3,0}$ &\\
NLO QCD & $\as^3\aem\,\Sigma_{4,0}$ &\\
NLO weak & $\as^2\aem^2\,\Sigma_{4,1}$ &
 no QED, no $b\bb\to t\bt H+X$, no $pp\to t\bt H+V$ \\
HBR & $\as^2\aem^2\,\Sigma_{4,1}$ &
 no QED, no $b\bb\to t\bt H+X$, only $pp\to t\bt H+V$\\
\end{tabular}
\end{center}
\caption{\label{tab:names}
Shorthand notation that we shall use in sect.~\protect\ref{sec:results}.
HBR is an acronym for Heavy Boson Radiation. $V$ stands for a Higgs,
a $W$, or a $Z$, and HBR understands the sum of the corresponding
three cross sections.
The reader is encouraged to check sect.~\protect\ref{sec:org} for the 
precise definitions of all the quantities involved.
}
\end{table}
In table~\ref{tab:names} we give the shorthand naming conventions
that we shall adopt in sect.~\ref{sec:results}. We use names
which are similar to those most often used in the context
of EW higher-order computations, so as to facilitate the reading
of the phenomenological results. As was discussed at length in
the present section, the contents of the various terms are more
involved than their names may suggest, and we refer the reader 
to such a section for the necessary definitions.

\section{Results\label{sec:results}}
In this section we present a sample of results obtained by simulating
$t\bt H$ production in $pp$ collisions at three different collider 
c.m.~energies: 8, 13, and 100~TeV. We have chosen the top-quark and 
Higgs masses as follows:
\beq
m_t=173.3~\gev\,,\;\;\;\;\;\;
m_H=125~\gev\,,
\eeq
and adopted the MSTWnlo2008~\cite{Martin:2009iq} PDFs
with the associated $\as(m_Z)$ for all NLO as well as LO predictions
(since we are chiefly interested in assessing effects of matrix-element
origin). In our default $\aem(m_Z)$-scheme, the EW coupling constant 
is~\cite{Jegerlehner:2001ca}:
\beq
\frac{1}{\aem(m_Z)}=128.93\,.
\label{aem}
\eeq
The central values of the renormalisation ($\muR$) and factorisation 
($\muF$) scales have been taken equal to the reference scale:
\beq
\mu=\frac{\Ht}{2}\equiv 
\frac{1}{2}\sum_i\sqrt{m_i^2+\pt^2(i)}\,,
\label{scref}
\eeq
where the sum runs over all final-state particles. The theoretical 
uncertainties due to the $\muR$ and $\muF$ dependencies that affect
the coefficient $\Sigma_{4,0}$ have been evaluated by varying these
scales independently in the range:
\beq
\frac{1}{2}\mu\le\muR,\muF\le 2\mu\,,
\label{scalevar}
\eeq
and by keeping the value of $\aem$ fixed. The calculation of
this theory systematics does not entail any independent runs, being
performed through the reweighting technique introduced in
ref.~\cite{Frederix:2011ss}, which is fully automated in \aNLO.
All the input parameters not explicitly mentioned here have been set
equal to their PDG values~\cite{Beringer:1900zz}.

We shall consider two scenarios: one where no final-state cuts are applied
(i.e.~fully inclusive), and a ``boosted'' one, generally helpful to reduce 
the contamination of light-Higgs signals due to background 
processes~\cite{Butterworth:2008iy,Plehn:2009rk}, where the following cuts
\beq
\pt(t)\ge 200~\gev\,,\;\;\;\;\;\;
\pt(\bt)\ge 200~\gev\,,\;\;\;\;\;\;
\pt(H)\ge 200~\gev\,,
\label{boost}
\eeq
are imposed; since these emphasise the role of the high-$\pt$ regions,
the idea is that of checking whether weak effects will have a bigger
impact there than in the whole of the phase space.
We shall report in sect.~\ref{sec:rates} our predictions
for total rates, for the three collider c.m.~energies and in both the 
fully inclusive and the boosted scenario. In sect.~\ref{sec:distr} 
several differential distributions will be shown, at a c.m.~of 13~TeV 
with and without the cuts of eq.~(\ref{boost}), and at a c.m.~of 100~TeV 
in the fully-inclusive case only.

Throughout this section, we shall make use of the shorthand notation
introduced at the end of sect.~\ref{sec:org} -- see in particular 
table~\ref{tab:names}.

\subsection{Inclusive rates\label{sec:rates}}
In this section we present our predictions for inclusive rates,
possibly within the cuts of eq.~(\ref{boost}). As was already stressed,
the results for the LO and NLO QCD contributions are computed in the
same way as has been done previously with \aNLO\ or its predecessor
{\sc\small aMC@NLO} in refs.~\cite{Alwall:2014hca,Frederix:2011zi}.
There are small numerical differences (${\cal O}(3\%)$) with 
ref.~\cite{Alwall:2014hca}, which are almost entirely due to 
the choice of the value of $\aem$, and to a very minor extent to
that of $m_t$. As far as ref.~\cite{Frederix:2011zi} is concerned, 
different choices had been made there for the top and Higgs masses,
and for the reference scale.

\begin{table}
\begin{center}
\begin{tabular}{l|ccc}
$\sigma$(pb)  & 8 TeV & 13 TeV & 100 TeV\\
\hline
LO & $1.001\mydot 10^{-1}(2.444\mydot 10^{-3})$ &
 $3.668\mydot 10^{-1}(1.385\mydot 10^{-2})$ & 
 $24.01(2.307)$\\
\hline
NLO QCD	& $2.56\mydot 10^{-2}(4.80\cdot 10^{-4})$ & 
 $1.076\mydot 10^{-1}(3.31\cdot 10^{-3})$ & 
 $9.69(0.902)$ \\
\hline
NLO weak & $-1.22\mydot 10^{-3}(-2.04\cdot 10^{-4})$ & 
 $-6.54\mydot 10^{-3}(-1.14\cdot 10^{-3})$ & 
 $-0.712(-0.181)$ \\
\end{tabular}
\end{center}
\caption{\label{tab:rates}
LO, NLO QCD, and NLO weak contributions to the total rate (in pb),
for three different collider energies. The results in parentheses 
are relevant to the boosted scenario, eq.~(\protect\ref{boost}).
}
\end{table}
\begin{table}
\begin{center}
\begin{tabular}{l|ccc}
$\delta_{\rm NLO}(\%)$ & 8 TeV & 13 TeV & 100 TeV\\
\hline
QCD & $+25.6^{+6.2}_{-11.8}~(+19.6^{+ 3.7}_{-11.0})$ & 
 $+29.3^{+7.4}_{-11.6}~(+23.9^{+ 5.4}_{-11.2}) $ &
 $+40.4 ^{+9.9}_{-11.6} ~(+39.1^{+ 9.7}_{-10.4})$ \\
\hline
weak &$-1.2 ~(-8.3)$ & $-1.8 ~(-8.2)$ & $-3.0~ (-7.8)$ \\
\end{tabular}
\end{center}
\caption{\label{corrQCDweak}
NLO QCD and weak contributions, as fractions of the corresponding
LO cross section. The results in parentheses 
are relevant to the boosted scenario, eq.~(\protect\ref{boost}).
In the case of QCD, the results of scale variations are also shown.
}
\end{table}
The predicted rates (in pb) are given in table~\ref{tab:rates};
the values outside parentheses are the fully-inclusive ones,
while those in parentheses are relevant to the boosted scenario;
in both cases, the NLO QCD contributions are sizable and positive.
As far as the NLO weak contributions are concerned, they are negative
and in absolute value rather small in the fully inclusive case, although 
their relative impact w.r.t.~that of QCD tends to increase with the 
collider energy. This picture is reversed (i.e.~the impact slightly
decreases) in the boosted 
scenario\footnote{Having said that, we also remark that the cuts
of eq.~(\ref{boost}) are imposed irrespective of the collider energy.
By increasing the c.m.~energy, one would have to increase the required 
minimal $\pt$'s in order to have similarly boosted configurations.}, 
where on the other hand the absolute values of the weak contributions
are non-negligible. These features can be understood more directly
by looking at the NLO contributions as fractions\footnote{The statistics
we have employed in the computation of the cross sections is such that
the typical error affecting such fractions, in the present and forthcoming
tables, is of the order of $0.1\%$.} of the corresponding LO cross
section; they are reported in this form in table~\ref{corrQCDweak}.
In that table, the entries of the first (second) row are the ratios 
of the entries in the second (third) row over those in the first row
of table~\ref{tab:rates}. One sees that the QCD contributions increase
the LO cross sections by 25\%(20\%) to 40\%, while the weak ones decrease
it by 1\% to 3\% in the fully-inclusive case, and by 8\% when the
cuts of eq.~(\ref{boost}) are applied. In the first row of
table~\ref{corrQCDweak} we also report (by using the usual ``error''
notation) the fractional scale uncertainty that affects the 
LO+NLO QCD rates. This is computed by taking the envelope of the 
cross sections that result from the scale variations as given
in eq.~(\ref{scalevar}), and by dividing it by the LO predictions
obtained with central scales. Note that this is not
the usual way of presenting the scale systematics (which entails
using the central LO+NLO prediction as a reference), and thus
the results of table~\ref{corrQCDweak} might seem, at the first 
glance, to be larger than those reported in ref.~\cite{Alwall:2014hca},
but are in fact perfectly consistent with those. Our choice here is
motivated by the fact that, by using the LO cross sections as 
references, we can compare NLO QCD and weak effects on a similar footing.
The main message of table~\ref{corrQCDweak} is, then, that in the
fully inclusive case the weak contributions are entirely negligible
in view of the scale uncertainties that affect the numerically-dominant
LO+NLO QCD cross sections. On the other hand, in the boosted scenario
they become comparable with the latter, and they must thus be taken
into account. This feature will also be evident when differential
distributions will be considered (see sect.~\ref{sec:distr}).

The impacts of the individual partonic channels on the 
NLO weak contributions are reported in table~\ref{corrchannel}, 
still as fractions of the LO cross sections -- hence, the sum of 
all the entries in a given column of table~\ref{corrchannel} is equal to 
the entry in the same column and in the last row of table~\ref{corrQCDweak}.
We point out that this breakdown into individual partonic contributions,
which is rather commonly shown in the context of EW calculations,
is unambiguous because QCD-induced singularities are only of
soft type (see sect.~\ref{sec:calc}), and thus real-emission matrix
elements and their associated Born-like counterparts have the same
initial-state partons. From table~\ref{corrchannel} we see, as is 
expected, that the dominance of the $gg$ channel, which is moderate
at 8~TeV, rapidly increases with the collider c.m.~energy. This trend
is mitigated when the cuts of eq.~(\ref{boost}) are applied, to the extent
that, at the LHC, the $u\bar{u}+d\bar{d}$ cross section is larger than
or comparable to the $gg$ one: the boosted scenario forces the Bjorken
$x$'s to assume larger values, where the quark densities are of similar
size as that of the gluon.
\begin{table}
\begin{center}
\begin{tabular}{l|ccc}
$\delta_{\rm NLO}(\%)$ & 8 TeV & 13 TeV & 100 TeV\\
\hline
$gg$ & $-0.67~(-2.9)$ & $-1.12~(-4.0)$ & $ -2.64 ~(-6.8)$ \\
\hline
$u\bar{u}$ & $-0.01~(-3.2)$ & $-0.15~(-2.3)$ & $-0.10~(-0.5)$ \\
\hline
$d\bar{d}$ & $-0.55~(-2.2)$ & $-0.52~(-1.9)$ & $-0.23~(-0.5)$ \\
\hline
$ug$ & $+0.03~(+0.02)$ & $+0.03~(+0.01)$ & $+0.01~(<0.01)$ \\
\hline
$dg$ & $-0.02~(-0.01)$ & $-0.02~(-0.01)$ & $-0.01~(>-0.01)$ \\
\end{tabular}
\end{center}
\caption{\label{corrchannel}
Breakdowns per partonic channel of the results of 
table~\protect\ref{corrQCDweak} for the NLO weak contributions.
The results in parentheses are relevant to the boosted scenario, 
eq.~(\protect\ref{boost}). By $u$ and $d$ we understand $c$ and $s$
as well, respectively. By $ug$ and $dg$ we understand $\bar{u}g$
and $\bar{d}g$ as well, respectively. 
}
\end{table}

We now turn to considering the contributions due to processes that
feature an extra weak boson in the final state, on top of the Higgs
which is present by definition; we remind the reader that these 
contributions have been denoted by HBR (see table~\ref{tab:names}).
The relevant results are shown in table~\ref{corrVrad}, as fractions of 
the corresponding LO cross section; hence, they are directly comparable 
to the last row of table~\ref{corrQCDweak}. Note that, in the case of 
the $t\bt HH$ final state, a kinematic configuration contributes to
the boosted scenario provided that the Higgs-$\pt$ cut of eq.~(\ref{boost})
is satisfied for at least one of the two Higgses. From tables~\ref{corrVrad}
and~\ref{corrQCDweak}, one sees that the HBR and NLO weak contributions,
in the case of the fully-inclusive cross sections, tend to cancel each
other to a good extent: at the 75\%, 50\%, and 30\% level 
at 8, 13 and 100~TeV respectively. This is not true in
the boosted scenario: although the HBR cross sections grow faster than
the LO ones (being 0.9\% of the latter in the fully-inclusive case,
and 1.7\% in the boosted one), their growth is slower than that of
their NLO-weak counterparts. Both contributions feature Sudakov 
logarithms, but we point out that the overall scaling behaviour
in hadronic collisions is determined, among other things, by the 
complicated interplay between that of the matrix elements, and the
parton luminosities; the latter are not the same in the case of the NLO-weak
and HBR contributions. This has several consequences. For example,
we note that the relative individual contributions to the HBR cross 
sections behave differently with the collider energy: the $W$-emission 
contribution decreases, while the $Z$- and $H$-emission ones  increase, owing
to the presence of $gg$-initiated partonic processes. Furthermore, the growth
of PDFs at small $x$'s implies that processes are closer to threshold
than the collider energy would naively imply, and thus the phase-space
suppression due to the presence of an extra massive particle in the HBR
processes is not negligible. Finally, this mass effect also implies that
the Bjorken $x$'s relevant to HBR are slightly larger than those relevant
to the NLO-weak contributions, and are thus associated on average with 
slightly smaller luminosity factors. As was already discussed in
sect.~\ref{sec:calc}, the results of table~\ref{corrVrad} are an upper bound 
for the HBR contributions when these are subject to extra boson-tagging 
conditions, which have not been considered here. On the other hand, 
nothing prevents one from defining the $t\bt H$ cross section inclusively in
any extra weak-boson radiation; given the opposite signs of the NLO-weak and
HBR cross sections, this may possibly be beneficial (for example, if
constraining or measuring $\tthc$). Such a definition is fully consistent
with perturbation theory, since both HBR and NLO-weak contributions are
of $\ord(\asa{2}{2})$.
\begin{table}
\begin{center}
\begin{tabular}{l|ccc}
$\delta_{\rm HBR}(\%)$ & 8 TeV & 13 TeV & 100 TeV\\
\hline
$W$ &$+0.42(+0.74)$ &$+0.37(+0.70)$ &$+0.14(+0.22)$ \\
\hline
$Z$ &$+0.29(+0.56)$ &$+0.34(+0.68)$ &$+0.51(+0.95)$ \\
\hline
$H$ &$+0.17(+0.43)$ &$+0.19(+0.48)$ &$+0.25(+0.53)$ \\
\hline\hline
sum &$+0.88(+1.73)$ &$+0.90(+1.86)$ &$+0.90(+1.70)$ \\
\end{tabular}
\end{center}
\caption{\label{corrVrad}
Contributions due to $W\!$, $Z$, and $H$ radiation, as fractions 
of the corresponding LO cross section. The results in parentheses 
are relevant to the boosted scenario, eq.~(\protect\ref{boost}).
}
\end{table}

All the results presented so far have been obtained in the
$\aem(m_Z)$ scheme. It is therefore interesting to check
what happens by considering the $G_\mu$ scheme, which entails
a different renormalisation procedure and different inputs.
In such a scheme we have (at the LO):
\beq
\frac{1}{\aem}=132.23\,.
\label{aemGmu}
\eeq
The LO results are presented in the first row of table~\ref{Gmu};
the second row displays the relative difference w.r.t.~their
$\aem(m_Z)$-counterparts of table~\ref{tab:rates}:
\beq
\Delta^{G_{\mu}}_\text{LO}=\frac{{\rm LO}-{\rm LO}^{G_{\mu}}}{{\rm LO}}\,.
\eeq
The latter figures constitute a simple cross check: given that the
LO cross section factorises $\as^2\aem$, at this perturbative order the 
difference can only be due to the values of the EW coupling constant,
and the $2.5\%$ reported in table~\ref{Gmu} is the difference\footnote{The
$\aem(m_Z)$- and $G_\mu$-scheme runs have been performed with different
statistics and seeds, so that other small differences are present.}
between the $\aem$'s of eqs.~(\ref{aemGmu}) and~(\ref{aem}).
Therefore, at the LO the EW-scheme dependence of the cross section is larger 
than (at the LHC) or comparable to (at 100~TeV) the NLO weak contribution
in the fully-inclusive case, while it is about a third of the latter
in the boosted scenario. When NLO corrections are included, however,
things do change. In the $G_\mu$ scheme, the NLO weak contributions
are positive for the fully-inclusive rates, at variance with the $\aem(m_Z)$ 
scheme; see the third row of table~\ref{Gmu}, where they are reported as 
fractions of the LO cross sections:
\beq
\delta^{G_{\mu}}_{{\rm weak}}=
\frac{{\rm NLO}_{{\rm weak}}^{G_{\mu}}}{{\rm LO}^{G_{\mu}}}\,,
\eeq
i.e.~they are the analogues of the quantities in the second row of
table~\ref{corrQCDweak}. More importantly, the differences between
the two schemes for the NLO-accurate weak cross sections are much
reduced w.r.t.~those at the LO. This is documented in the fourth row
of table~\ref{Gmu}, where we show the values of:
\beq
\Delta^{G_{\mu}}_\text{LO+NLO}=
\frac{{\rm LO}+{\rm NLO}_{{\rm weak}}-
({\rm LO}^{G_{\mu}}+{\rm NLO}_{{\rm weak}}^{G_{\mu}})}
{{\rm LO}+{\rm NLO}_{{\rm weak}}}\,,
\eeq
which are smaller in absolute value 
than their LO counterparts, and whose independence
of the collider energy is remarkable. Thus, in the boosted case one sees
that the fact that weak contributions have a significant impact on 
NLO-accurate cross sections is a conclusion that holds true 
in both of the EW schemes adopted in this paper. 
\begin{table}
\begin{center}
\begin{tabular}{r|ccc}
 & 8 TeV & 13 TeV & 100 TeV\\
\hline
LO$^{G_{\mu}}$(pb) & $9.758\cdot 10^{-2}(2.382\cdot 10^{-3})$ & 
 $3.575\cdot 10^{-1}(1.351\cdot 10^{-2})$ & 23.41(2.249) \\
\hline
$\Delta^{G_{\mu}}_\text{LO}$(\%) & $+2.5(+2.5)$ &
 $+2.5(+2.5)$ & $+2.5(+2.5)$ \\
\hline\hline
$\delta^{G_{\mu}}_{{\rm weak}}$(\%) & $+1.8(-5.1)$ &  
 $+1.3(-4.9)$ & $+0.1(-4.5)$ \\
\hline
$\Delta^{G_{\mu}}_\text{LO+NLO}$(\%) & $-0.5(-0.9)$ &
 $-0.5(-1.1)$ & $-0.6(-1.0)$ \\
\end{tabular}
\end{center}
\caption{\label{Gmu}
Results in $G_\mu$-scheme: Born cross sections, relative differences ($\Delta$) 
w.r.t.~those obtained in the $\aem(m_Z)$ scheme, and fractional NLO-weak 
contribution ($\delta$). The results in parentheses are relevant to the boosted 
scenario, eq.~(\protect\ref{boost}).
}
\end{table}

We conclude this section with a brief discussion on the impact of
the $b\bb$-initiated contributions, which we have ignored in our
NLO-accurate results, for the reasons explained in sect.~\ref{sec:calc}.
In table~\ref{bborders} we present the contributions to the Born
cross sections (again for the input parameters relevant to the 
$\aem(m_Z)$ scheme) due to all the relevant coupling-constant factors (see
eq.~(\ref{SigB})); we remind the reader than only the $\as^2\aem$ term
is included in the LO predictions shown so far. The $b\bb$ contribution
to $\Sigma_{3,0}$ appears in fact to be quite irrelevant (being at most 
$0.36\%$ at 100~TeV); those to $\Sigma_{3,1}$ and $\Sigma_{3,2}$
are comparable or slightly larger in absolute value, and furthermore
they tend to cancel each other. Given that there is no mechanism at the
NLO that could enhance the $b\bb$-initiated cross section in a much
stronger way than for the other partonic contributions at the same order, 
our assumption appears to be perfectly safe. It is thus of academic
interest the fact that the results for the $b\bb$-induced
$\Sigma_{3,q}$ coefficients do not obey the numerical hierarchy
suggested by their corresponding coupling-constant factors
(which hierarchy is violated owing to the opening of $t$-channel
diagrams, such as the one on the right of fig.~\ref{fig:diagtree2}).
When the mixed-coupling expansion will be fully automated in
\aNLO, one will easily verify whether such a feature survives
NLO corrections.
\begin{table}
\begin{center}
\begin{tabular}{c|ccc}
$\sigma_{b\bar{b}\to\tth}$(pb) & 8 TeV & 13 TeV & 100 TeV\\
\hline
$\as^2\aem\,\Sigma_{3,0}$ & $1.8\cdot 10^{-4}$ & $9.1\cdot 10^{-4}$ & 
 $8.6\cdot 10^{-2}$ \\
\hline
$\as\aem^2\,\Sigma_{3,1}$ & $-1.3\cdot 10^{-4}$ & $-1.5\cdot 10^{-3}$ & 
 $-1.3\cdot 10^{-1}$ \\
\hline
$\aem^3\,\Sigma_{3,2}$ & $3.1\cdot 10^{-4}$ & $1.6\cdot 10^{-3}$ &
 $1.9\cdot 10^{-1}$ \\
\end{tabular}
\end{center}
\caption{\label{bborders}
Leading, second-leading, and third-leading Born contributions due to 
the $b\bar{b}$ initial state.}
\end{table}

\subsection{Differential distributions\label{sec:distr}}
\begin{figure*}[]
 \center 
 \includegraphics[width=0.49\textwidth]{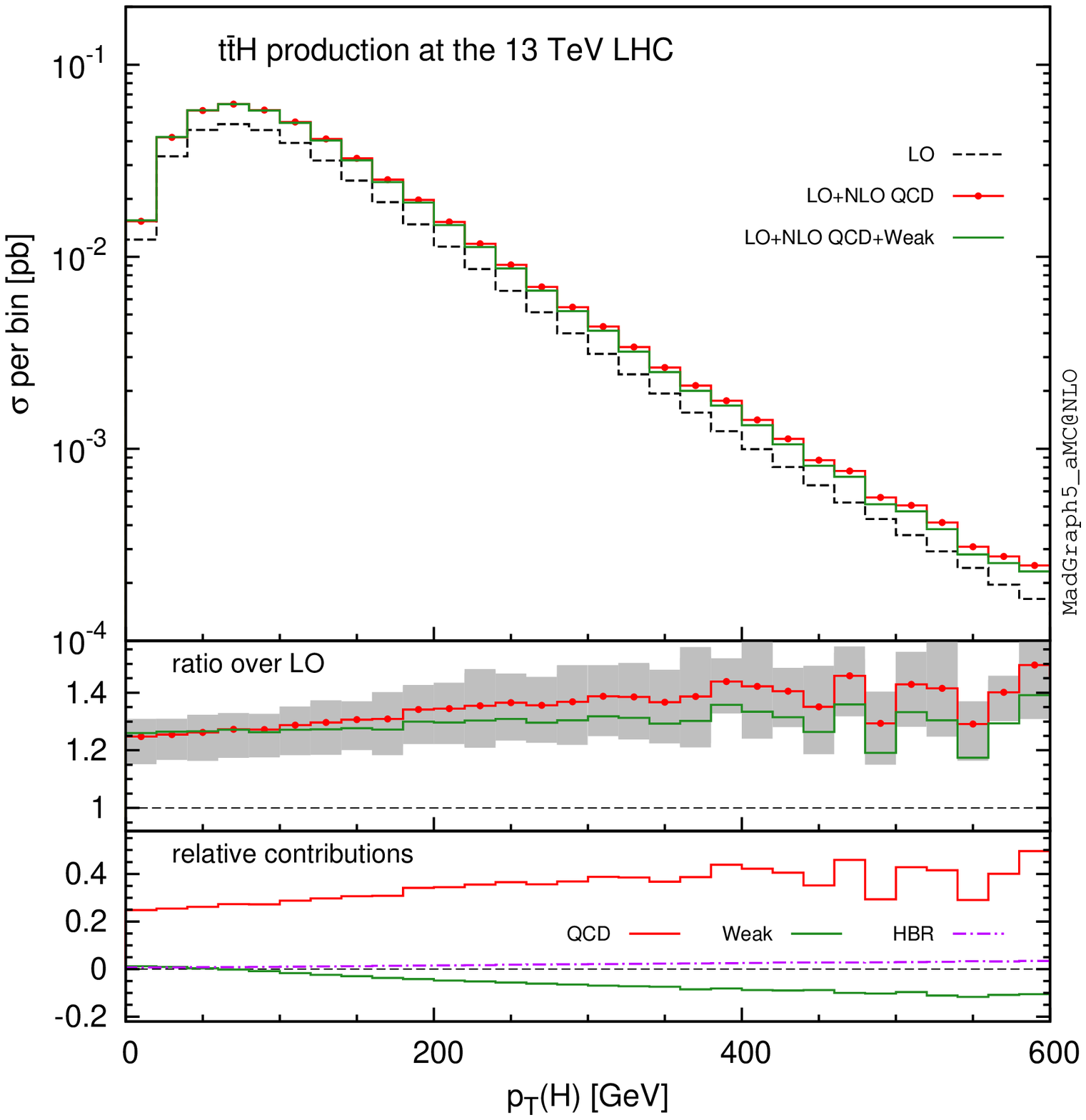}
  \includegraphics[width=0.49\textwidth]{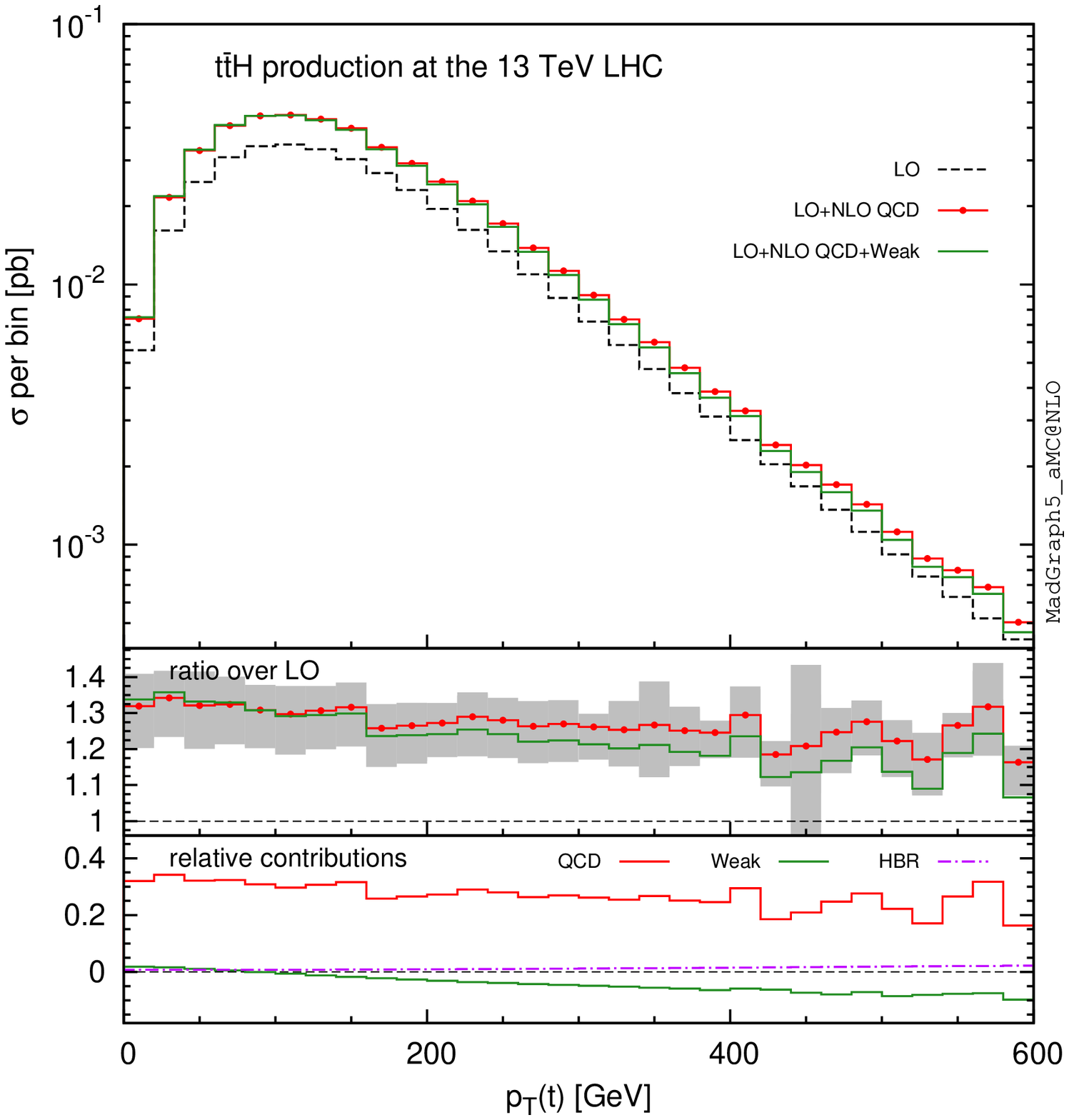}
  \includegraphics[width=0.49\textwidth]{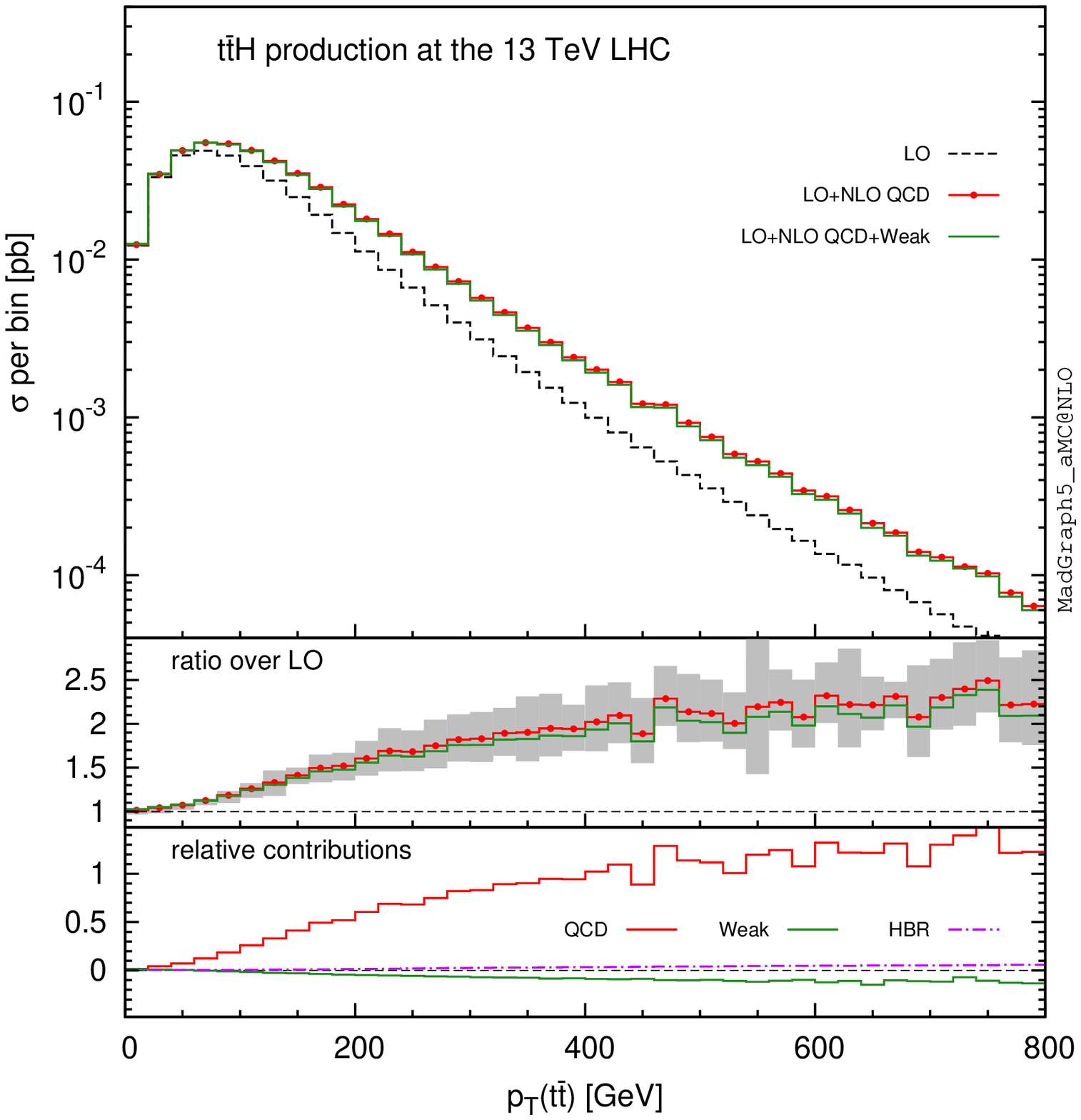}
  \includegraphics[width=0.49\textwidth]{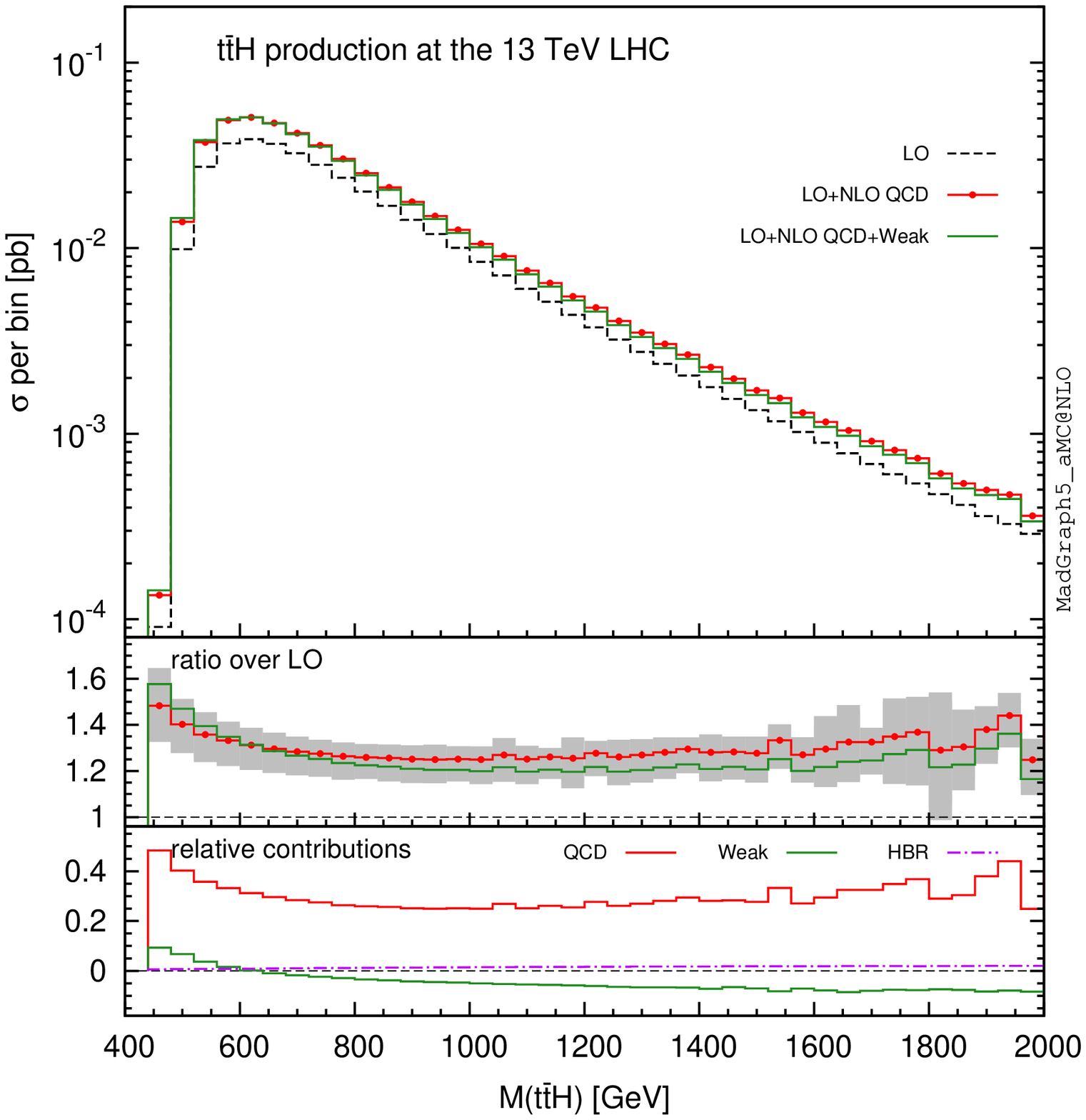}
  \includegraphics[width=0.49\textwidth]{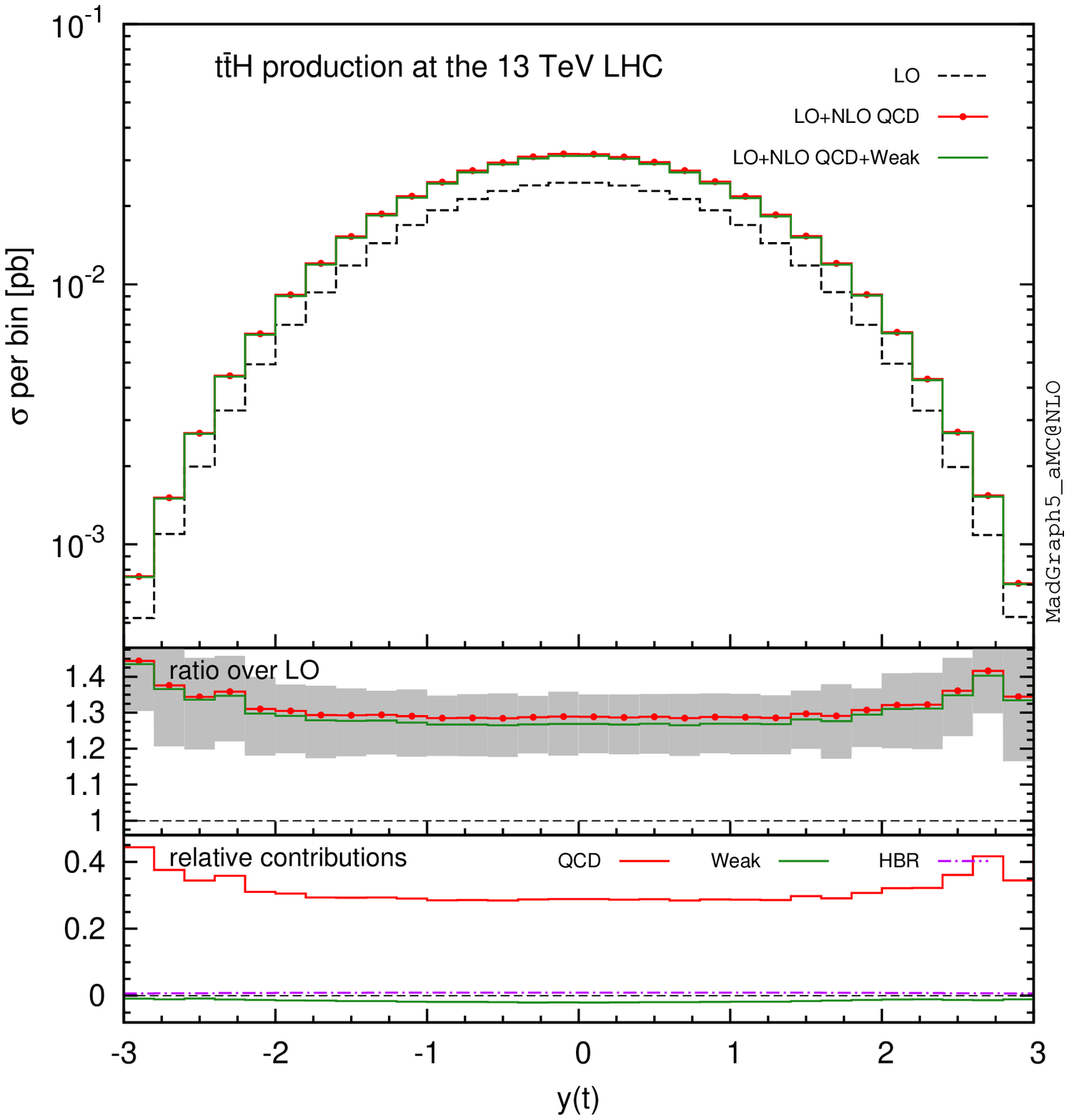}
  \includegraphics[width=0.49\textwidth]{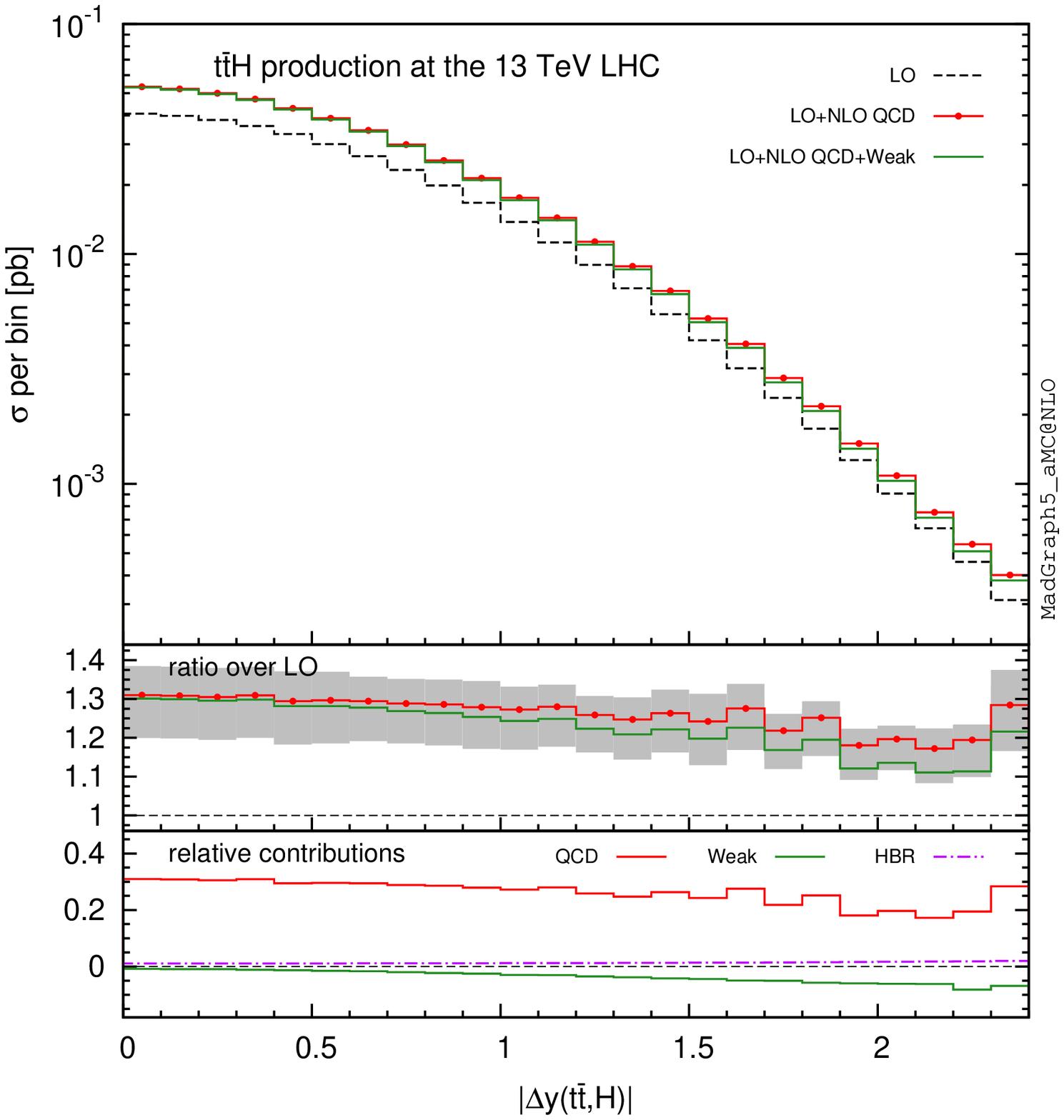}
\caption{\label{fig:plot-13} LO- and NLO-accurate results at 13 TeV.}
\end{figure*}

\begin{figure*}[]
 \center 
 \includegraphics[width=0.49\textwidth]{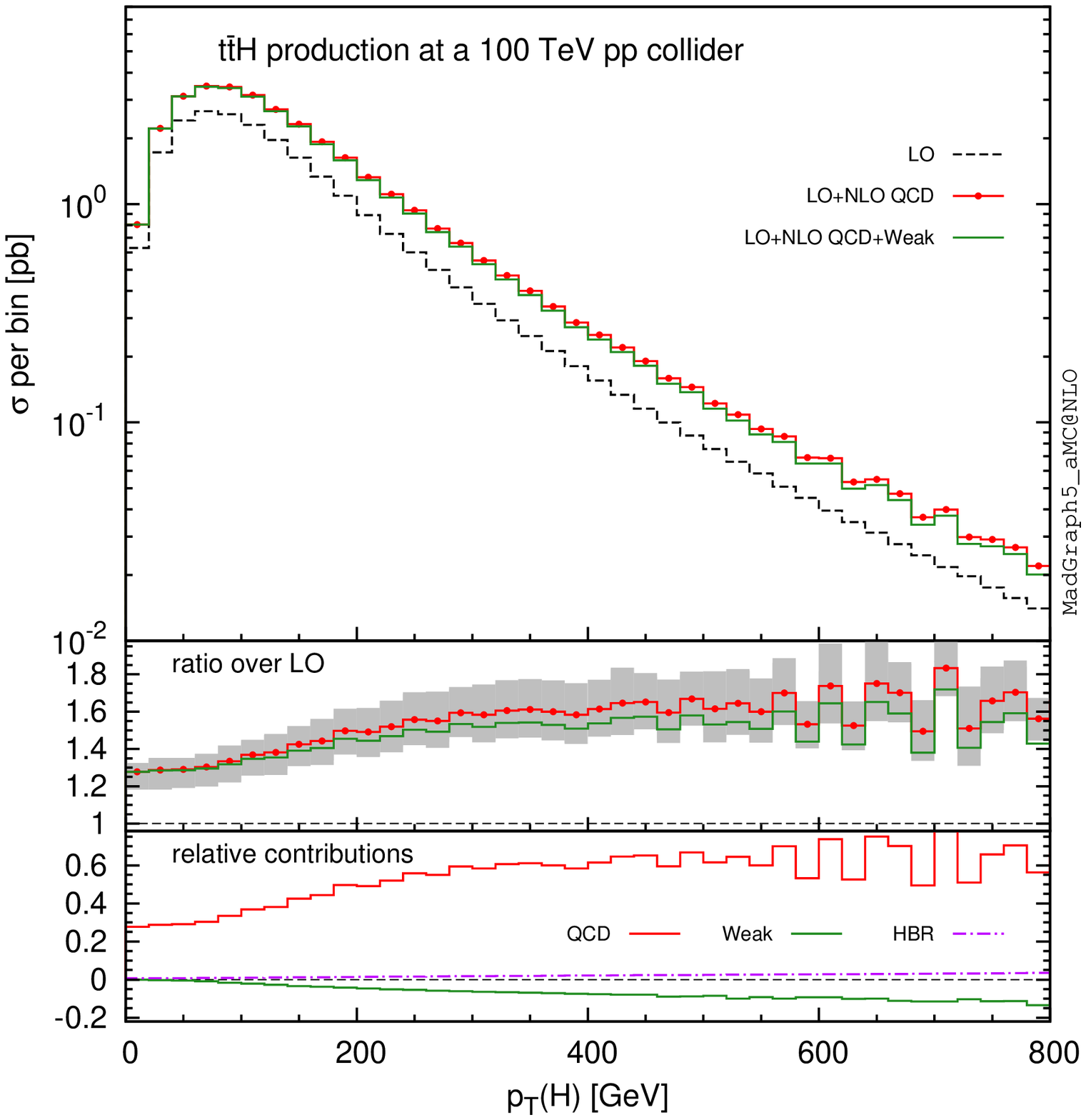}
  \includegraphics[width=0.49\textwidth]{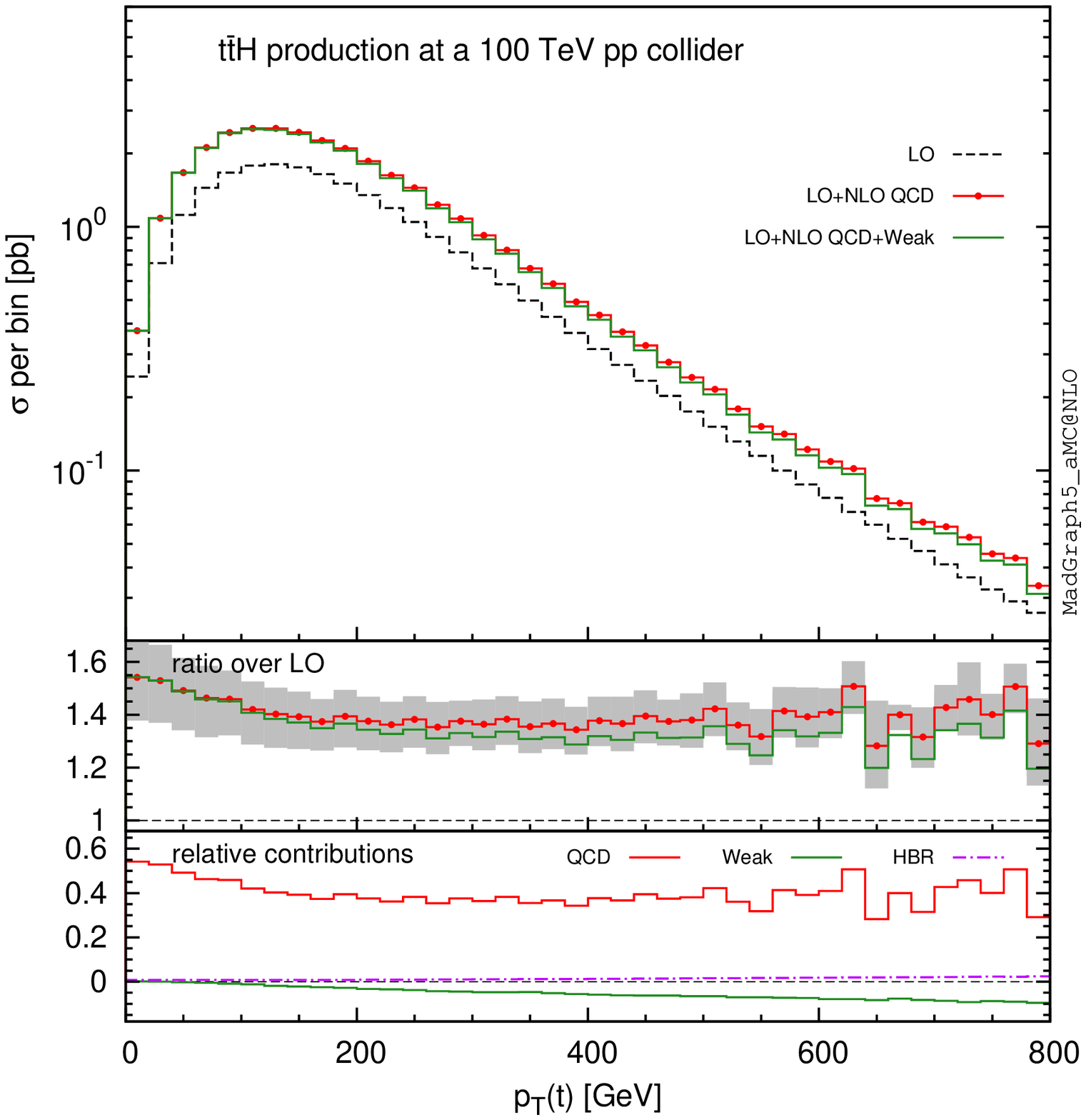}
  \includegraphics[width=0.49\textwidth]{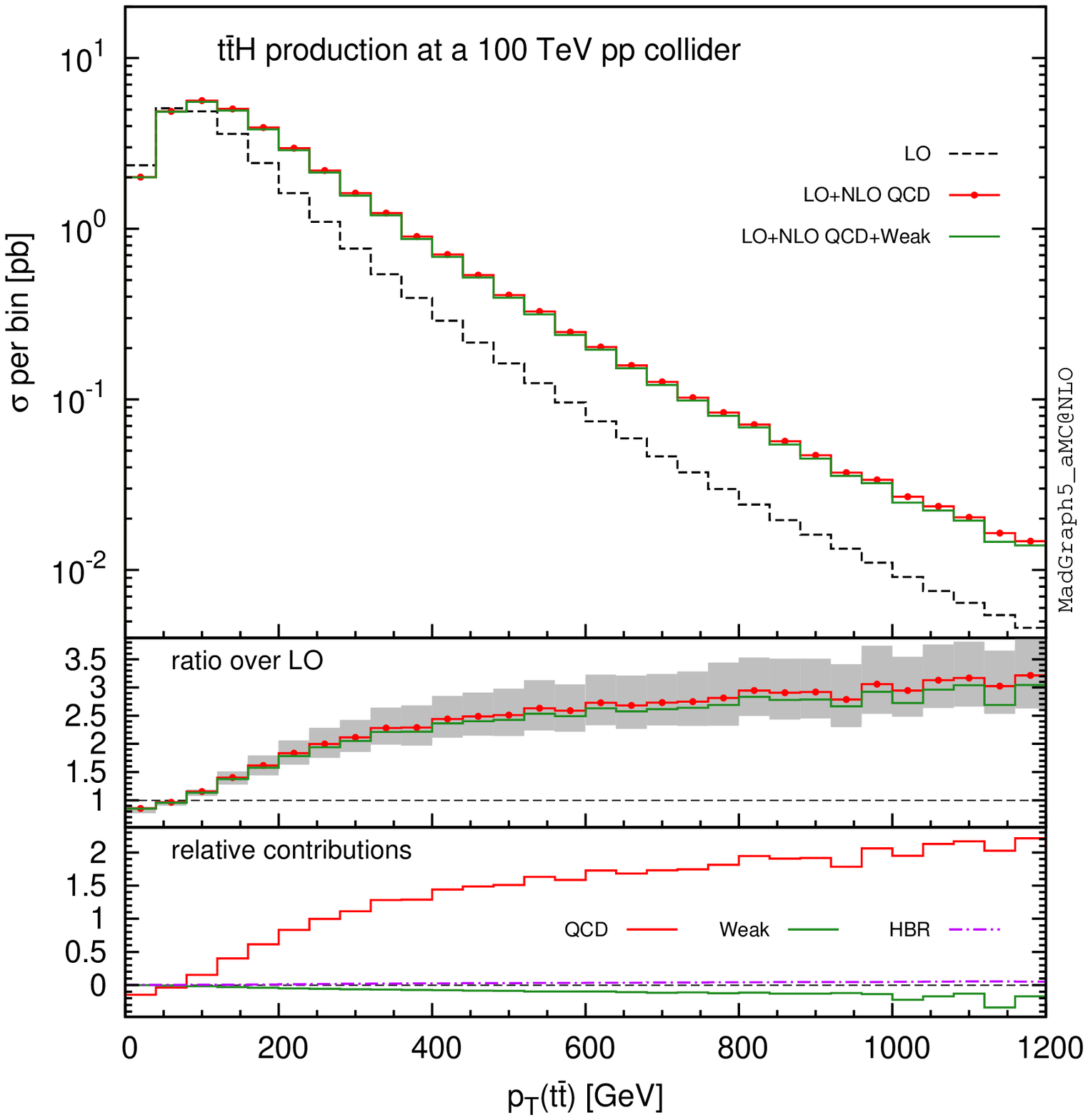}
  \includegraphics[width=0.49\textwidth]{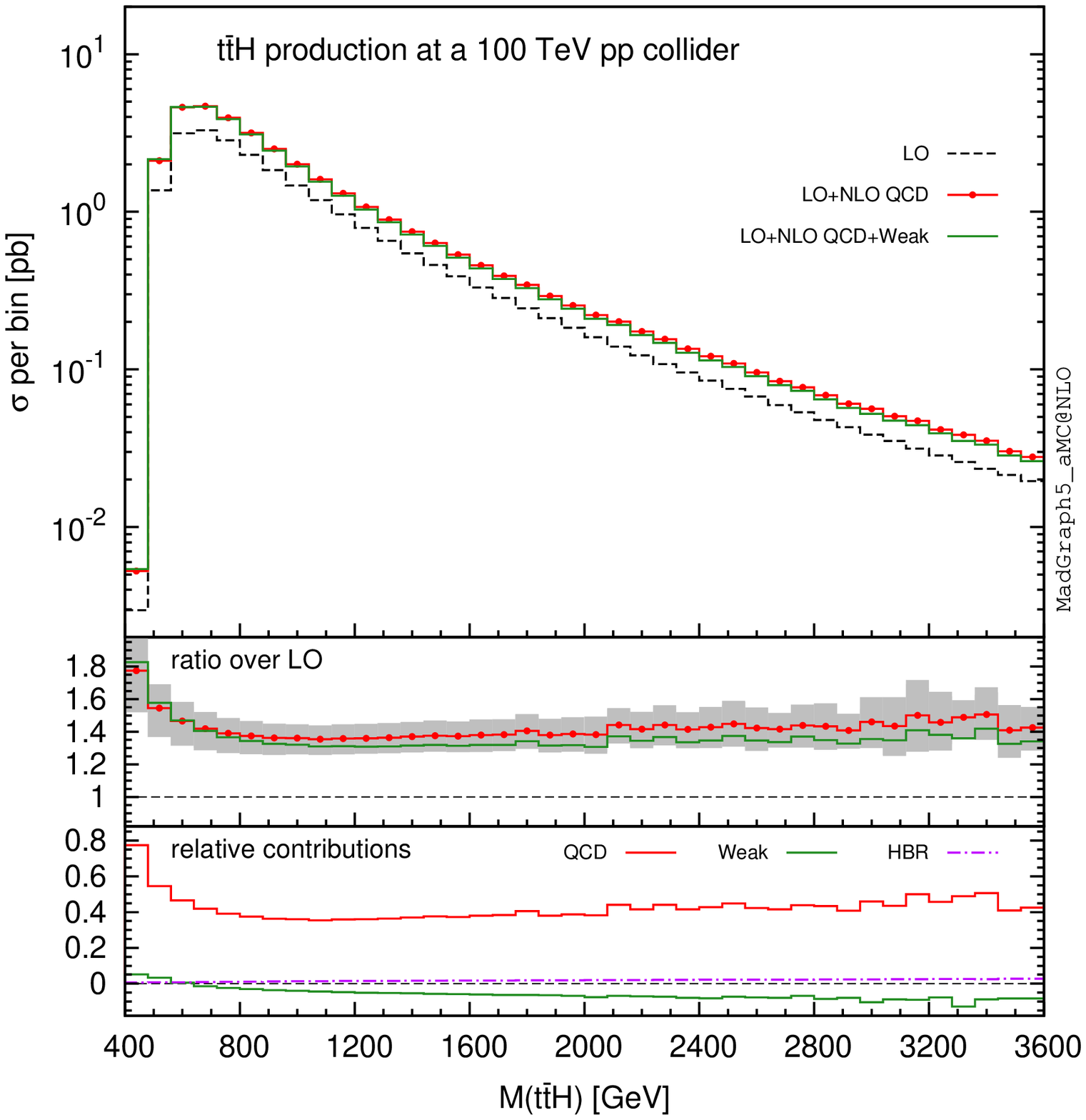}
  \includegraphics[width=0.49\textwidth]{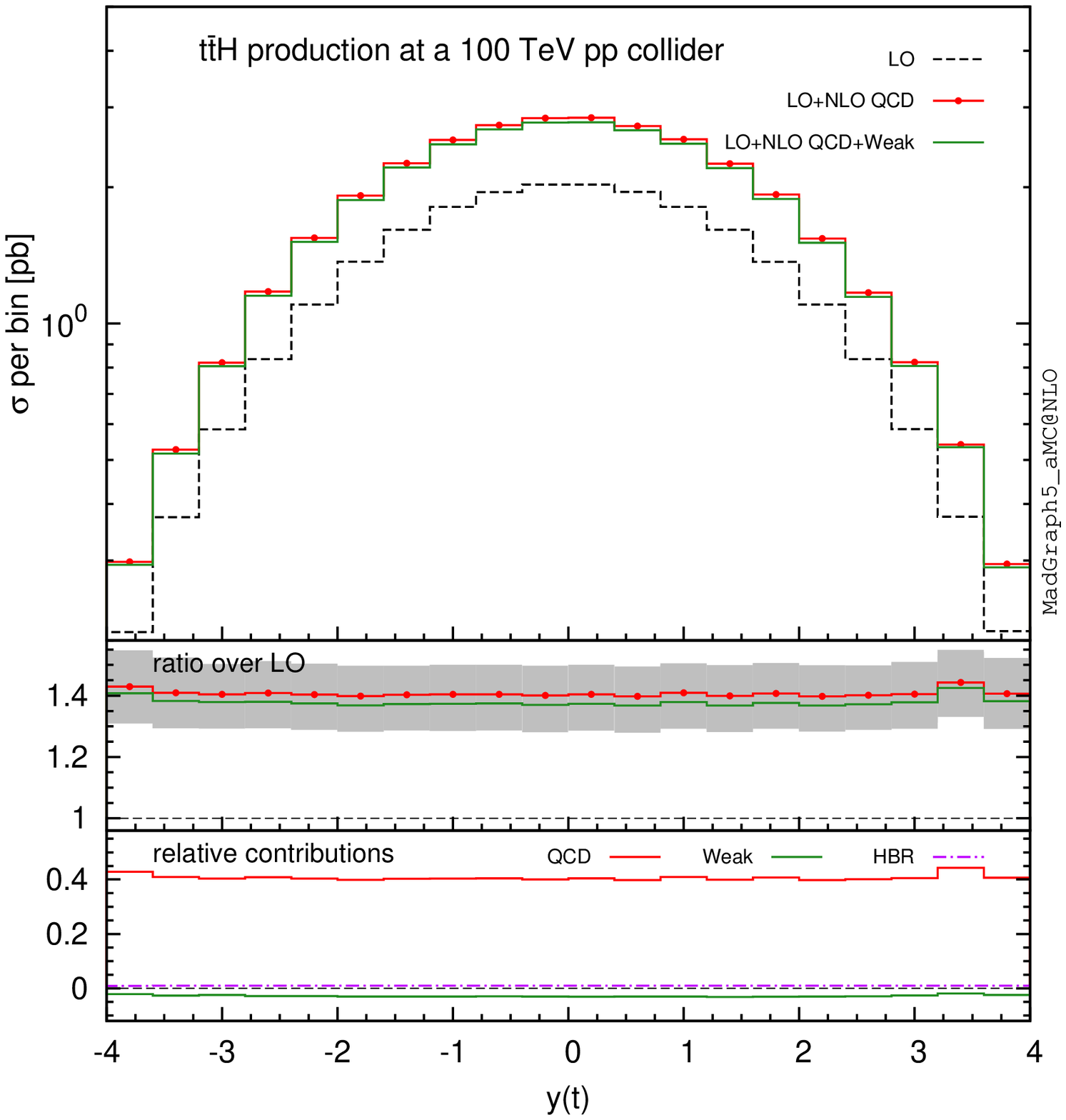}
  \includegraphics[width=0.49\textwidth]{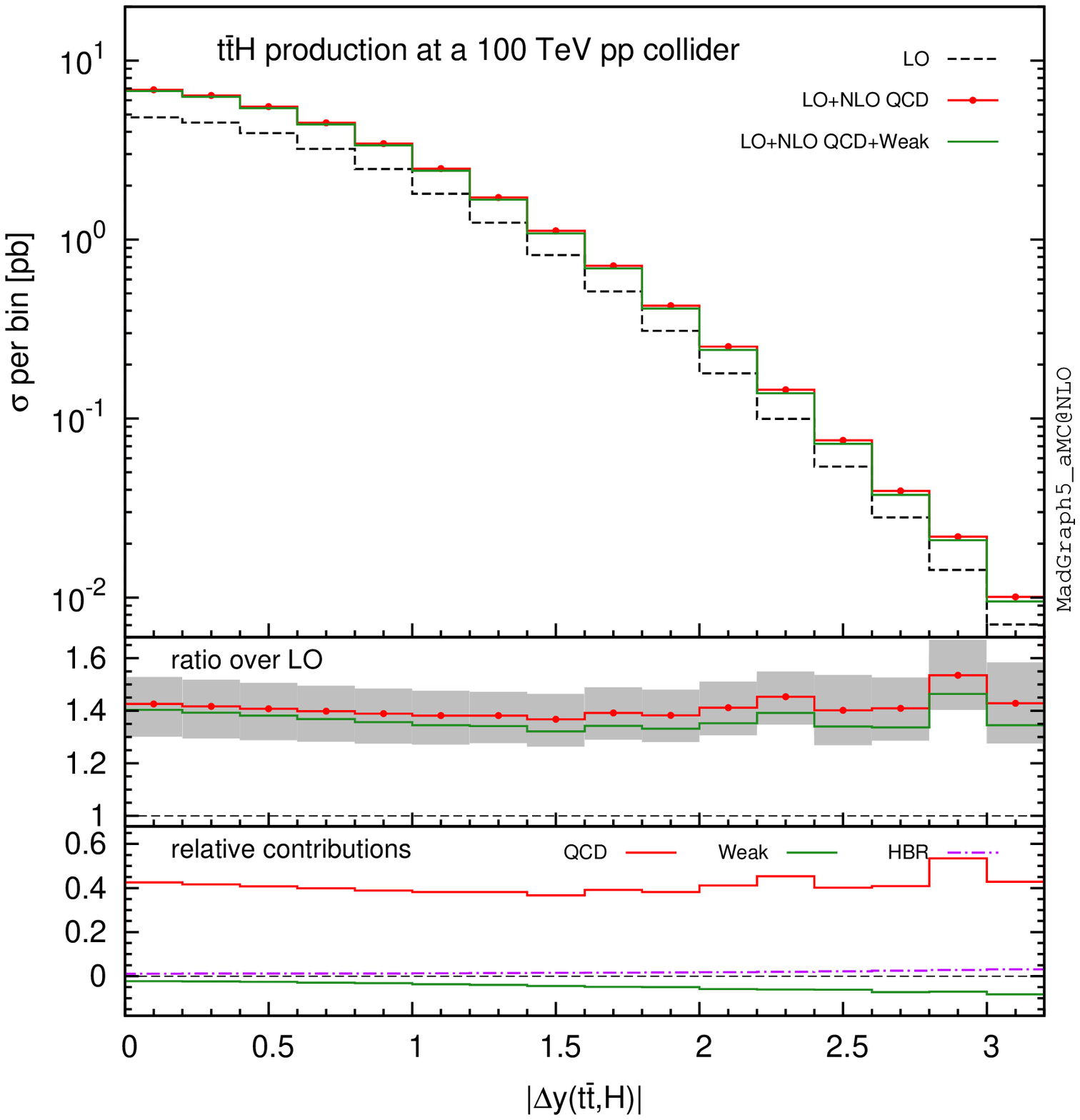}
\caption{\label{fig:plot-100} LO- and NLO-accurate results at 100 TeV.}
\end{figure*}

\begin{figure*}[]
 \center 
 \includegraphics[width=0.49\textwidth]{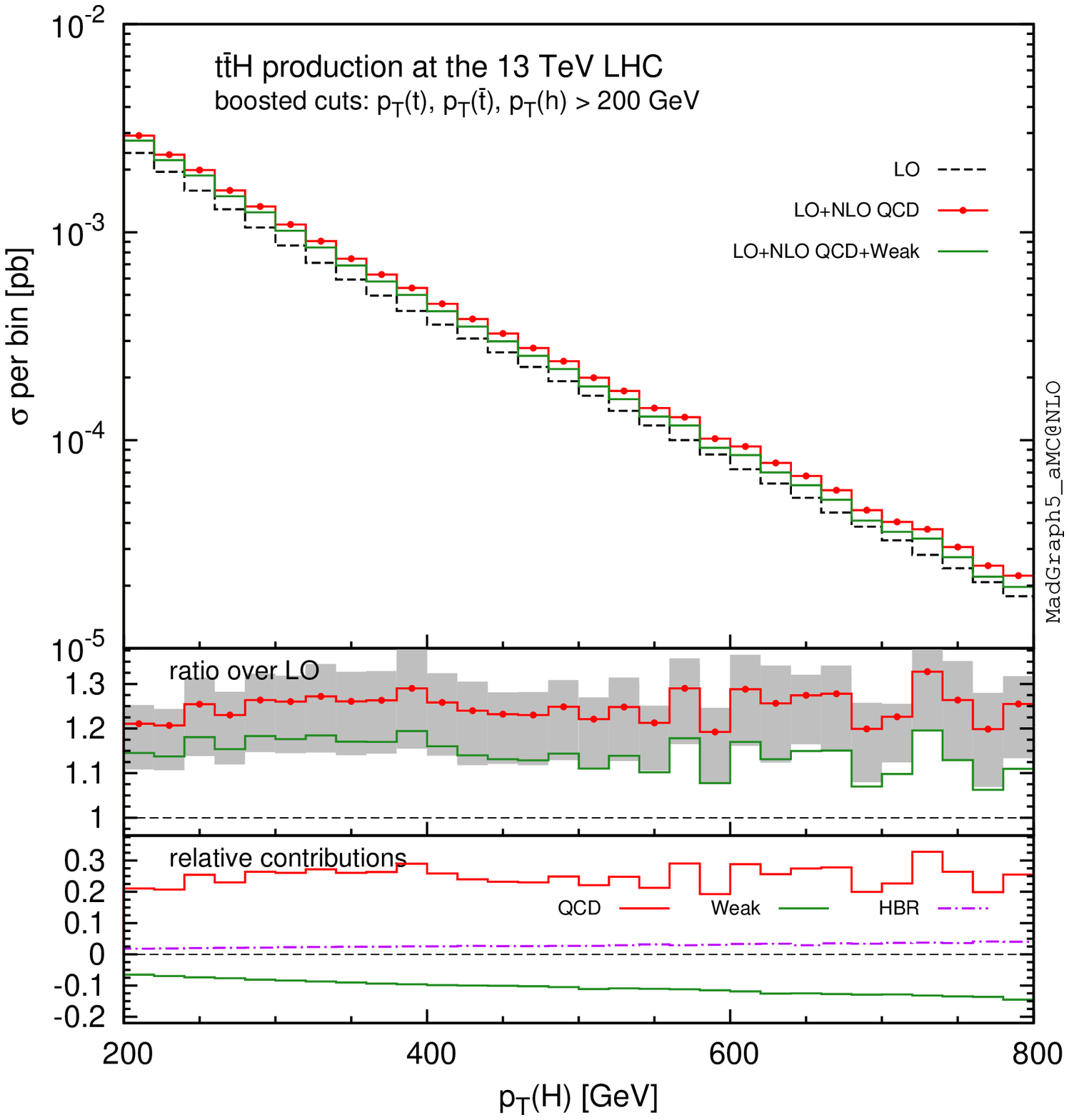}
  \includegraphics[width=0.49\textwidth]{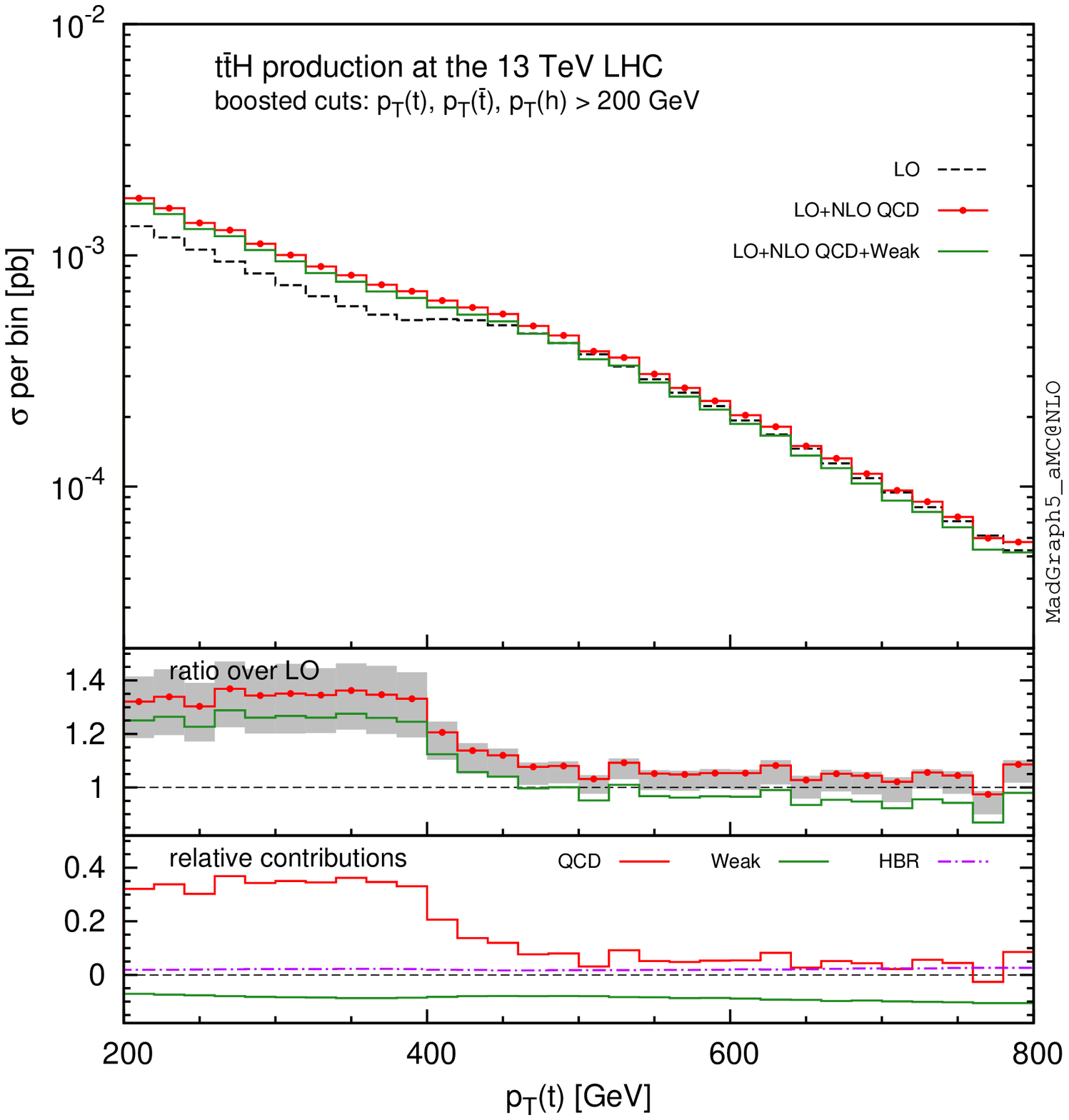}
  \includegraphics[width=0.49\textwidth]{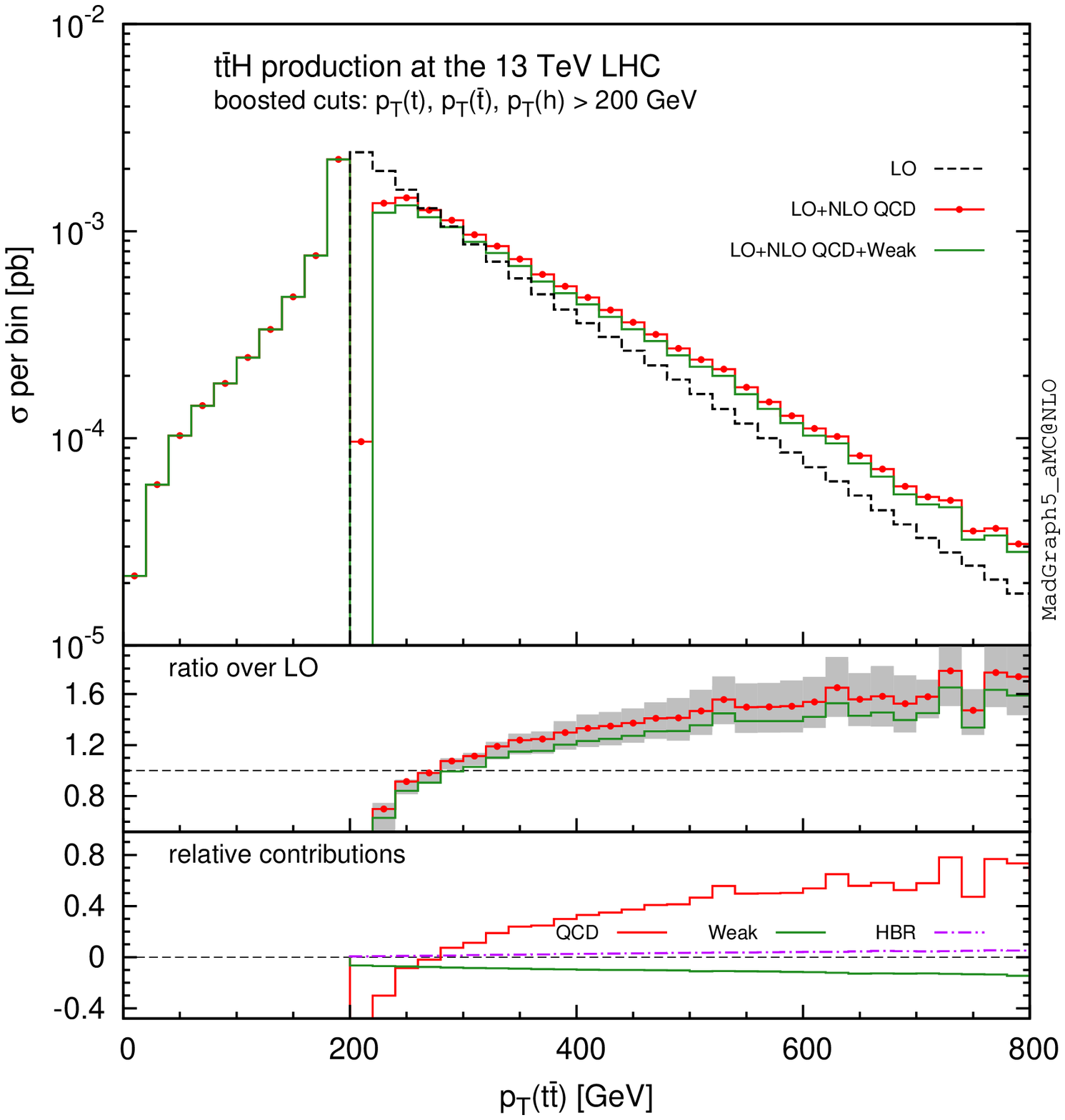}
  \includegraphics[width=0.49\textwidth]{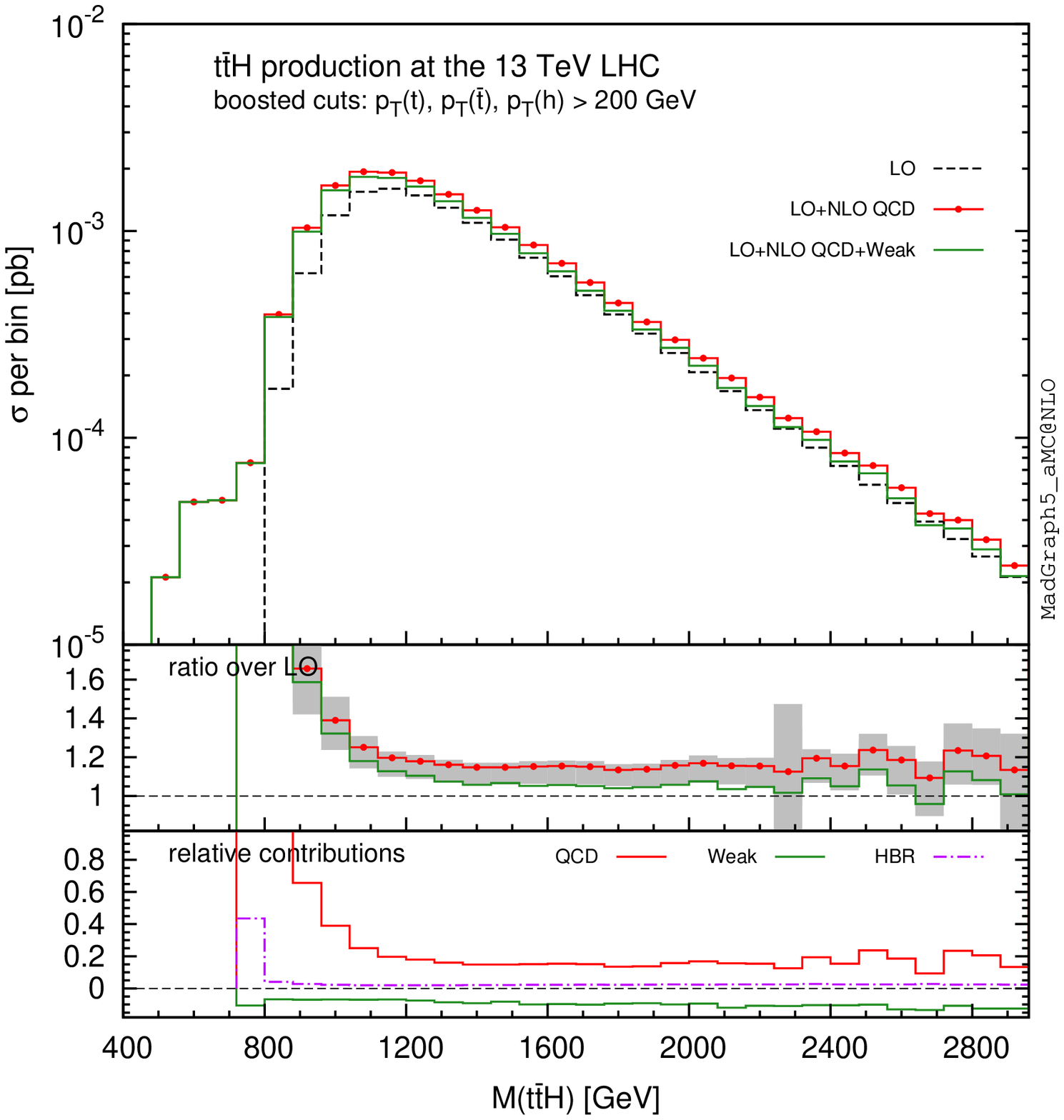}
  \includegraphics[width=0.49\textwidth]{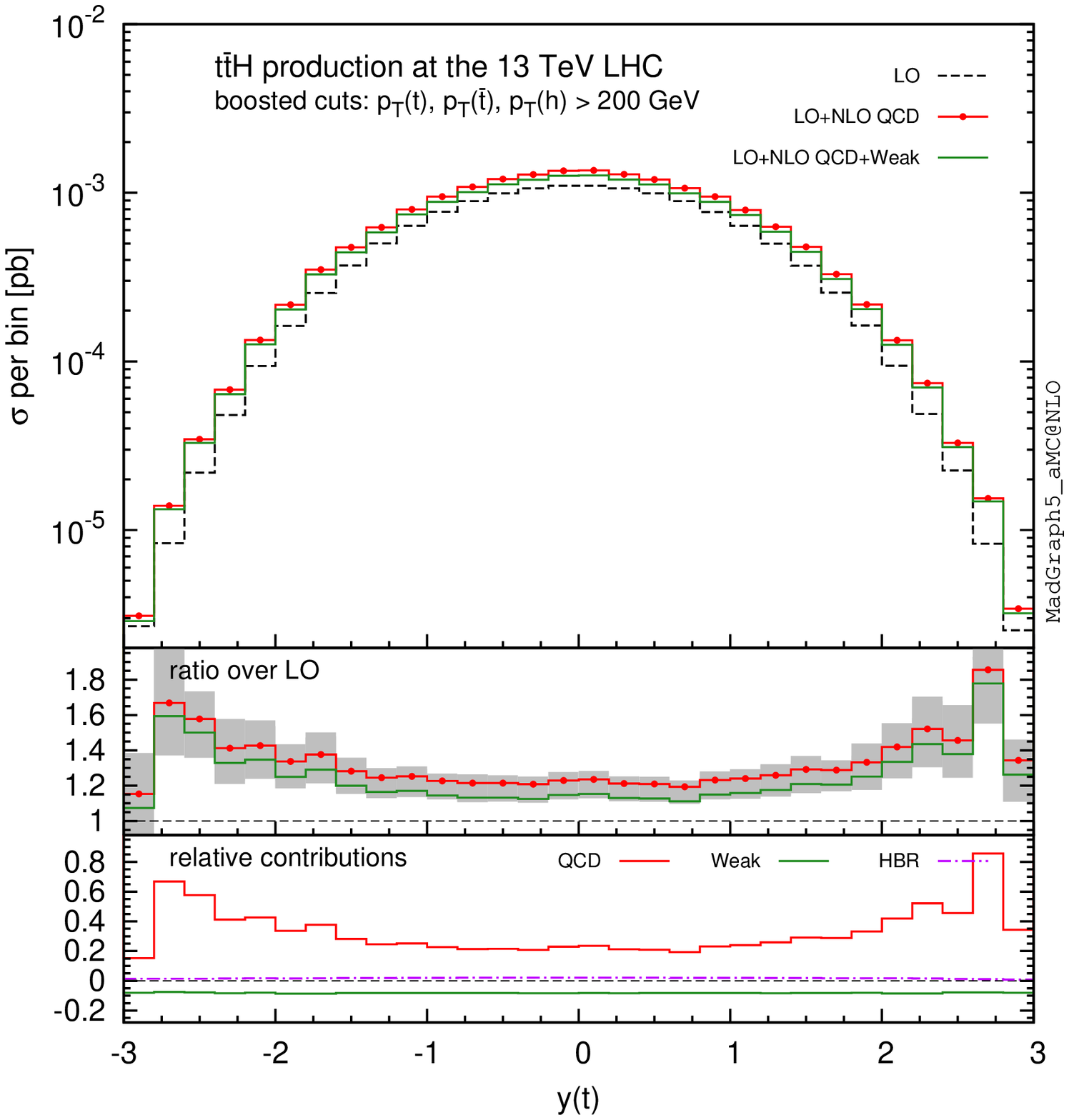}
  \includegraphics[width=0.49\textwidth]{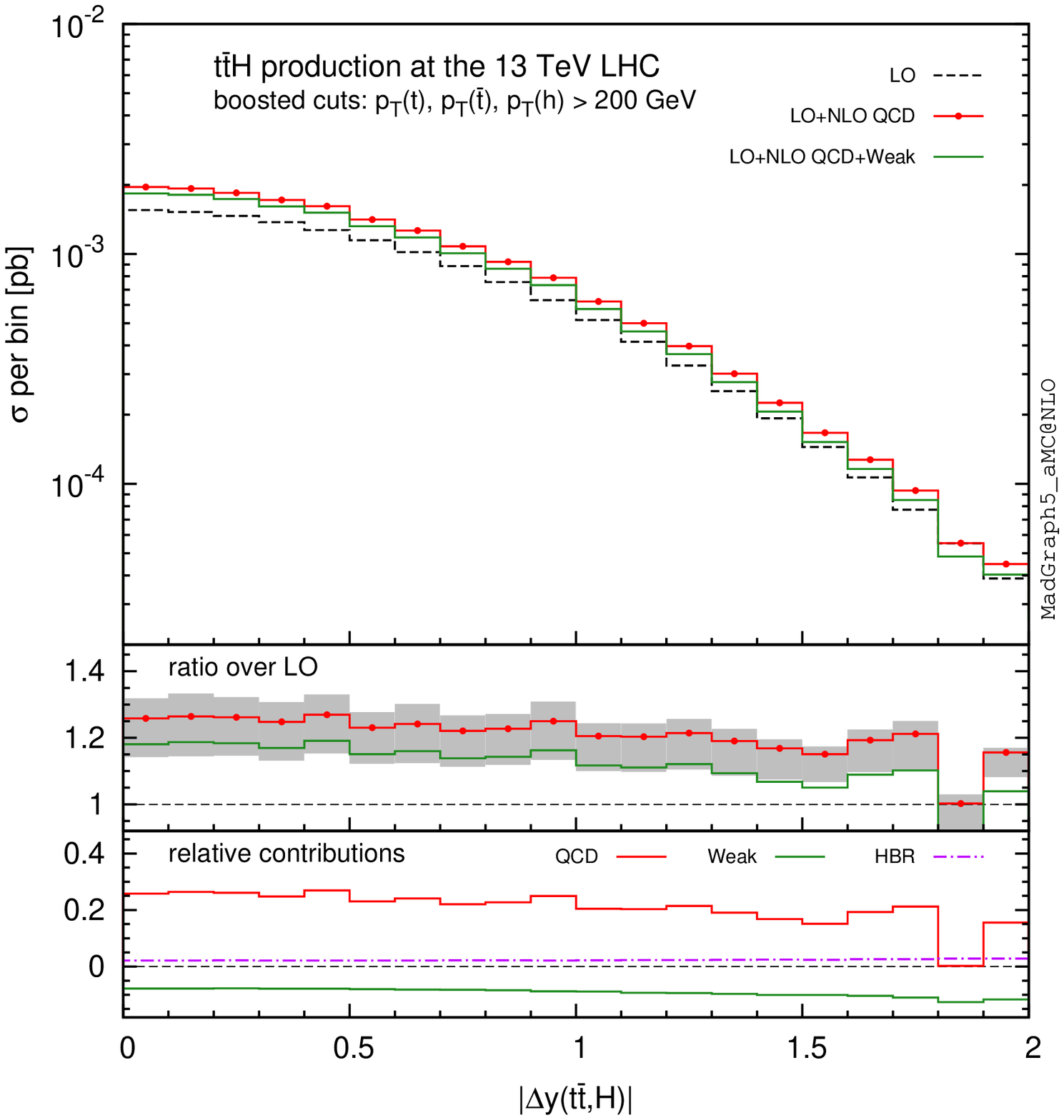}
\caption{\label{fig:plot-13-cuts} LO- and NLO-accurate results at 13 TeV,
in the boosted scenario.}
\end{figure*}

We now turn to presenting results for differential distributions.
In order to be definite, we have considered the following observables:
the transverse momenta of the Higgs ($\pt(H)$), top quark ($\pt(t)$),
and $t\bt$ pair ($\pt(t\bt)$), the invariant mass of the $t\bt H$
system ($M(t\bt H)$), the rapidity of the top quark ($y(t)$),
and the difference in rapidity between the $t\bt$ pair and the
Higgs boson ($\Delta y(t\bt,H)$). The corresponding six distributions are 
shown at a collider energy of 13~TeV (fig.~\ref{fig:plot-13}), 100~TeV
(fig.~\ref{fig:plot-100}), and 13~TeV in the boosted scenario of
eq.~(\ref{boost}) (fig.~\ref{fig:plot-13-cuts}). In the 
case of the HBR process $pp\to t\bt HH$, owing to the inclusive 
(in the two Higgses) definition of the latter the histograms relevant to 
the observables that depend explicitly on the Higgs four-momentum (i.e., 
$\pt(H)$, $M(t\bt H)$, and $\Delta y(t\bt,H)$) may receive up to two 
entries per event. 

Figures~\ref{fig:plot-13}, \ref{fig:plot-100}, 
and~\ref{fig:plot-13-cuts} have identical layouts. The main frame
displays three distributions, which correspond to the LO (black dashed),
LO+NLO QCD (red solid, superimposed with full circles), and LO+NLO QCD+NLO
weak (green solid) cross sections. The latter two distributions are
therefore the bin-by-bin analogues of the sum of the upper two entries
and of the sum of the three entries, respectively, in a given column of 
table~\ref{tab:rates}. The middle inset presents the ratios of the two
NLO-accurate predictions over the corresponding LO one -- 
these are therefore the
$K$ factors. Centered around the NLO QCD $K$ factor we show a mouse-grey
band, which represents the fractional scale uncertainty, defined in
full analogy to what has been done in table~\ref{corrQCDweak}. Finally,
the lower inset displays the ratios of the NLO QCD, NLO weak, and HBR
(dot-dashed magenta) contributions over the LO cross section -- these
are therefore the analogues of the first two lines of table~\ref{corrQCDweak}
and of the last line of table~\ref{corrVrad}, respectively.

\begin{figure*}[]
 \center 
 \includegraphics[width=0.49\textwidth]{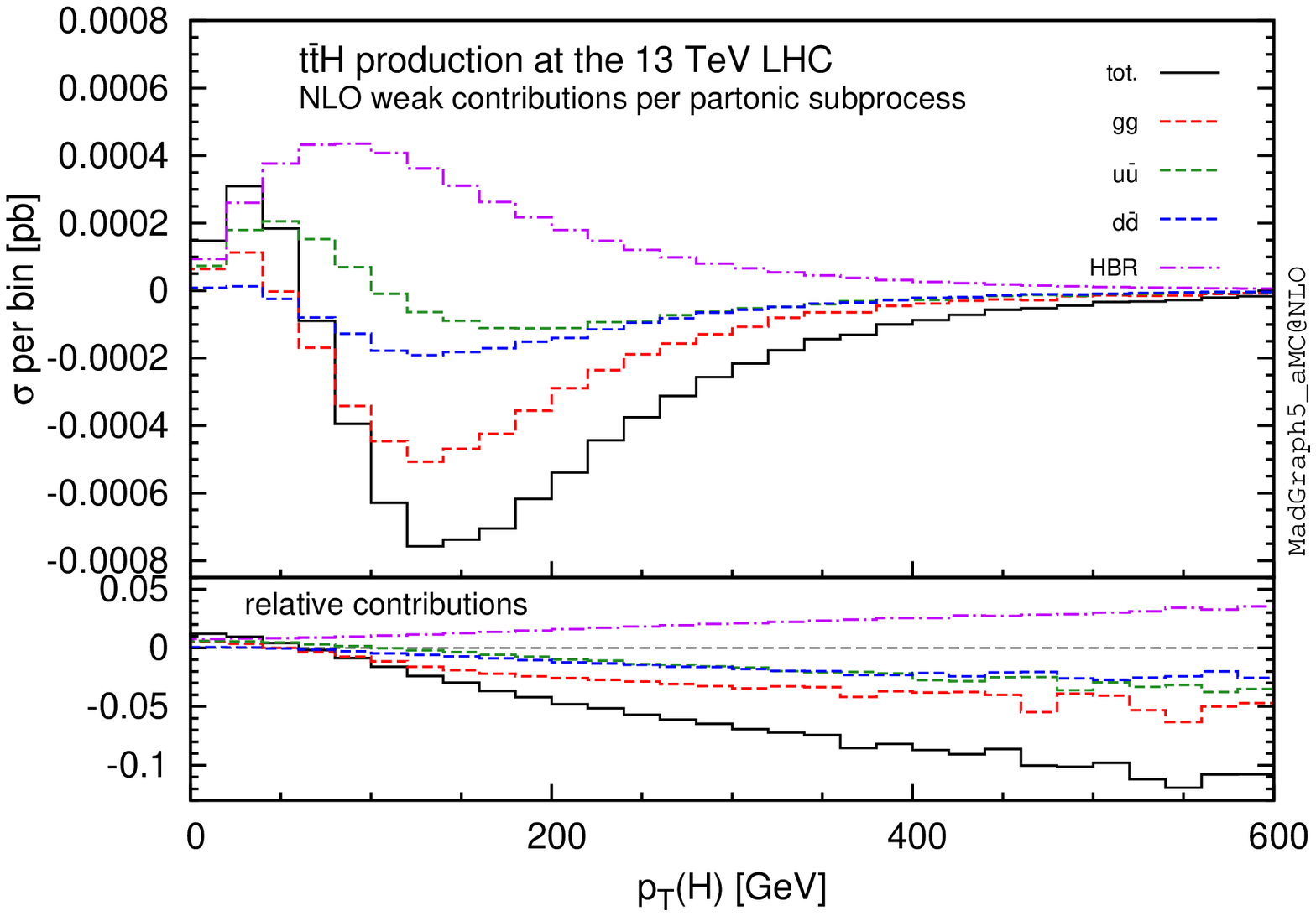}
  \includegraphics[width=0.49\textwidth]{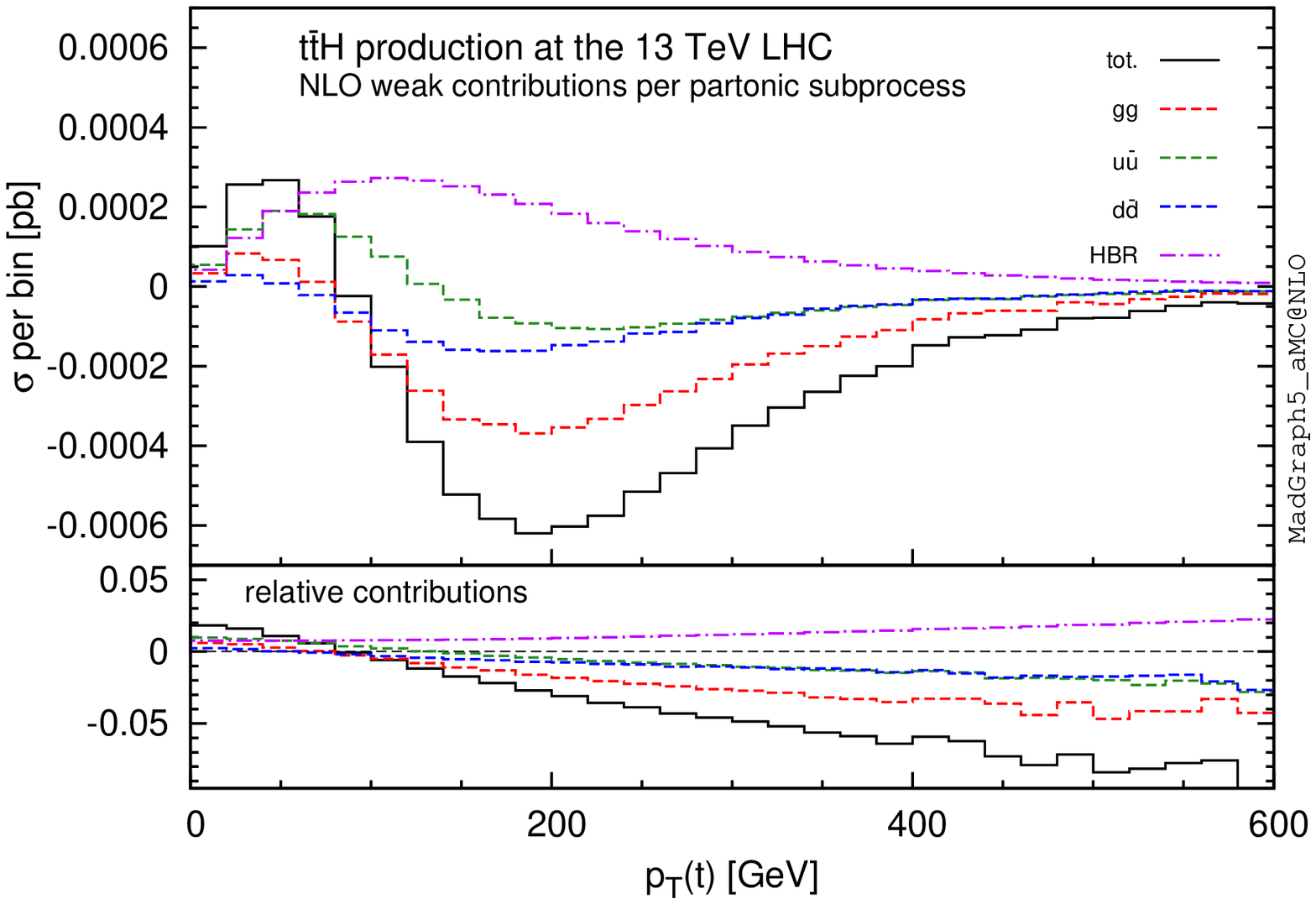}
  \includegraphics[width=0.49\textwidth]{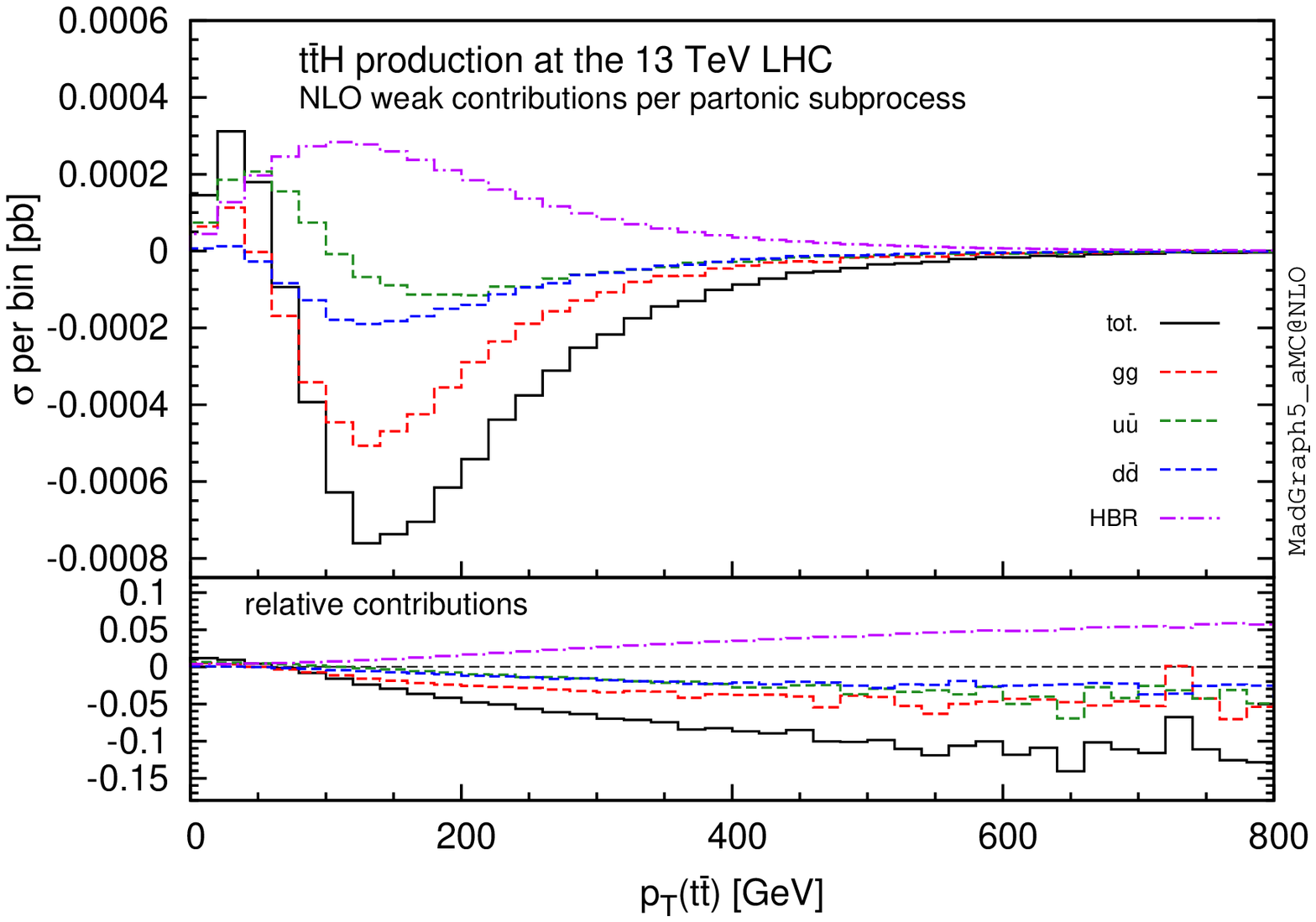}
  \includegraphics[width=0.49\textwidth]{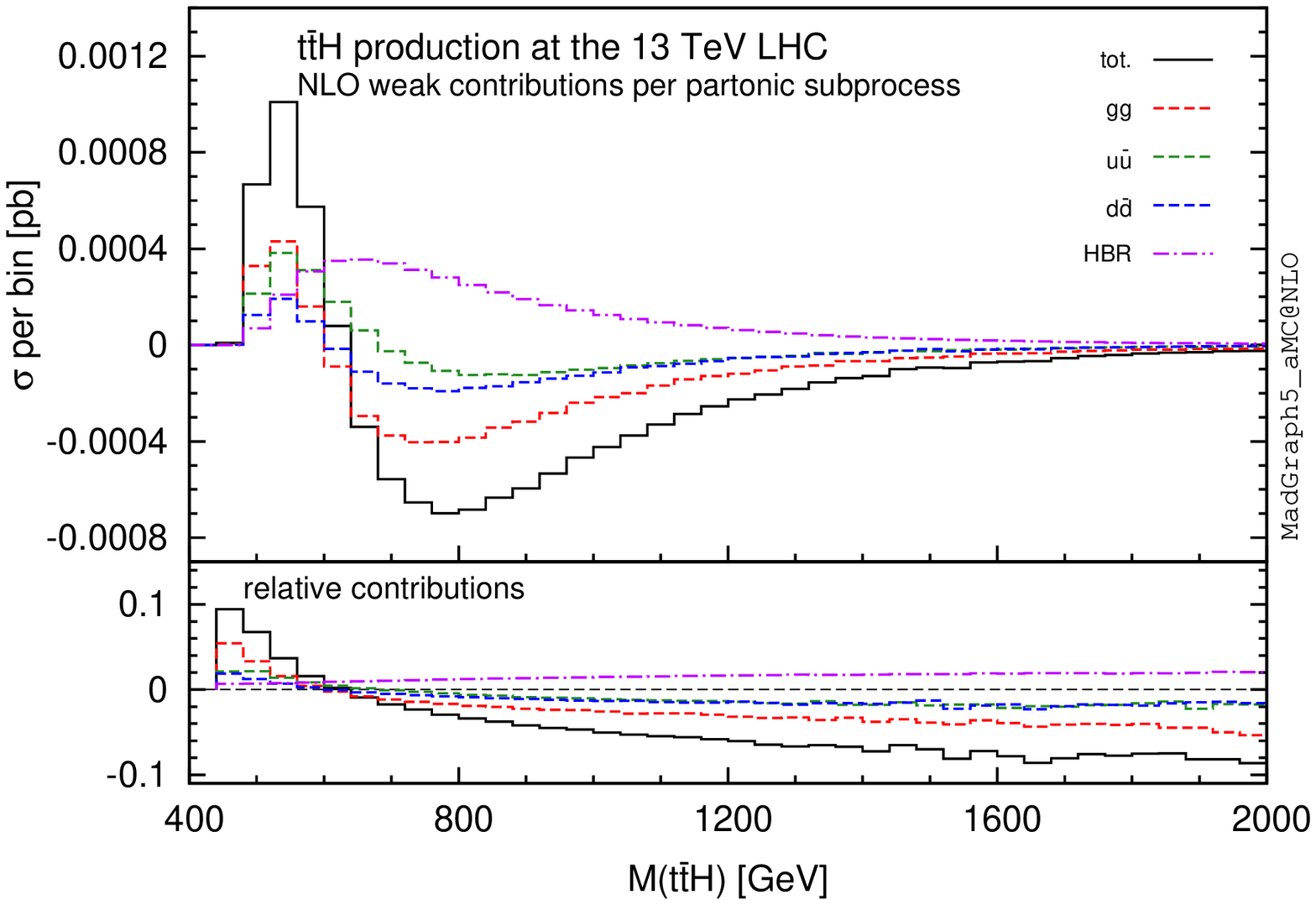}
  \includegraphics[width=0.49\textwidth]{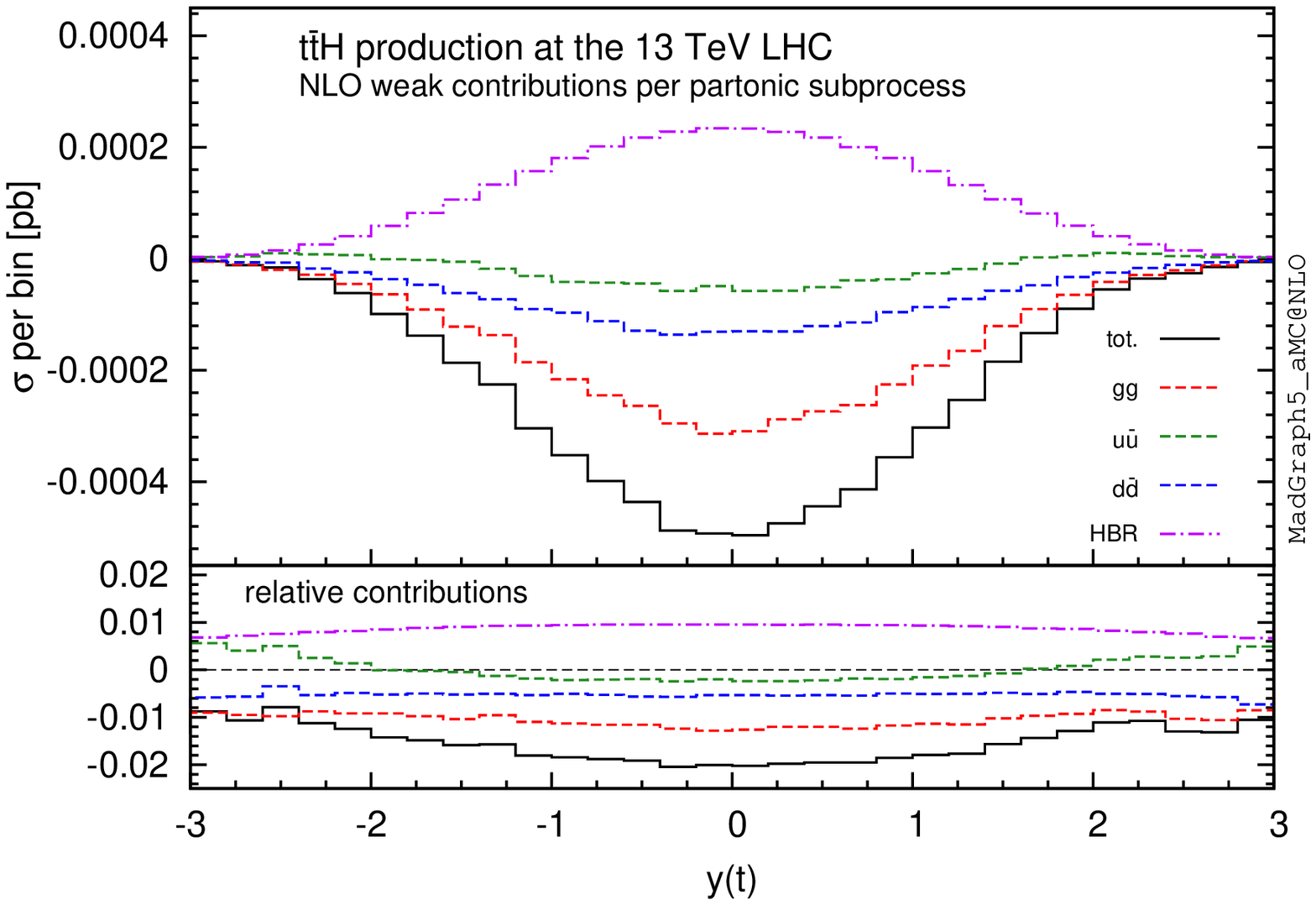}
  \includegraphics[width=0.49\textwidth]{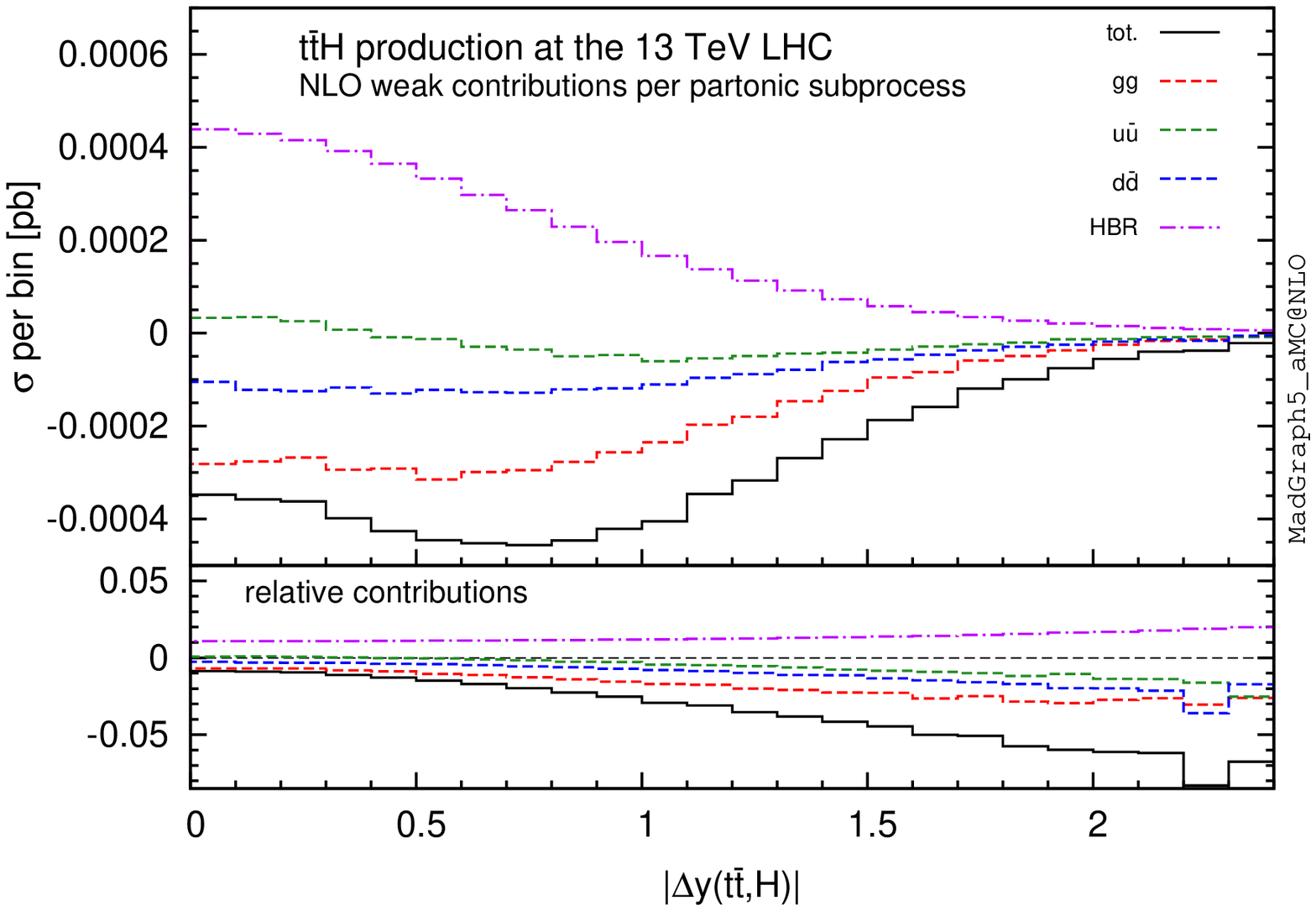}
\caption{\label{fig:plot-part13} Individual contributions to the NLO weak
cross section, at 13 TeV.}
\end{figure*}

\begin{figure*}[]
 \center 
 \includegraphics[width=0.49\textwidth]{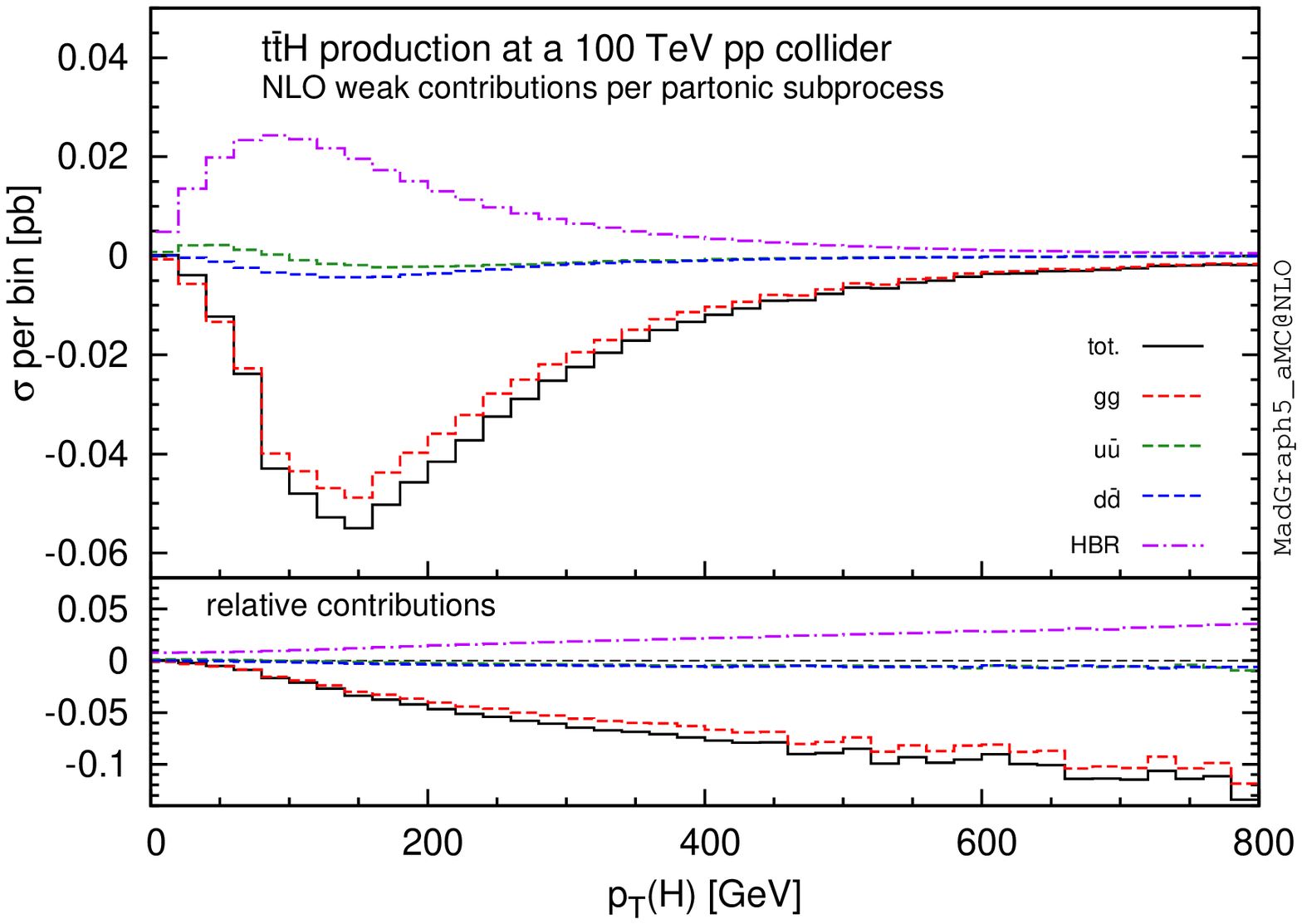}
  \includegraphics[width=0.49\textwidth]{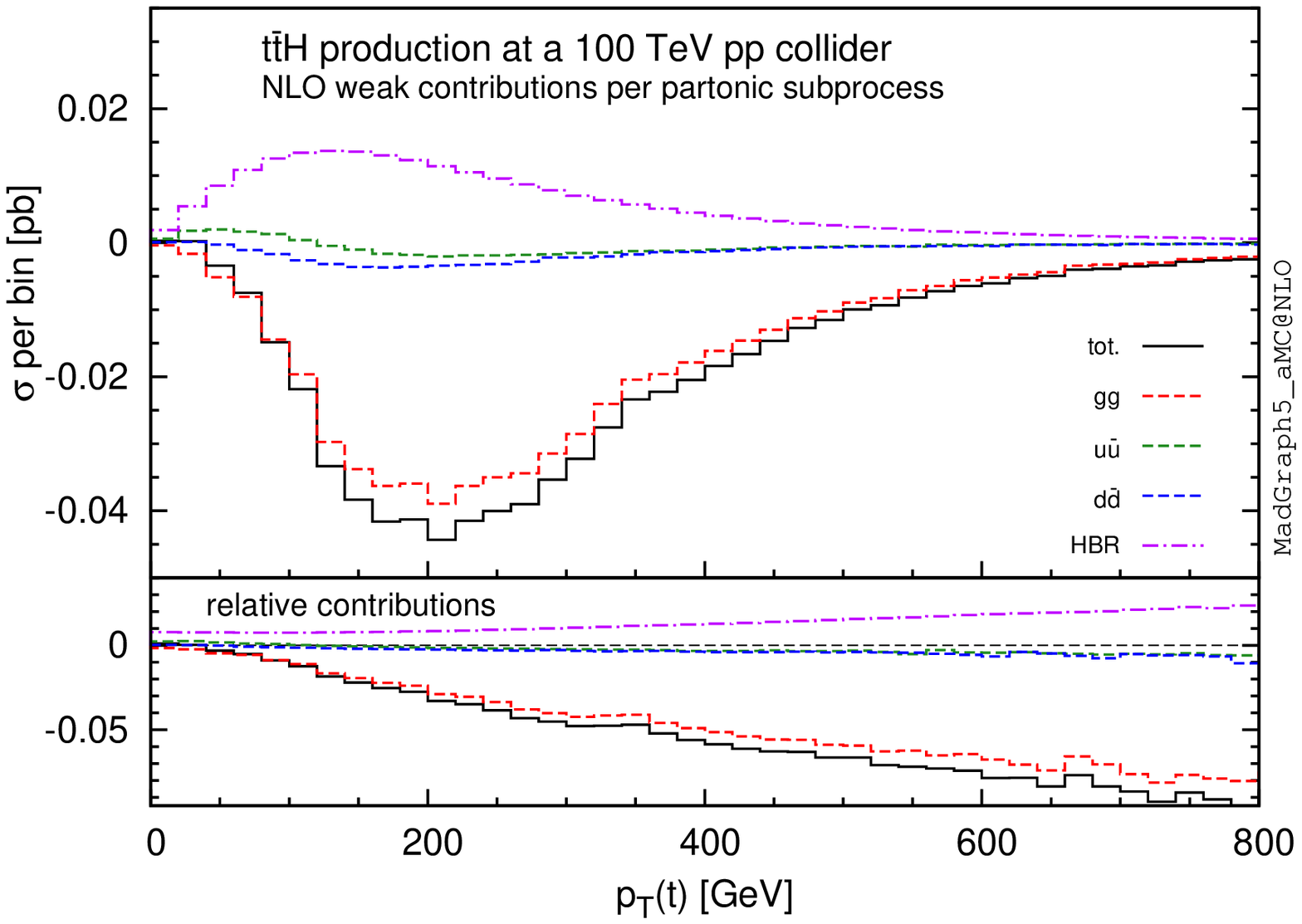}
  \includegraphics[width=0.49\textwidth]{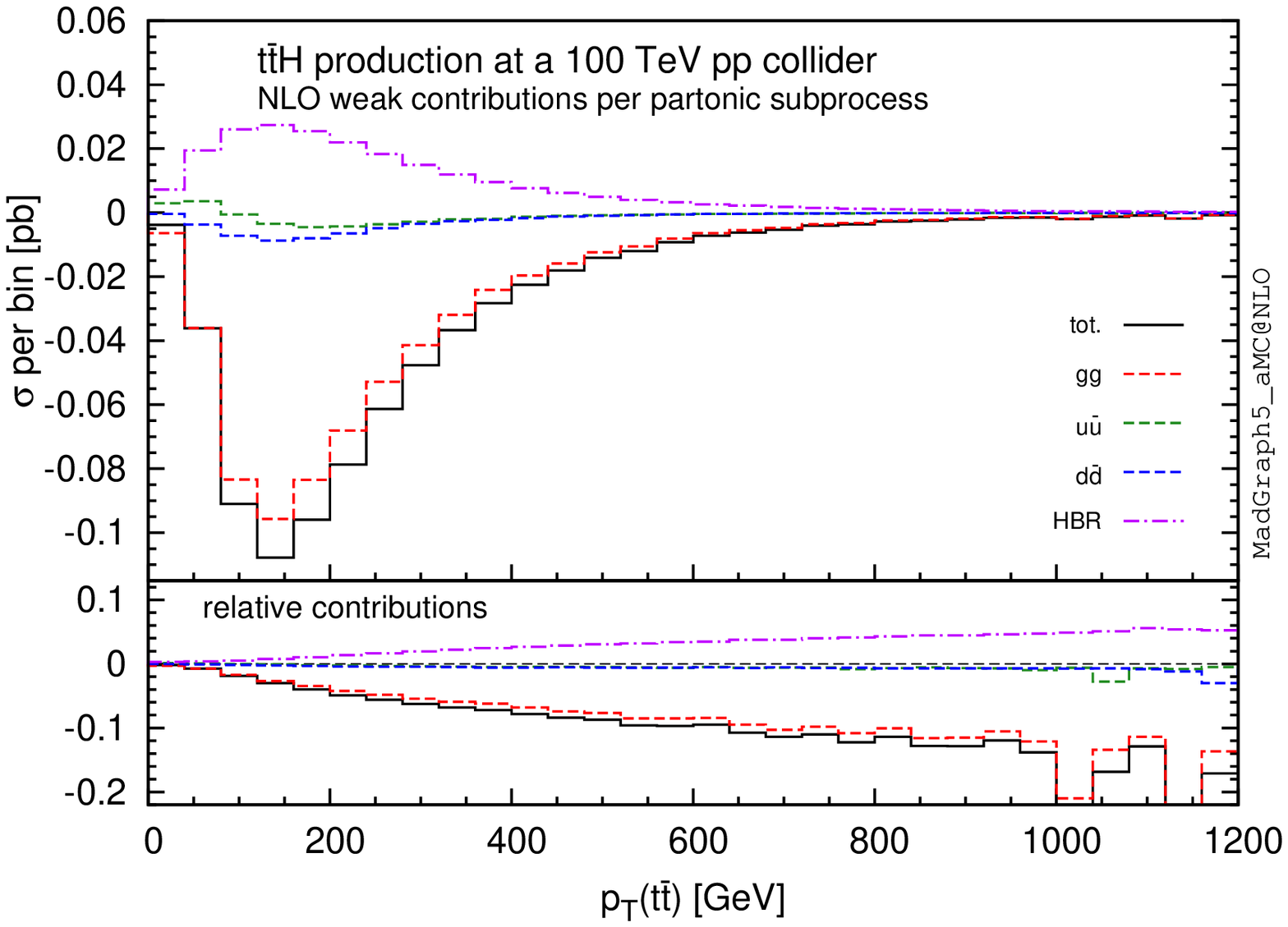}
  \includegraphics[width=0.49\textwidth]{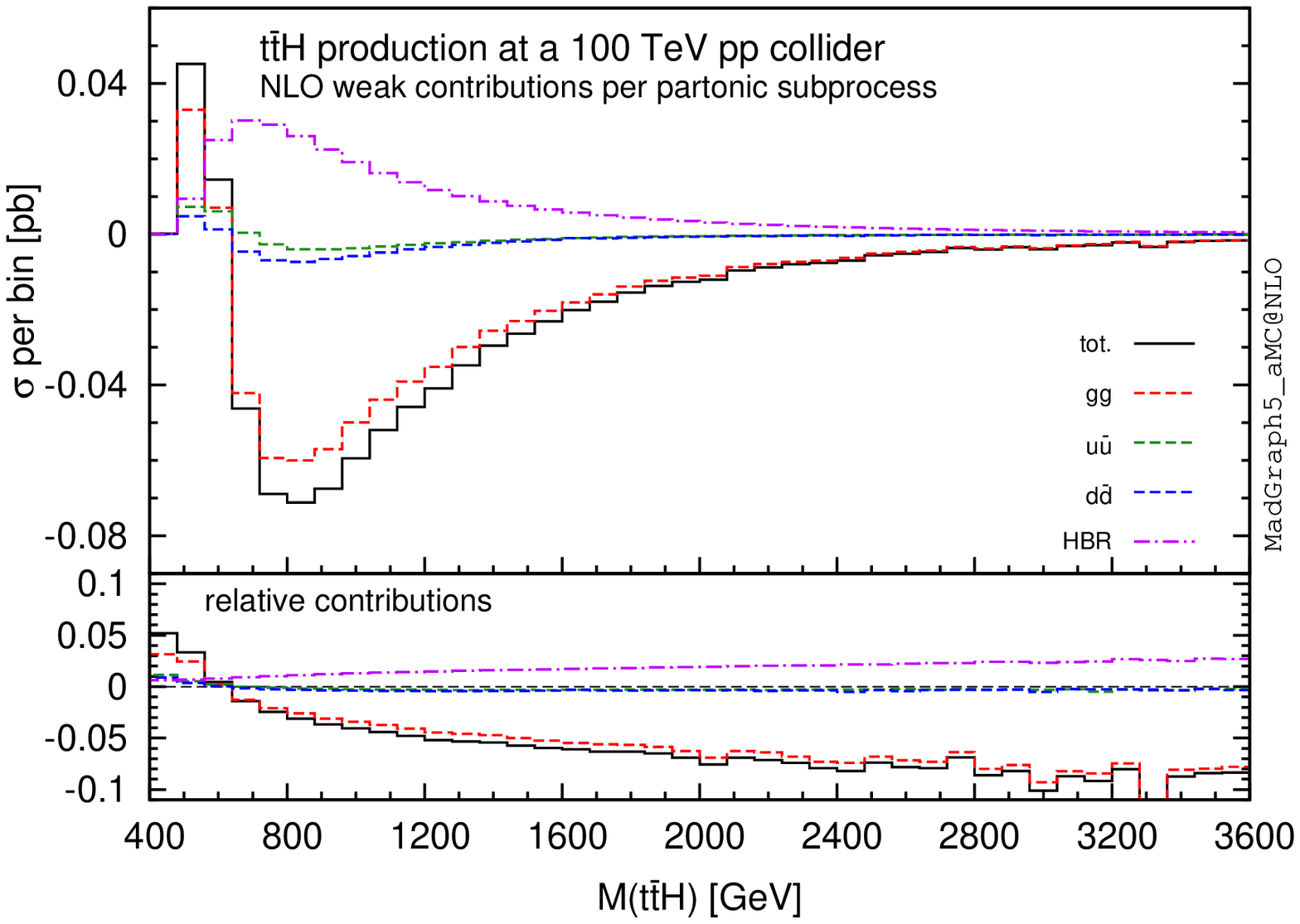}
  \includegraphics[width=0.49\textwidth]{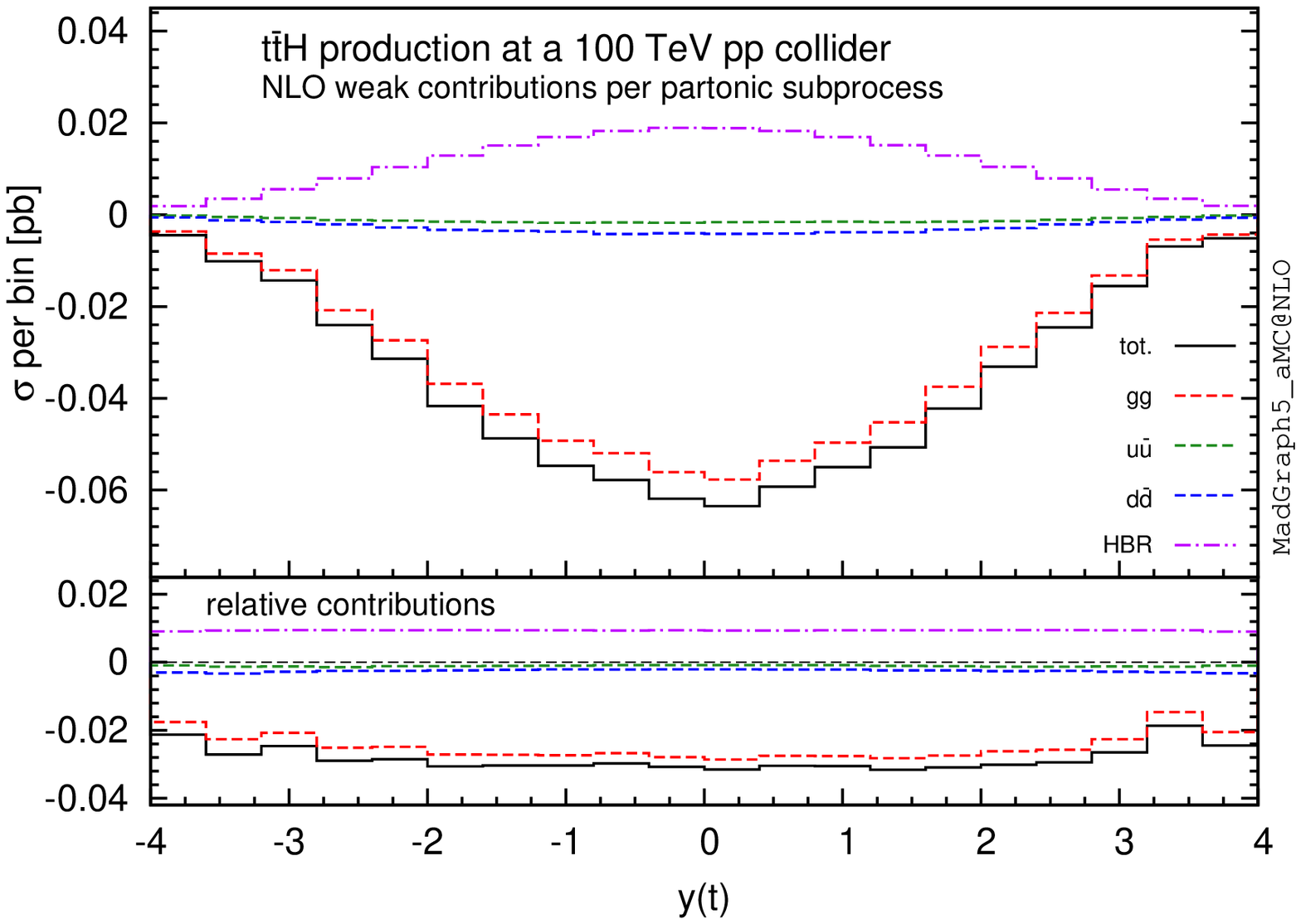}
  \includegraphics[width=0.49\textwidth]{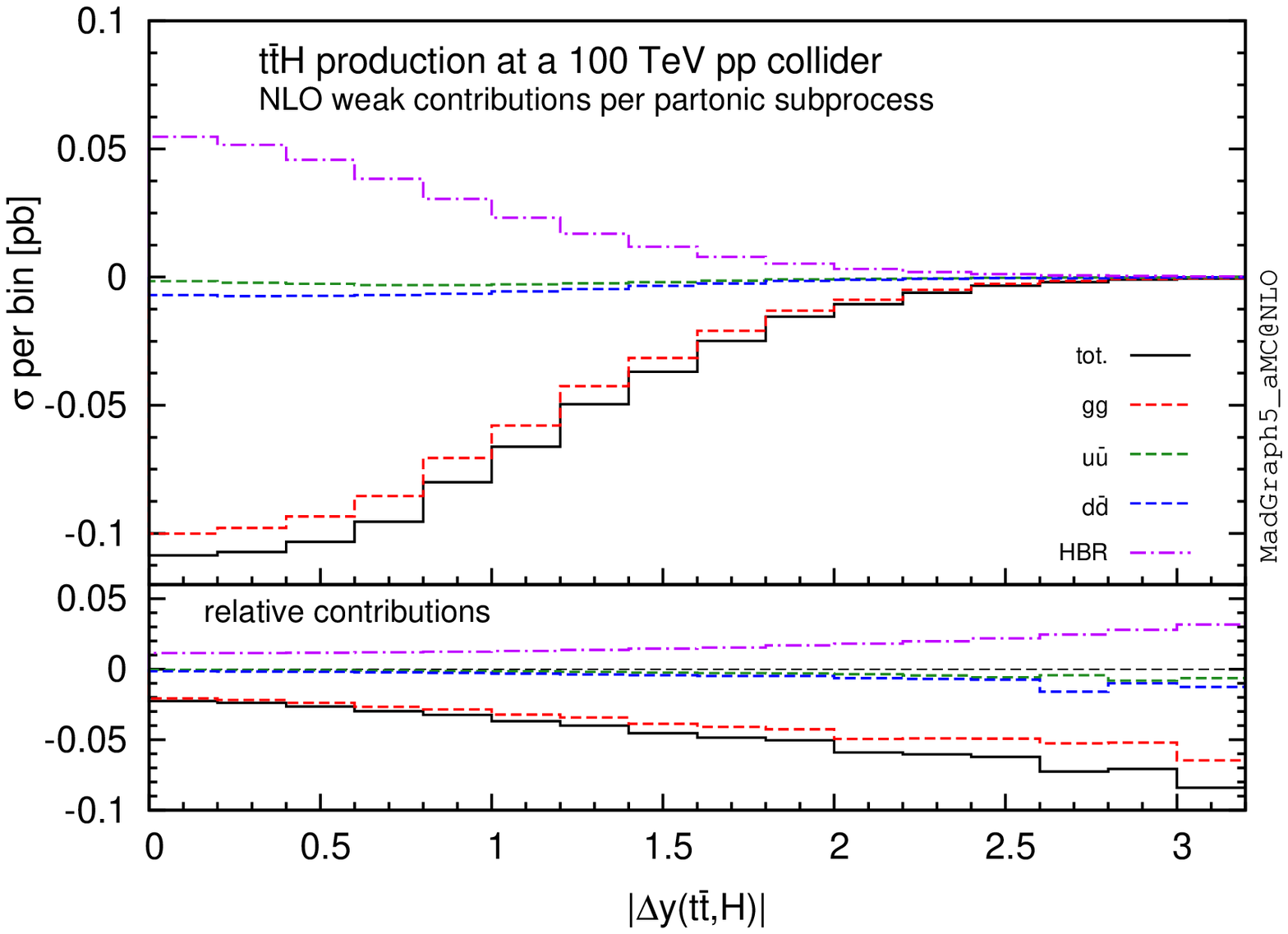}
\caption{\label{fig:plot-part100} Individual contributions to the NLO weak
cross section, at 100 TeV.}
\end{figure*}

Further details on the NLO weak and HBR results relevant
to figs.~\ref{fig:plot-13} and~\ref{fig:plot-100} are given in
figs.~\ref{fig:plot-part13} and~\ref{fig:plot-part100}, respectively.
The main frames display the cross sections, and in the case of the
NLO weak contributions the individual results for the three dominant
partonic channels (namely, $gg$, $d\bar{d}$, and $u\bar{u}$) are
also shown. The lower insets contain the same information, but
in the form of fractions over the relevant LO cross sections; these
are thus the differential analogues of tables~\ref{corrchannel}
and~\ref{corrVrad}.

As far as QCD and weak effects are concerned,
figs.~\ref{fig:plot-13} and~\ref{fig:plot-100} show rather similar
patterns. NLO QCD contributions are dominant everywhere
in the phase space, and their size increase with the collider energy
in a manner which is, in the first approximation, rather independent 
of the observable or the range considered (however, a closer inspection
reveals some minor differences in the shapes of the relative contributions 
to several observables). In other words, there is no single phase-space
region associated with the growth with energy of the relative NLO QCD 
contribution observed in table~\ref{corrQCDweak}. At a given collider
energy, the NLO QCD $K$ factors are generally not flat, with the
exception of $y(t)$ and, to a good extent, of $\Delta y(t\bt,H)$
at 100~TeV; the $K$ factors also tend to flatten out at large transverse
momenta or invariant masses. The case of NLO weak effects is interesting
because they become significant only in certain regions of the phase
space (we remind the reader that we are discussing here the analogue 
of the fully inclusive case of sect.~\ref{sec:rates}, for which at the
level of rates weak contributions are smaller than QCD scale
uncertainties, as documented by the entries not included in round brackets
in table~\ref{corrQCDweak}). In particular, the histograms that 
include the NLO weak contributions lie at the lower end of the
QCD scale-uncertainty band at large $\pt(H)$, $\pt(t)$, and
(to a somewhat lesser extent) $\Delta y(t\bt,H)$. Weak effects induce 
therefore a significant distortion of the spectra in those regions, 
and cannot be neglected. The above regions are rather directly related
with those relevant to the boosted scenario; it is therefore consistent
with the behaviour of the rates within the cuts of eq.~(\ref{boost})
shown in table~\ref{corrQCDweak} that we observe that the relative
importance of NLO weak vs NLO QCD contributions is greater at 13~TeV
than at 100~TeV.

One has to keep in mind that the impact of the NLO weak effects discussed 
above can be partly compensated by that of the HBR contributions, since 
the relative importance of the latter tends to increase (in absolute value) 
in the same regions where the NLO weak corrections are most significant, at 
both 13 and 100~TeV, as shown by the insets of figs.~\ref{fig:plot-part13} 
and~\ref{fig:plot-part100}. From these figures, we also see the differential
counterpart of table~\ref{corrchannel}: at 13~TeV, the interplay of the
$gg$ with the $d\bar{d}$ and $u\bar{u}$ channels is involved, while
at 100~TeV one is dominated everywhere in the phase space by the
$gg$-initiated process.

We conclude this section by presenting in fig.~\ref{fig:plot-13-cuts}
the results for our six reference differential distributions obtained
by imposing the cuts of eq.~(\ref{boost}). As expected, the effect of
such cuts is that of further enhancing the impact of the NLO weak 
contributions, which become competitive with the QCD ones,
and non-negligible even close to the $\pt$ thresholds
(compare e.g.~the insets of the upper two panels of 
figs.~\ref{fig:plot-13} and~\ref{fig:plot-13-cuts}).
Note that this conclusion is not modified when the
HBR contributions are taken into account, as was already observed
for the predictions of the total rates. We finally comment on a 
few visible features that appear in the differential $\pt(t)$,
$\pt(t\bt)$, and $M(t\bt H)$ distributions in the boosted scenario.
These are all due to the cuts eq.~(\ref{boost}), which at the LO result 
in a sharp threshold in the case of $\pt(t\bt)$ and $M(t\bt H)$.
Such a threshold, which disappears when extra particles may be
radiated, is thus a critical point~\cite{Catani:1997xc} inside the 
phase space, and therefore a source of instabilities in perturbation
theory. We point out that, although hardly visible in 
fig.~\ref{fig:plot-13-cuts}, below threshold the two NLO-accurate
results which differ by the presence of the NLO weak effects are
not identical; note that, in this region, the latter are solely due
to the QCD-type radiation that is responsible for real-emission
corrections (i.e.~to the term $R_{{\rm\sss QCD},0}\mydot R_{{\rm\sss QCD},1}$
that appears in eq.~(\ref{SvsVR1})). In the case 
of $\pt(t)$, the knee around 400~GeV
is due to the fact that the Higgs prefers to stay closer to
either the top or the antitop than away from both of them (as we have
verified by studying the relevant $(\eta,\varphi)$ distances).
This results effectively in a non-sharp threshold, which largely disappears
when NLO contributions are included; since this threshold is not
due to a tight kinematic constraint, the effects are much milder than 
for $\pt(t\bt)$ and $M(t\bt H)$.

\section{Conclusions\label{sec:concl}}
In this paper we have presented the first calculation of the 
${\cal O}(\as^2\aem^2)$ next-to-leading order contribution
to $t\bt H$ hadroproduction that includes all weak and QCD 
effects. The computation is performed
in the \aNLO\ framework, and constitutes the first step towards
the complete automation, in that framework, of NLO-accurate cross section 
calculations in theories other than QCD. These are relevant to the 
perturbative expansion in terms of either a coupling constant different 
from $\as$ (e.g.~the EW one $\aem$), or simultaneously in more than
one coupling constant (e.g.~$\as$ and $\aem$, as in the case
studied here). Weak corrections have more interesting physics
implications than those of QED origin, which have not been considered
in this paper, because of their potential significant impact in phase-space 
regions characterised by large invariants (as we have documented), and
because of the different Higgs couplings they involve. Furthermore,
from the point of view of an automated approach which is already
capable of computing NLO QCD effects, the case of weak corrections
poses a vastly different challenge, and thus offers a better testing 
context, than that of QED corrections.

We have used $t\bt H$ production as a case study to discuss two 
issues that will become increasingly frequent in the near future.
Firstly, for processes that feature several classes of
amplitudes, which differ by the coupling-constant
combination they factorise, the intuitive classification of 
the two dominant next-to-leading order contributions as QCD
and EW corrections may become a source of confusion,
since in general at the level of cross sections these two
kind of corrections mix. Therefore, it is best avoided;
we have suggested an alternative terminology, which can
be applied to arbitrary processes and perturbative series.
Note that this is no longer an academic matter,
in view of the fact that automated codes will soon be able to
evaluate all contributions to Born and NLO cross sections, regardless
of their hierarchy in terms of coupling constants.
Secondly, weak contributions due to the emission of potentially
resolvable massive EW vector bosons need to be taken into account,
at least when one is not able to discard them in the context of
a fully realistic analysis at the level of final states.
We have shown that, in the case of $t\bt H$ inclusive production, 
these processes may in fact not be entirely negligible in precision
phenomenology studies.

We have compared the ${\cal O}(\as^2\aem^2)$ predictions with
those of ${\cal O}(\as^3\aem)$, which constitute the dominant
(in terms of coupling hierarchy) contribution to NLO effects.
We have found that such a hierarchy, established a priori on the
basis of the coupling-constant behaviour, is amply respected at
the level of fully-inclusive cross sections, for which the scale
uncertainty of the latter contribution is significantly larger
than the whole ${\cal O}(\as^2\aem^2)$ result. This picture does
change, however, when one emphasises the role of phase-space regions
characterised by some large scale (typically related to a high-$\pt$
configuration), which can be done by either looking directly at the
relevant kinematics, or at the inclusive level by applying suitable cuts;
both options have been considered here. The main conclusion is that,
in these regions, effects of weak origin play an important role,
and that ${\cal O}(\as^2\aem^2)$ results may be numerically of the
same order as the ${\cal O}(\as^3\aem)$ ones. Therefore, $t\bt H$
production appears to follow the same pattern as other processes,
where Sudakov logarithms can induce significant distortions of
spectra. This implies that the computation of weak contributions 
is a necessary ingredient for precision phenomenology at large 
transverse momenta.

\section*{Acknowledgements}
We thank Fabio Maltoni and Michelangelo Mangano for having encouraged
us to pursue the present project. This work is supported in part (DP)
by, and performed in the framework of,
the ERC grant 291377 ``LHCtheory: Theoretical predictions 
and analyses of LHC physics: advancing the precision frontier". 
The work of VH is supported by the SNF with grant PBELP2 146525.
The work of MZ is supported by the Research Executive Agency
(REA) of the European Union under the Grant Agreement number 
PITN-GA-2010-264564 (LHCPhenoNet), and by the ILP LABEX (ANR-10-LABX-63),
in turn supported by French state funds managed by the ANR within the
``Investissements d'Avenir'' programme under reference ANR-11-IDEX-0004-02.

\bibliographystyle{JHEP}
\bibliography{tthEW}

\providecommand{\href}[2]{#2}\begingroup\raggedright\begin{thebibliography}{10}

\bibitem{Aad:2012tfa}
{\bf ATLAS Collaboration}, G.~Aad et~al., {\it {Observation of a new particle
  in the search for the Standard Model Higgs boson with the ATLAS detector at
  the LHC}},  {\em Phys.Lett.} {\bf B716} (2012) 1--29,
  [\href{http://xxx.lanl.gov/abs/1207.7214}{{\tt arXiv:1207.7214}}].

\bibitem{Chatrchyan:2012ufa}
{\bf CMS Collaboration}, S.~Chatrchyan et~al., {\it {Observation of a new boson
  at a mass of 125 GeV with the CMS experiment at the LHC}},  {\em Phys.Lett.}
  {\bf B716} (2012) 30--61, [\href{http://xxx.lanl.gov/abs/1207.7235}{{\tt
  arXiv:1207.7235}}].

\bibitem{Englert:1964et}
F.~Englert and R.~Brout, {\it {Broken Symmetry and the Mass of Gauge Vector
  Mesons}},  {\em Phys.Rev.Lett.} {\bf 13} (1964) 321--323.

\bibitem{Higgs:1964ia}
P.~W. Higgs, {\it {Broken symmetries, massless particles and gauge fields}},
  {\em Phys.Lett.} {\bf 12} (1964) 132--133.

\bibitem{Higgs:1964pj}
P.~W. Higgs, {\it {Broken Symmetries and the Masses of Gauge Bosons}},  {\em
  Phys.Rev.Lett.} {\bf 13} (1964) 508--509.

\bibitem{Chatrchyan:2012jja}
{\bf CMS Collaboration}, S.~Chatrchyan et~al., {\it {Study of the Mass and
  Spin-Parity of the Higgs Boson Candidate Via Its Decays to Z Boson Pairs}},
  {\em Phys.Rev.Lett.} {\bf 110} (2013) 081803,
  [\href{http://xxx.lanl.gov/abs/1212.6639}{{\tt arXiv:1212.6639}}].

\bibitem{Aad:2013xqa}
{\bf ATLAS Collaboration}, G.~Aad et~al., {\it {Evidence for the spin-0 nature
  of the Higgs boson using ATLAS data}},  {\em Phys.Lett.} {\bf B726} (2013)
  120--144, [\href{http://xxx.lanl.gov/abs/1307.1432}{{\tt arXiv:1307.1432}}].

\bibitem{ATLAS:2013sla}
{\bf ATLAS Collaboration}, G.~Aad et~al., {\it {Combined coupling measurements
  of the Higgs-like boson with the ATLAS detector using up to 25 fb$^{-1}$ of
  proton-proton collision data}},
  \href{http://xxx.lanl.gov/abs/ATLAS-CONF-2013-034,
  ATLAS-COM-CONF-2013-035}{{\tt ATLAS-CONF-2013-034, ATLAS-COM-CONF-2013-035}}.

\bibitem{Chatrchyan:2013lba}
{\bf CMS Collaboration}, S.~Chatrchyan et~al., {\it {Observation of a new boson
  with mass near 125 GeV in pp collisions at $\sqrt{s}$ = 7 and 8 TeV}},  {\em
  JHEP} {\bf 1306} (2013) 081, [\href{http://xxx.lanl.gov/abs/1303.4571}{{\tt
  arXiv:1303.4571}}].

\bibitem{TheATLAScollaboration:2013mia}
{\bf ATLAS Collaboration}, G.~Aad et~al., {\it {Search for ttH production in
  the $H\to\gamma\gamma$ channel at sqrt(s) = 8 TeV with the ATLAS detector}},
  \href{http://xxx.lanl.gov/abs/ATLAS-CONF-2013-080,
  ATLAS-COM-CONF-2013-089}{{\tt ATLAS-CONF-2013-080, ATLAS-COM-CONF-2013-089}}.

\bibitem{ATLAS:2012cpa}
{\bf ATLAS Collaboration}, G.~Aad et~al., {\it {Search for the Standard Model
  Higgs boson produced in association with top quarks in proton-proton
  collisions at √s = 7 TeV using the ATLAS detector}},
  \href{http://xxx.lanl.gov/abs/ATLAS-CONF-2012-135,
  ATLAS-COM-CONF-2012-162}{{\tt ATLAS-CONF-2012-135, ATLAS-COM-CONF-2012-162}}.

\bibitem{Chatrchyan:2013yea}
{\bf CMS Collaboration}, S.~Chatrchyan et~al., {\it {Search for the standard
  model Higgs boson produced in association with a top-quark pair in pp
  collisions at the LHC}},  {\em JHEP} {\bf 1305} (2013) 145,
  [\href{http://xxx.lanl.gov/abs/1303.0763}{{\tt arXiv:1303.0763}}].

\bibitem{CMS:2013tfa}
{\bf CMS Collaboration}, S.~Chatrchyan et~al., {\it {Search for the standard
  model Higgs boson produced in association with top quarks in multilepton
  final states}},  \href{http://xxx.lanl.gov/abs/CMS-PAS-HIG-13-020}{{\tt
  CMS-PAS-HIG-13-020}}.

\bibitem{CMS:2013sea}
{\bf CMS Collaboration}, S.~Chatrchyan et~al., {\it {Search for Higgs Boson
  Production in Association with a Top-Quark Pair and Decaying to Bottom Quarks
  or Tau Leptons}},  \href{http://xxx.lanl.gov/abs/CMS-PAS-HIG-13-019}{{\tt
  CMS-PAS-HIG-13-019}}.

\bibitem{CMS:2013fda}
{\bf CMS Collaboration}, S.~Chatrchyan et~al., {\it {Search for ttH production
  in events where H decays to photons at 8 TeV collisions}},
  \href{http://xxx.lanl.gov/abs/CMS-PAS-HIG-13-015}{{\tt CMS-PAS-HIG-13-015}}.

\bibitem{CMS:2012qaa}
{\bf CMS Collaboration}, S.~Chatrchyan et~al., {\it {Search for Higgs boson
  production in association with top quark pairs in pp collisions}},
  \href{http://xxx.lanl.gov/abs/CMS-PAS-HIG-12-025}{{\tt CMS-PAS-HIG-12-025}}.

\bibitem{Beenakker:2001rj}
W.~Beenakker, S.~Dittmaier, M.~Kramer, B.~Plumper, M.~Spira, et~al., {\it
  {Higgs radiation off top quarks at the Tevatron and the LHC}},  {\em
  Phys.Rev.Lett.} {\bf 87} (2001) 201805,
  [\href{http://xxx.lanl.gov/abs/hep-ph/0107081}{{\tt hep-ph/0107081}}].

\bibitem{Beenakker:2002nc}
W.~Beenakker, S.~Dittmaier, M.~Kramer, B.~Plumper, M.~Spira, et~al., {\it {NLO
  QCD corrections to t anti-t H production in hadron collisions}},  {\em
  Nucl.Phys.} {\bf B653} (2003) 151--203,
  [\href{http://xxx.lanl.gov/abs/hep-ph/0211352}{{\tt hep-ph/0211352}}].

\bibitem{Dawson:2002tg}
S.~Dawson, L.~Orr, L.~Reina, and D.~Wackeroth, {\it {Associated top quark Higgs
  boson production at the LHC}},  {\em Phys.Rev.} {\bf D67} (2003) 071503,
  [\href{http://xxx.lanl.gov/abs/hep-ph/0211438}{{\tt hep-ph/0211438}}].

\bibitem{Dawson:2003zu}
S.~Dawson, C.~Jackson, L.~Orr, L.~Reina, and D.~Wackeroth, {\it {Associated
  Higgs production with top quarks at the large hadron collider: NLO QCD
  corrections}},  {\em Phys.Rev.} {\bf D68} (2003) 034022,
  [\href{http://xxx.lanl.gov/abs/hep-ph/0305087}{{\tt hep-ph/0305087}}].

\bibitem{Frederix:2011zi}
R.~Frederix, S.~Frixione, V.~Hirschi, F.~Maltoni, R.~Pittau, and P.~Torielli,
  {\it {Scalar and pseudoscalar Higgs production in association with a
  top-antitop pair}},  {\em Phys.Lett.} {\bf B701} (2011) 427--433,
  [\href{http://xxx.lanl.gov/abs/1104.5613}{{\tt arXiv:1104.5613}}].

\bibitem{Garzelli:2011vp}
M.~Garzelli, A.~Kardos, C.~Papadopoulos, and Z.~Trocsanyi, {\it {Standard Model
  Higgs boson production in association with a top anti-top pair at NLO with
  parton showering}},  {\em Europhys.Lett.} {\bf 96} (2011) 11001,
  [\href{http://xxx.lanl.gov/abs/1108.0387}{{\tt arXiv:1108.0387}}].

\bibitem{Bredenstein:2009aj}
A.~Bredenstein, A.~Denner, S.~Dittmaier, and S.~Pozzorini, {\it {NLO QCD
  corrections to pp $\rightarrow$ t anti-t b anti-b + X at the LHC}},  {\em
  Phys.Rev.Lett.} {\bf 103} (2009) 012002,
  [\href{http://xxx.lanl.gov/abs/0905.0110}{{\tt arXiv:0905.0110}}].

\bibitem{Bevilacqua:2009zn}
G.~Bevilacqua, M.~Czakon, C.~Papadopoulos, R.~Pittau, and M.~Worek, {\it
  {Assault on the NLO Wishlist: pp $\to$ t anti-t b anti-b}},  {\em JHEP} {\bf
  0909} (2009) 109, [\href{http://xxx.lanl.gov/abs/0907.4723}{{\tt
  arXiv:0907.4723}}].

\bibitem{Bredenstein:2010rs}
A.~Bredenstein, A.~Denner, S.~Dittmaier, and S.~Pozzorini, {\it {NLO QCD
  Corrections to Top Anti-Top Bottom Anti-Bottom Production at the LHC: 2. full
  hadronic results}},  {\em JHEP} {\bf 1003} (2010) 021,
  [\href{http://xxx.lanl.gov/abs/1001.4006}{{\tt arXiv:1001.4006}}].

\bibitem{Kardos:2013vxa}
A.~Kardos and Z.~Tr{\'o}cs{\'a}nyi, {\it {Hadroproduction of t anti-t pair with
  a b anti-b pair using PowHel}},  {\em J.Phys.} {\bf G41} (2014) 075005,
  [\href{http://xxx.lanl.gov/abs/1303.6291}{{\tt arXiv:1303.6291}}].

\bibitem{Cascioli:2013era}
F.~Cascioli, P.~Maierhoefer, N.~Moretti, S.~Pozzorini, and F.~Siegert, {\it
  {NLO matching for ttbb production with massive b-quarks}},
  \href{http://xxx.lanl.gov/abs/1309.5912}{{\tt arXiv:1309.5912}}.

\bibitem{Djouadi:1994ge}
A.~Djouadi and P.~Gambino, {\it {Leading electroweak correction to Higgs boson
  production at proton colliders}},  {\em Phys.Rev.Lett.} {\bf 73} (1994)
  2528--2531, [\href{http://xxx.lanl.gov/abs/hep-ph/9406432}{{\tt
  hep-ph/9406432}}].

\bibitem{Aglietti:2004nj}
U.~Aglietti, R.~Bonciani, G.~Degrassi, and A.~Vicini, {\it {Two loop light
  fermion contribution to Higgs production and decays}},  {\em Phys.Lett.} {\bf
  B595} (2004) 432--441, [\href{http://xxx.lanl.gov/abs/hep-ph/0404071}{{\tt
  hep-ph/0404071}}].

\bibitem{Degrassi:2004mx}
G.~Degrassi and F.~Maltoni, {\it {Two-loop electroweak corrections to Higgs
  production at hadron colliders}},  {\em Phys.Lett.} {\bf B600} (2004)
  255--260, [\href{http://xxx.lanl.gov/abs/hep-ph/0407249}{{\tt
  hep-ph/0407249}}].

\bibitem{Actis:2008ug}
S.~Actis, G.~Passarino, C.~Sturm, and S.~Uccirati, {\it {NLO Electroweak
  Corrections to Higgs Boson Production at Hadron Colliders}},  {\em
  Phys.Lett.} {\bf B670} (2008) 12--17,
  [\href{http://xxx.lanl.gov/abs/0809.1301}{{\tt arXiv:0809.1301}}].

\bibitem{Ciccolini:2007ec}
M.~Ciccolini, A.~Denner, and S.~Dittmaier, {\it {Electroweak and QCD
  corrections to Higgs production via vector-boson fusion at the LHC}},  {\em
  Phys.Rev.} {\bf D77} (2008) 013002,
  [\href{http://xxx.lanl.gov/abs/0710.4749}{{\tt arXiv:0710.4749}}].

\bibitem{Ciccolini:2007jr}
M.~Ciccolini, A.~Denner, and S.~Dittmaier, {\it {Strong and electroweak
  corrections to the production of Higgs + 2jets via weak interactions at the
  LHC}},  {\em Phys.Rev.Lett.} {\bf 99} (2007) 161803,
  [\href{http://xxx.lanl.gov/abs/0707.0381}{{\tt arXiv:0707.0381}}].

\bibitem{Ciccolini:2003jy}
M.~Ciccolini, S.~Dittmaier, and M.~Kramer, {\it {Electroweak radiative
  corrections to associated WH and ZH production at hadron colliders}},  {\em
  Phys.Rev.} {\bf D68} (2003) 073003,
  [\href{http://xxx.lanl.gov/abs/hep-ph/0306234}{{\tt hep-ph/0306234}}].

\bibitem{Beenakker:1993yr}
W.~Beenakker, A.~Denner, W.~Hollik, R.~Mertig, T.~Sack, et~al., {\it
  {Electroweak one loop contributions to top pair production in hadron
  colliders}},  {\em Nucl.Phys.} {\bf B411} (1994) 343--380.

\bibitem{Bernreuther:2005is}
W.~Bernreuther, M.~Fuecker, and Z.~Si, {\it {Mixed QCD and weak corrections to
  top quark pair production at hadron colliders}},  {\em Phys.Lett.} {\bf B633}
  (2006) 54--60, [\href{http://xxx.lanl.gov/abs/hep-ph/0508091}{{\tt
  hep-ph/0508091}}].

\bibitem{Kuhn:2005it}
J.~H. Kuhn, A.~Scharf, and P.~Uwer, {\it {Electroweak corrections to top-quark
  pair production in quark-antiquark annihilation}},  {\em Eur.Phys.J.} {\bf
  C45} (2006) 139--150, [\href{http://xxx.lanl.gov/abs/hep-ph/0508092}{{\tt
  hep-ph/0508092}}].

\bibitem{Bernreuther:2006vg}
W.~Bernreuther, M.~Fuecker, and Z.-G. Si, {\it {Weak interaction corrections to
  hadronic top quark pair production}},  {\em Phys.Rev.} {\bf D74} (2006)
  113005, [\href{http://xxx.lanl.gov/abs/hep-ph/0610334}{{\tt
  hep-ph/0610334}}].

\bibitem{Kuhn:2006vh}
J.~H. Kuhn, A.~Scharf, and P.~Uwer, {\it {Electroweak effects in top-quark pair
  production at hadron colliders}},  {\em Eur.Phys.J.} {\bf C51} (2007) 37--53,
  [\href{http://xxx.lanl.gov/abs/hep-ph/0610335}{{\tt hep-ph/0610335}}].

\bibitem{Hollik:2007sw}
W.~Hollik and M.~Kollar, {\it {NLO QED contributions to top-pair production at
  hadron collider}},  {\em Phys.Rev.} {\bf D77} (2008) 014008,
  [\href{http://xxx.lanl.gov/abs/0708.1697}{{\tt arXiv:0708.1697}}].

\bibitem{Bernreuther:2010ny}
W.~Bernreuther and Z.-G. Si, {\it {Distributions and correlations for top quark
  pair production and decay at the Tevatron and LHC.}},  {\em Nucl.Phys.} {\bf
  B837} (2010) 90--121, [\href{http://xxx.lanl.gov/abs/1003.3926}{{\tt
  arXiv:1003.3926}}].

\bibitem{Hollik:2011ps}
W.~Hollik and D.~Pagani, {\it {The electroweak contribution to the top quark
  forward-backward asymmetry at the Tevatron}},  {\em Phys.Rev.} {\bf D84}
  (2011) 093003, [\href{http://xxx.lanl.gov/abs/1107.2606}{{\tt
  arXiv:1107.2606}}].

\bibitem{Kuhn:2011ri}
J.~H. Kuhn and G.~Rodrigo, {\it {Charge asymmetries of top quarks at hadron
  colliders revisited}},  {\em JHEP} {\bf 1201} (2012) 063,
  [\href{http://xxx.lanl.gov/abs/1109.6830}{{\tt arXiv:1109.6830}}].

\bibitem{Alwall:2014hca}
J.~Alwall, R.~Frederix, S.~Frixione, V.~Hirschi, F.~Maltoni, et~al., {\it {The
  automated computation of tree-level and next-to-leading order differential
  cross sections, and their matching to parton shower simulations}},
  \href{http://xxx.lanl.gov/abs/1405.0301}{{\tt arXiv:1405.0301}}.

\bibitem{Ciafaloni:1998xg}
P.~Ciafaloni and D.~Comelli, {\it {Sudakov enhancement of electroweak
  corrections}},  {\em Phys.Lett.} {\bf B446} (1999) 278--284,
  [\href{http://xxx.lanl.gov/abs/hep-ph/9809321}{{\tt hep-ph/9809321}}].

\bibitem{Ciafaloni:2000df}
M.~Ciafaloni, P.~Ciafaloni, and D.~Comelli, {\it {Bloch-Nordsieck violating
  electroweak corrections to inclusive TeV scale hard processes}},  {\em
  Phys.Rev.Lett.} {\bf 84} (2000) 4810--4813,
  [\href{http://xxx.lanl.gov/abs/hep-ph/0001142}{{\tt hep-ph/0001142}}].

\bibitem{Denner:2000jv}
A.~Denner and S.~Pozzorini, {\it {One loop leading logarithms in electroweak
  radiative corrections. 1. Results}},  {\em Eur.Phys.J.} {\bf C18} (2001)
  461--480, [\href{http://xxx.lanl.gov/abs/hep-ph/0010201}{{\tt
  hep-ph/0010201}}].

\bibitem{Denner:2001gw}
A.~Denner and S.~Pozzorini, {\it {One loop leading logarithms in electroweak
  radiative corrections. 2. Factorization of collinear singularities}},  {\em
  Eur.Phys.J.} {\bf C21} (2001) 63--79,
  [\href{http://xxx.lanl.gov/abs/hep-ph/0104127}{{\tt hep-ph/0104127}}].

\bibitem{Dittmaier:2003ej}
S.~Dittmaier, .~Kramer, Michael, and M.~Spira, {\it {Higgs radiation off bottom
  quarks at the Tevatron and the CERN LHC}},  {\em Phys.Rev.} {\bf D70} (2004)
  074010, [\href{http://xxx.lanl.gov/abs/hep-ph/0309204}{{\tt
  hep-ph/0309204}}].

\bibitem{Hirschi:2011pa}
V.~Hirschi, R.~Frederix, S.~Frixione, M.~V. Garzelli, F.~Maltoni, et~al., {\it
  {Automation of one-loop QCD corrections}},  {\em JHEP} {\bf 1105} (2011) 044,
  [\href{http://xxx.lanl.gov/abs/1103.0621}{{\tt arXiv:1103.0621}}].

\bibitem{Degrande:2011ua}
C.~Degrande, C.~Duhr, B.~Fuks, D.~Grellscheid, O.~Mattelaer, et~al., {\it {UFO
  - The Universal FeynRules Output}},  {\em Comput.Phys.Commun.} {\bf 183}
  (2012) 1201--1214, [\href{http://xxx.lanl.gov/abs/1108.2040}{{\tt
  arXiv:1108.2040}}].

\bibitem{Ossola:2008xq}
G.~Ossola, C.~G. Papadopoulos, and R.~Pittau, {\it {On the Rational Terms of
  the one-loop amplitudes}},  {\em JHEP} {\bf 0805} (2008) 004,
  [\href{http://xxx.lanl.gov/abs/0802.1876}{{\tt arXiv:0802.1876}}].

\bibitem{Dittmaier:2001ay}
S.~Dittmaier and .~Kramer, Michael, {\it {Electroweak radiative corrections to
  W boson production at hadron colliders}},  {\em Phys.Rev.} {\bf D65} (2002)
  073007, [\href{http://xxx.lanl.gov/abs/hep-ph/0109062}{{\tt
  hep-ph/0109062}}].

\bibitem{Denner:1991kt}
A.~Denner, {\it {Techniques for calculation of electroweak radiative
  corrections at the one loop level and results for W physics at LEP-200}},
  {\em Fortsch.Phys.} {\bf 41} (1993) 307--420,
  [\href{http://xxx.lanl.gov/abs/0709.1075}{{\tt arXiv:0709.1075}}].

\bibitem{Garzelli:2009is}
M.~Garzelli, I.~Malamos, and R.~Pittau, {\it {Feynman rules for the rational
  part of the Electroweak 1-loop amplitudes}},  {\em JHEP} {\bf 1001} (2010)
  040, [\href{http://xxx.lanl.gov/abs/0910.3130}{{\tt arXiv:0910.3130}}].

\bibitem{Garzelli:2010qm}
M.~Garzelli, I.~Malamos, and R.~Pittau, {\it {Feynman rules for the rational
  part of the Electroweak 1-loop amplitudes in the $R_xi$ gauge and in the
  Unitary gauge}},  {\em JHEP} {\bf 1101} (2011) 029,
  [\href{http://xxx.lanl.gov/abs/1009.4302}{{\tt arXiv:1009.4302}}].

\bibitem{Shao:2011tg}
H.-S. Shao, Y.-J. Zhang, and K.-T. Chao, {\it {Feynman Rules for the Rational
  Part of the Standard Model One-loop Amplitudes in the 't Hooft-Veltman
  $\gamma_5$ Scheme}},  {\em JHEP} {\bf 09} (2011) 048,
  [\href{http://xxx.lanl.gov/abs/1106.5030}{{\tt arXiv:1106.5030}}].

\bibitem{Hahn:2000kx}
T.~Hahn, {\it {Generating Feynman diagrams and amplitudes with FeynArts 3}},
  {\em Comput.Phys.Commun.} {\bf 140} (2001) 418--431,
  [\href{http://xxx.lanl.gov/abs/hep-ph/0012260}{{\tt hep-ph/0012260}}].

\bibitem{Agrawal:2011tm}
S.~Agrawal, T.~Hahn, and E.~Mirabella, {\it {FormCalc 7}},  {\em
  J.Phys.Conf.Ser.} {\bf 368} (2012) 012054,
  [\href{http://xxx.lanl.gov/abs/1112.0124}{{\tt arXiv:1112.0124}}].

\bibitem{Hahn:1998yk}
T.~Hahn and M.~Perez-Victoria, {\it {Automatized one loop calculations in
  four-dimensions and D-dimensions}},  {\em Comput.Phys.Commun.} {\bf 118}
  (1999) 153--165, [\href{http://xxx.lanl.gov/abs/hep-ph/9807565}{{\tt
  hep-ph/9807565}}].

\bibitem{Altarelli:1977zs}
G.~Altarelli and G.~Parisi, {\it {Asymptotic Freedom in Parton Language}},
  {\em Nucl.Phys.} {\bf B126} (1977) 298.

\bibitem{Frixione:1995ms}
S.~Frixione, Z.~Kunszt, and A.~Signer, {\it {Three jet cross-sections to
  next-to-leading order}},  {\em Nucl.Phys.} {\bf B467} (1996) 399--442,
  [\href{http://xxx.lanl.gov/abs/hep-ph/9512328}{{\tt hep-ph/9512328}}].

\bibitem{Frixione:1997np}
S.~Frixione, {\it {A General approach to jet cross-sections in QCD}},  {\em
  Nucl.Phys.} {\bf B507} (1997) 295--314,
  [\href{http://xxx.lanl.gov/abs/hep-ph/9706545}{{\tt hep-ph/9706545}}].

\bibitem{Frederix:2009yq}
R.~Frederix, S.~Frixione, F.~Maltoni, and T.~Stelzer, {\it {Automation of
  next-to-leading order computations in QCD: The FKS subtraction}},  {\em JHEP}
  {\bf 0910} (2009) 003, [\href{http://xxx.lanl.gov/abs/0908.4272}{{\tt
  arXiv:0908.4272}}].

\bibitem{Ossola:2006us}
G.~Ossola, C.~G. Papadopoulos, and R.~Pittau, {\it {Reducing full one-loop
  amplitudes to scalar integrals at the integrand level}},  {\em Nucl.Phys.}
  {\bf B763} (2007) 147--169,
  [\href{http://xxx.lanl.gov/abs/hep-ph/0609007}{{\tt hep-ph/0609007}}].

\bibitem{Ossola:2007ax}
G.~Ossola, C.~G. Papadopoulos, and R.~Pittau, {\it {CutTools: A Program
  implementing the OPP reduction method to compute one-loop amplitudes}},  {\em
  JHEP} {\bf 0803} (2008) 042, [\href{http://xxx.lanl.gov/abs/0711.3596}{{\tt
  arXiv:0711.3596}}].

\bibitem{Cascioli:2011va}
F.~Cascioli, P.~Maierhofer, and S.~Pozzorini, {\it {Scattering Amplitudes with
  Open Loops}},  {\em Phys.Rev.Lett.} {\bf 108} (2012) 111601,
  [\href{http://xxx.lanl.gov/abs/1111.5206}{{\tt arXiv:1111.5206}}].

\bibitem{Martin:2009iq}
A.~Martin, W.~Stirling, R.~Thorne, and G.~Watt, {\it {Parton distributions for
  the LHC}},  {\em Eur.Phys.J.} {\bf C63} (2009) 189--285,
  [\href{http://xxx.lanl.gov/abs/0901.0002}{{\tt arXiv:0901.0002}}].

\bibitem{Jegerlehner:2001ca}
F.~Jegerlehner, {\it {The Effective fine structure constant at TESLA
  energies}},  \href{http://xxx.lanl.gov/abs/hep-ph/0105283}{{\tt
  hep-ph/0105283}}.

\bibitem{Frederix:2011ss}
R.~Frederix, S.~Frixione, V.~Hirschi, F.~Maltoni, R.~Pittau, et~al., {\it
  {Four-lepton production at hadron colliders: aMC@NLO predictions with
  theoretical uncertainties}},  {\em JHEP} {\bf 1202} (2012) 099,
  [\href{http://xxx.lanl.gov/abs/1110.4738}{{\tt arXiv:1110.4738}}].

\bibitem{Beringer:1900zz}
{\bf Particle Data Group}, J.~Beringer et~al., {\it {Review of Particle Physics
  (RPP)}},  {\em Phys.Rev.} {\bf D86} (2012) 010001.

\bibitem{Butterworth:2008iy}
J.~M. Butterworth, A.~R. Davison, M.~Rubin, and G.~P. Salam, {\it {Jet
  substructure as a new Higgs search channel at the LHC}},  {\em
  Phys.Rev.Lett.} {\bf 100} (2008) 242001,
  [\href{http://xxx.lanl.gov/abs/0802.2470}{{\tt arXiv:0802.2470}}].

\bibitem{Plehn:2009rk}
T.~Plehn, G.~P. Salam, and M.~Spannowsky, {\it {Fat Jets for a Light Higgs}},
  {\em Phys.Rev.Lett.} {\bf 104} (2010) 111801,
  [\href{http://xxx.lanl.gov/abs/0910.5472}{{\tt arXiv:0910.5472}}].

\bibitem{Catani:1997xc}
S.~Catani and B.~Webber, {\it {Infrared safe but infinite: Soft gluon
  divergences inside the physical region}},  {\em JHEP} {\bf 9710} (1997) 005,
  [\href{http://xxx.lanl.gov/abs/hep-ph/9710333}{{\tt hep-ph/9710333}}].

\end{thebibliography}\endgroup
\end{document}